\documentclass[12pt,a4paper]{article}
\usepackage[a4paper, left=3cm,right=3cm,top=3cm,bottom=3cm]{geometry}
\usepackage{float}
\usepackage{array}
\usepackage{lscape}
\usepackage{enumitem}
\usepackage{tikz}
\usetikzlibrary{arrows,shapes,positioning,shadows,trees, calc}
\usetikzlibrary{matrix,decorations.pathreplacing, calc, positioning,fit}
\usetikzlibrary{shapes.misc}
\usetikzlibrary{arrows.meta}
\tikzset{
	basic/.style  = {draw, text width=2cm, drop shadow, font=\sffamily, rectangle},
	root/.style   = {basic, rounded corners=2pt, thin, align=center,
		fill=green!30},
	level 2/.style = {basic, rounded corners=6pt, thin,align=center, fill=green!60,
		text width=4em},
	level 3/.style = {basic, thin, align=left, fill=pink!60, text width=1.5em}
}
\newcommand{\relation}[3]
{
	\draw (#3.south) -- +(0,-#1) -| ($ (#2.north) $)
}

\usepackage{graphicx}
\usepackage{amsmath}
\usepackage{amssymb}
\usepackage[linesnumbered,ruled]{algorithm2e}
\usepackage{cite}
\usepackage{float}
\usepackage{color}
\usepackage{geometry}
\usepackage{longtable}
\usepackage[authoryear,round]{natbib}
\usepackage{rotating}
\usepackage[nodisplayskipstretch]{setspace}
\usepackage{titletoc}
\usepackage{booktabs}
\usepackage{xcolor}

\usepackage{bm}

\newcommand{\avet}{\bm{a}}
\newcommand{\bvet}{\bm{b}}

\newcommand{\evet}{\bm{e}}

\newcommand{\yvet}{\bm{y}}

\newcommand{\Avet}{\bm{A}}

\newcommand{\Cvet}{\bm{C}}

\newcommand{\Gvet}{\bm{G}}

\newcommand{\Ivet}{\bm{I}}

\newcommand{\Mvet}{\bm{M}}

\newcommand{\Svet}{\bm{S}}

\newcommand{\Wvet}{\bm{W}}

\newcommand{\Zerovet}{\bm{0}}

\usepackage{caption}
\DeclareCaptionStyle{italic}[justification=centering]
 {labelfont={bf},textfont={it},labelsep=colon}
\usepackage[hidelinks]{hyperref}

\usepackage[bottom, flushmargin,hang]{footmisc}

\usepackage[affil-it, blocks]{authblk}
\newcommand{\email}[1]{\affil{Email: {\upshape\href{mailto:#1}{#1}}}}

\makeatletter
\renewenvironment{abstract}{%
    \if@twocolumn
      \section*{\abstractname}%
    \else %
      \begin{center}%
        {\bfseries \large\abstractname\vspace{\z@}}%
      \end{center}%
      \quotation
    \fi}
    {\if@twocolumn\else\endquotation\fi}
\makeatother

\title{\textbf{Exploiting Intraday Decompositions in Realized Volatility Forecasting:}\\
	A Forecast Reconciliation Approach}

\author{Massimiliano Caporin}
\affil{Department of Statistical Sciences, University of Padova}
\email{massimiliano.caporin@unipd.it}
\author{Tommaso Di Fonzo}
\affil{Department of Statistical Sciences, University of Padova}
\email{tommaso.difonzo@unipd.it}
\author{Daniele Girolimetto}
\affil{Department of Statistical Sciences, University of Padova}
\email{daniele.girolimetto@phd.unipd.it}

\begin{document}

\maketitle

\begin{abstract} 
\noindent We address the construction of Realized Variance ($RV$) forecasts by exploiting the hierarchical structure implicit in available decompositions of $RV$. By using data referred to the Dow Jones Industrial Average Index and to its constituents  we show that exploiting the informative content of hierarchies improves the forecast accuracy. Forecasting performance is evaluated out-of-sample based on the empirical $MSE$ and $QLIKE$ criteria as well as using the Model Confidence Set approach.
\end{abstract}

\newpage
\doublespacing
\section{Introduction}
\label{sec:intro}

Volatility forecasting has attracted a relevant amount of interest in the financial econometrics literature since the seminal contribution of \cite{Engle1982}. The reasons are well-known and ground on the importance of volatility in several areas, from risk management to asset allocation, from hedging to pricing. In the last two decades the interest has shifted from conditional variance models to the modeling and forecasting of Realized Variances, $RV$ \citep{Andersenetal2001a,Andersenetal2001b,Andersenetal2003}. In this case, starting from the work of \cite{Corsi2009}, based on the introduction of a simple specification capable of capturing the strong serial correlation of $RV$ sequences, several additional specifications have been introduced. These include models including price jumps in the variance dynamic \citep{Andersenetal2007a}, controlling for residual heteroskedasticity \citep{CorsiReno2012}, dealing with measurement errors \citep{Bollerslevetal2016}, disentangling the role of positive and negative returns \citep{PattonSheppard2015}, and considering a quantile-based intraday decomposition of $RV$ (\citealp{Bollerslevetal2022}). From a pure forecasting perspective, despite all models might provide statistical and/or economic advantages compared to simpler specifications, there is no clear evidence that a model clearly superior to all competitors exists \citep{Caporin2022}. 

A few contributions share a common feature from the modeling perspective: to predict the $RV$ they extract information from a decomposition of lagged $RV$. This holds, in particular, when separating the continuous and discontinuous variance components, as in \cite{Andersenetal2007a}, or when `Good and Bad' volatilities are used, as in \cite{PattonSheppard2015}, or when the two approaches are combined (\citealp{Caporin2022}), or finally when a more flexible decomposition according to conditional and time-varying intraday returns quantiles is considered (\citealp{Bollerslevetal2022}). In all of these cases, the decomposition provides what is known as a \textit{hierarchy} in the hierarchical forecasting literature (\citealp{Hyndman2011}), that is a structure in which an aggregate series (e.g., daily $RV$) can be seen at the top over its constituents series (e.g., intraday $RV$ decompositions).

Therefore, when dealing with intraday data, sometimes the observed returns may be grouped based on some criterion of similarity, such as the sign, or the occurrence in portions of the intraday returns density support. This decomposition may be employed in forecasting the daily $RV$, through segment-level forecasting %is then employed to represent the population
within each segment. Challenges associated with successfully applying intraday decompositions include how to create segments %when descriptive market information is lacking
and how to combine the segment-level $RV$ forecasts to recover a daily $RV$ forecast. The current paper proposes a method to exploit existing and to create new decompositions of the daily $RV$ based on high-frequency intraday data, create segment-level forecasts, and then combine these forecasts to improve the daily $RV$ forecasts. 
We present a combined-aggregative forecasting method for daily $RV$ that allows to obtain a global prognosis by summing up/combining the forecasts of the compounding individual components. We detail a bottom-up (indirect) and a regression-based forecast reconciliation (\citealp{Hyndman2011}, \citealp{Wickramasuriya2019}) approach, and study their forecasting performance \textit{vis-\`a-vis} the daily direct $RV$ forecasts produced by the classic $HAR$ model (\citealp{Corsi2009}), and two variants that take into account intraday $RV$ decompositions (\citealp{PattonSheppard2015}, \citealp{Bollerslevetal2022}).
At this end, we have devised a forecasting experiment to evaluate the new proposed forecasting approaches on the high-frequency data of a few assets.
The proposed method utilizes standard forecasting tools, but applies them in a unique combination that results in a higher level of daily $RV$ forecast accuracy than other traditional methods.

A common technique used to forecast an aggregate involve bottom-up method. This procedure starts from forecasting all bottom-level components and then obtaining the top-level forecast by simply summing these bottom-level forecasts. By contrast, the direct approach simply produces forecasts at the top level.
Nevertheless, realized volatility in different segments of the day usually have quite different patterns, hence the trivial approach of only forecasting the bottom-level series is unlikely to provide very accurate forecasts for the top-level series.
In addition, the different behaviour of $RV$ at different time periods, suggests %considering a
to group the observed volatility according to different time intervals. This may be considered either alone or in conjunction with other grouping schemes related to the nature of the volatility itself (i.e., `Good \& Bad' volatility, \citealp{PattonSheppard2015}).
This gives raise to a hierarchical/grouped time series, where daily $RV$ may be seen as the top-level series of a hierarchy, whose bottom level consists of the components obtained by crossing time periods and volatility decompositions.
In this case, besides bottom-up, another (hopefully) more accurate method for hierarchical forecasting is to independently generate $RV$ forecasts at all levels of the hierarchy. The advantage of independently generating the forecasts at each level is that each level can customize its forecasting model according to the varying characteristic of the $RV$ at its own level. Thus, such approach could provide more accurate top-level forecasts than traditional direct or bottom-up approaches. However, these independently-made forecasts have the undesirable consequence that the lower-level forecasts cannot add up exactly to the higher-level forecasts. Thus, it is necessary to carry out some adjustments to ensure that hierarchical forecasts meet the constraints introduced by the hierarchical structure in the same way as their measurement data, i.e., in each day the sum of the lower-level intraday $RV$ components forecasts should be equal to the higher-level daily $RV$ forecast.

The methodology we put forward comprises two steps:
\begin{enumerate}

\item \textit{Independent forecasting of daily $RV$ and its components}, generating the so called ‘base forecasts’. For the daily $RV$ series and for all its intraday components from a specific decomposition we issue a forecast as accurate as possible, using three different HAR-based forecasting models proposed in the literature. In general, the base forecasts are not coherent with the additive decomposition law linking the observed daily $RV$s to its observed intraday components.

\item \textit{Aggregation post-process}. We then combine these forecasting results to forecast the $h$-day-ahead $RV$, where $h \ge 1$ is the forecast horizon. In this paper, we consider a postprocess aggregation method suggested in the vast literature on regression-based cross-sectional forecast reconciliation (\citealp{Hyndman2011}, \citealp{Wickramasuriya2019}).

\end{enumerate}

We compare the accuracy of the aggregate (direct) forecasting with the disaggregate (indirect) bottom-up and the regression-based forecast-reconciliation approach for the daily $RV$ of the Dow Jones Industrial Average index and 26 of its constituents assets. Most of the existing studies on forecasting daily $RV$ did not consider possible hierarchical structures deriving from intraday decompositions of the $RV$, and often missed the coherent relationships between individual components. An exception is \cite{SohnLim2007}, who evaluated aggregate \textit{vs.} disaggregate forecasting of 30 simulated coherent components of the DJIA index based on the AR(2)-GARCH(1,1) model. However, the results of this experiment were quite inconclusive, as it was found
that the accuracy of the indirect forecasting method varied depending on the correlation degree of the coherent components.
Instead, and contrary to \cite{Sevy2014}, our results indicate that considering the various components of the realized variance do represent a significant improvement in an out-of-sample forecast evaluation framework.

The paper proceeds as follows. The forecast reconciliation methodology is reviewed in \autoref{FRrecap}. \autoref{sec:RVmm} briefly describes the volatility modeling and introduces the hierarchies. The empirical setup of the out-of-sample forecasting experiment is described in \autoref{sec:forecexperiment}, and \autoref{sec:results} shows the results. A robustness analysis is performed in \autoref{sec:robust}. Finally, conclusive remarks and indications for future developments are proposed in \autoref{sec:conclusion}.

\section{Forecast reconciliation: a recap}
\label{FRrecap}

\begin{figure}[!t]
	\centering
	%\resizebox{0.34\linewidth}{!}{
		\begin{tikzpicture}[baseline=(current  bounding  box.center),
			every node/.append style={shape=circle, 
				draw=black},
			minimum width=1.25cm,
			minimum height=1.25cm]
			
			\node at (0, 0) (AA){$AA$};
			\node at (2, 0) (AB){$AB$};
			\node at (4, 0) (BA){$BA$};
			\node at (6, 0) (BB){$BB$};
			\node at (8, 0) (BC){$BC$};
			\node at (1, 1.8) (A){$A$};
			\node at (6, 1.8) (B){$B$};
			\node at (4, 3.6) (T){Tot};
			
			\relation{0.2}{AA}{A};
			\relation{0.2}{AB}{A};
			\relation{0.2}{BA}{B};
			\relation{0.2}{BB}{B};
			\relation{0.2}{BC}{B};
			\relation{0.2}{A}{T};
			\relation{0.2}{B}{T};
		\end{tikzpicture}
	%}
	\caption{A simple three-level hierarchical structure.}
	\label{hts1}
\end{figure}
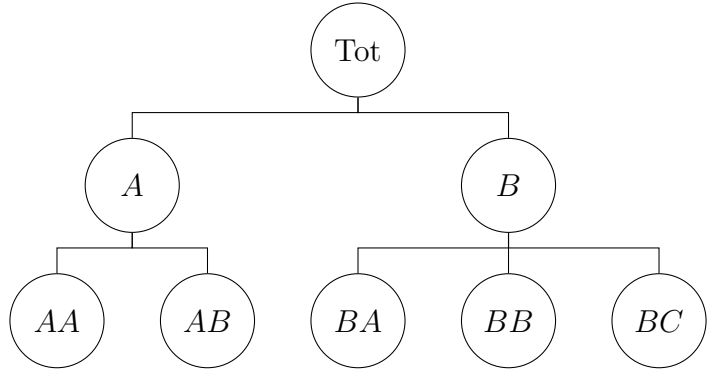

Forecast reconciliation is a post-forecasting process aimed to improve the quality of the {\em base} forecasts for a system of hierarchical/grouped, and more generally linearly constrained, time series  by exploiting the constraints that the series in the system must fulfil, whereas in general the base forecasts do not; see, among others, \cite{Hyndman2011} and \cite{Girolimetto2023}. Following \cite{Panagiotelis2021}, a linearly constrained time series $\yvet_t$ is defined as a $n$-dimensional time series such that all observed values $\yvet_1 , \ldots , \yvet_T$ and all future values $\yvet_{T+1}, \yvet_{T+2} , \ldots$ lie in the coherent linear subspace ${\cal{S}} \subset {\mathbb R}^{n}$, that is: $\yvet_t \in \cal{S}$, $\forall t$. In many cases, the linear constraints can be represented as a hierarchy, where the time series are linked through summation constraints. \autoref{hts1} shows an example of a hierarchical time series with eight variables and three levels: the top-variable at level 0, two variables (A and B) at level 1, and five variables at level 2 (AA, AB, BA, BB, BB, BC). The 3 aggregated upper time series %(uts) 
are linked to the bottom-level variables %(bts) 
through summation:
 $$
 \begin{aligned}
 	&y_{Tot, t} = y_{A,t} + y_{B,t}\\
 	&y_{A, t} = y_{AA,t} + y_{AB,t}\\
 	&y_{B, t} = y_{BA,t} + y_{BB,t} + y_{BC,t}\\
 \end{aligned} \qquad\forall t = 1,\dots, T .
 $$

The bottom-level series can be thought of as building blocks that cannot be obtained as sum of other series in the hierarchy, while all the series at upper levels can be expressed by appropriately summing part or all of them.
In details, let $\bvet_t$ and $\avet_t$ be the vectors of bottom level and upper level time series at time $t$, respectively. For example,
$\bvet_t = \left[y_{AA,t} \quad
y_{AB,t} \quad
y_{BA,t} \quad
y_{BA,t} \quad
y_{BA,t}\right]',$ %\qquad
$\avet_t = \left[y_{Tot,t} \quad
y_{A,t} \quad
y_{B,t}\right]'$.
Denoting by $\yvet_t$ the vector
$\yvet_t = \left[\avet_t' \; \; \bvet_t'\right]'$, 
the relationships linking bottom and upper time series can be equivalently expressed as:
\begin{equation}
\label{contemp}
\avet_t = \Avet\bvet_t, \quad \yvet_t = \Svet\bvet_t, \quad
\Cvet\yvet_t = \Zerovet_{(n_a \times 1)}, \quad    t=1,\ldots,T.
\end{equation}
where $\Avet$ is the $(n_a \times n_b)$ aggregation matrix, $\Svet = \left[\begin{array}{c}
	\Avet \\
	\Ivet_{n_b}
\end{array}\right]$ is the $(n \times n_b)$ structural matrix and $\Cvet = \left[\Ivet_{n_a}\quad -\Avet\right]$ is the $(n_a \times n)$ zero constraints matrix.
We call \emph{structural representation} of series $\yvet_t$ the formulation $\yvet_t = \Svet\bvet_t$, $t=1,\ldots, T$, and \emph{zero-constrained representation} of series $\yvet_t$ the equivalent expression $\Cvet\yvet_t = \Zerovet$, $t=1,\ldots, T$.

\begin{figure}[tb]
	\centering
	%\resizebox{0.34\linewidth}{!}{
		\begin{tikzpicture}[baseline=(current  bounding  box.center),
			every node/.append style={shape=circle, 
				draw=black},
			minimum width=1.25cm,
			minimum height=1.25cm]
			
			\node[fill = red, fill opacity = 0.2, text opacity = 1] at (0, 0) (AX1){$AX$};
			\node[fill = red, fill opacity = 0.2, text opacity = 1] at (2, 0) (BX1){$BX$};
			\node[fill = red, fill opacity = 0.2, text opacity = 1] at (4, 0) (AY1){$AY$};
			\node[fill = red, fill opacity = 0.2, text opacity = 1] at (6, 0) (BY1){$BY$};
			\node at (1, 1.8) (X){$X$};
			\node at (5, 1.8) (Y){$Y$};
			\node[fill = blue, fill opacity = 0.2, text opacity = 1] at (3, 3.6) (T1){$Tot$};
			
			\node[fill = red, fill opacity = 0.2, text opacity = 1] at (8, 0) (AX2){$AX$};
			\node[fill = red, fill opacity = 0.2, text opacity = 1] at (10, 0) (AY2){$AY$};
			\node[fill = red, fill opacity = 0.2, text opacity = 1] at (12, 0) (BX2){$BX$};
			\node[fill = red, fill opacity = 0.2, text opacity = 1] at (14, 0) (BY2){$BY$};
			\node[fill = green, fill opacity = 0.2, text opacity = 1] at (9, 1.8) (A){$A$};
			\node[fill = green, fill opacity = 0.2, text opacity = 1] at (13, 1.8) (B){$B$};
			\node[fill = blue, fill opacity = 0.2, text opacity = 1] at (11, 3.6) (T2){$Tot$};
			
			\relation{0.2}{AX1}{X};
			\relation{0.2}{BX1}{X};
			\relation{0.2}{AY1}{Y};
			\relation{0.2}{BY1}{Y};
			\relation{0.2}{X}{T1};
			\relation{0.2}{Y}{T1};
			
			\relation{0.2}{AX2}{A};
			\relation{0.2}{BX2}{B};
			\relation{0.2}{AY2}{A};
			\relation{0.2}{BY2}{B};
			\relation{0.2}{A}{T2};
			\relation{0.2}{B}{T2};
		\end{tikzpicture}
	%}
	\caption{A simple grouped structure.}
	\label{gts1}
\end{figure}
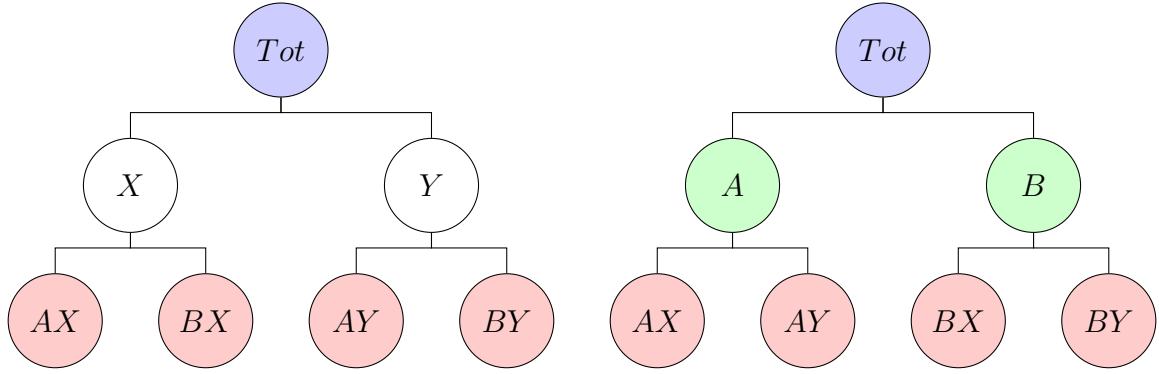

A linearly constrained time series formed by two or more hierarchical time series sharing the same top level series, and the same bottom level series, is called \textit{grouped time series} \citep{Hyndman2011}. An example is shown in Figure \ref{gts1}, where the Total variable can be described as two different hierarchies with intermediate variables $(X,Y)$ and $(A,B)$, respectively, which share the same four bottom-level variables $(AX,BX,AY,BY)$.
Provided matrix $\Avet$ is appropriately designed, the definitions of matrices $\Svet$ and $\Cvet$ remain unchanged.

Now, suppose we have the $(n \times 1)$ vector $\widehat{\yvet}_h$ of unbiased base forecasts for the $n$ variables of the linearly constrained series $\yvet_t$ for the forecast horizon $h$. If the base forecasts have been independently computed, generally, they do not fulfil the cross-sectional aggregation constraints, that is, $\Cvet\widehat{\yvet}_h \ne \Zerovet_{(n \times 1)}$. The aim of forecast reconciliation is to adjust the base forecast $\widehat{\yvet}_{h}$ by using a mapping $\psi: \mathbb{R}^{n} \rightarrow \mathcal{S}$ to obtain the reconciled forecast vector $\widetilde{\yvet}_{h} = \psi\left(\widehat{\yvet}_{h}\right)$, where $\widetilde{\yvet}_{h} \in \cal{S}$.
The mapping $\psi$ can be defined as a projection onto $\cal{S}$ %(\citealp{vanerven2015}, 
(\citealp{vanerven2015}, \citealp{Panagiotelis2021}, \citealp{DifonzoGirolimetto2023}):
%, given by
\begin{equation}
\label{eq:Mvet}
\widetilde{\yvet}_{h} = \Mvet \widehat{\yvet}_h,
\end{equation}
where $\Mvet = \Ivet_{n} - \Wvet\Cvet'\left(\Cvet\Wvet\Cvet'\right)^{-1}\Cvet$, with %a positive definite matrix 
$\Wvet$  error covariance matrix of the base forecasts $\widehat{\yvet}$.
Another way to obtain the reconciled forecasts is through the structural approach proposed by \cite{Hyndman2011}, such that
\begin{equation}\label{eq:SGy}
\widetilde{\yvet}_{h} = \Svet \Gvet \widehat{\yvet}_{h},
\end{equation}
where $\Gvet = (\Svet' \Wvet^{-1}\Svet)^{-1} \Svet'\Wvet^{-1}$, and it can be shown that $\Mvet = \Svet \Gvet$ (\citealp{Wickramasuriya2019}).
Several alternatives have been provided in the literature to approximate the covariance matrix $\Wvet$ \citep{Hyndman2011, Hyndman2016, Wickramasuriya2019}. In this work, we will consider the state of the art shrinkage covariance matrix approximation proposed by \citet{Wickramasuriya2019}, 
$$\Wvet = \hat{\lambda} \widehat{\Wvet}_{D} + (1 - \hat{\lambda})\widehat{\Wvet}_1 , $$
where
$\widehat{\Wvet}_1= \displaystyle\frac{1}{T}\sum_{t=1}^{T}\hat{\evet}_t\hat{\evet}_t'$ is the covariance matrix of the one-step ahead in-sample forecast errors ($\widehat{\evet}_t = \yvet_t - \widehat{\yvet}_t$, $t=1,\ldots,T$), $\widehat{\Wvet}_{D} = \Ivet_{n} \odot \widehat{\Wvet}_1$, and $\odot$ denotes the Hadamard product.

In the next section, after briefly reviewing the estimation of $RV$, we link the hierarchical forecasting literature to the $RV$ modeling one.

\section{$RV$ modeling: a hierarchical perspective}
\label{sec:RVmm}

The measurement of daily $RV$ builds on the availability of data at a frequency higher than the day. If we denote by $t=1,\ldots,T$, the daily time index, and by $i=1,2,\ldots N$, the intraday time index, the prices of a financial instrument observed in high frequency are denoted by $P_{i,t}$. From the prices we move to log-returns $r_{i,t}$ and to the estimation of $RV$ in a given day as follows:

\begin{equation}
RV_t = \sum_{i=1}^N \left[\log\left(P_{i,t}\right)-\log\left(P_{i-1,t}\right)\right]^2 = \sum_{i=1}^N r_{i,t}^2,
\end{equation} 

\noindent where prices at the intraday level are assumed to be observed on an equally spaced time grid (e.g., every minute), and for $i=1$ the lagged price corresponds to the opening price of the day, thus excluding the overnight return from the evaluation. The financial econometrics literature has extensively discussed the issue of estimation of $RV$ in the presence of microstructure noise and of price jumps; see, among many others, \cite{AitSahaliaJacod2014} and therein cited references. In this work, we refer to the simplest approach reported above. Moreover, as our final purpose is to adopt hierarchical forecast reconciliation approaches starting from the forecast of bottom time series, we do not consider the decomposition of $RV$ into its continuous and discontinuous component, since it is known that the discontinuous component is not predictable by means of relatively simple linear models; among the possible approaches, see \cite{Andersenetal2011} and \cite{AitSahaliaetal2015}.

As mentioned in the introduction, several authors have focused on decompositions of $RV$. The most known example is given by the use of signed variations, as in \cite{PattonSheppard2015}, such that $RV_t = SV_t^{+} + SV_t^{-}$, with
\begin{equation*}
	SV_t^{+} = \sum_{i=1}^N r_{i,t}^2 I\left(r_{i,t}\geq 0\right) \quad \text{and} \quad SV_t^{-} = \sum_{i=1}^N r_{i,t}^2 I\left(r_{i,t}< 0\right)
\end{equation*}
where $I\left(a\right)$ is an indicator function taking unit value when condition $a$ is true and zero otherwise. The signed variations are also know as \textit{Semi-Variances} (\citealp{Barndorffetal2010}), or as \textit{Good} ($SV_t^{+}$) and \textit{Bad} ($SV_t^{-}$) volatility, respectively, and separate the contribution to $RV$ coming from upside and downside price movements.

More recently, \cite{Bollerslevetal2022} introduced a quantile-based decomposition, generalizing the signed variation approach, $RV_t = \displaystyle\sum_{l=1}^p PV_t^{(l)}$, with
\begin{equation}
PV_t^{(l)}=\sum_{i=1}^N r_{i,t}^2 I\left(\mathcal{Q}_{r,t}\left(\alpha_{l-1}\right) < r_{i,t} \leq \mathcal{Q}_{r,t}\left(\alpha_{l}\right)\right),\nonumber
\end{equation}
where $\mathcal{Q}_{r,t}\left(\tau\right)=\sqrt{N^{-1}RV_t}\mathcal{Q}_{z,t}\left(\tau\right)$, $z^{i,t}=\displaystyle\frac{r_{i,t}}{\sqrt{N^{-1}RV_t}}$ is the standardized intraday return, $\mathcal{Q}_{z,t}\left(\tau\right)$ is the empirical $\tau$-quantile of the intraday standardized returns distribution in day $t$, $\mathcal{Q}_{r,t}\left(\alpha_0\right)=-\infty$ and $\mathcal{Q}_{r,t}\left(\alpha_p\right)=+\infty$, and $0<\alpha_1<\ldots<\alpha_{p-1}<1$ is a sequence of probabilities. The $PV^{(l)}_t$ components, also called \textit{Partial Variances}, allow separating the contribution to $RV$ coming from intraday returns according to both their sign and size.

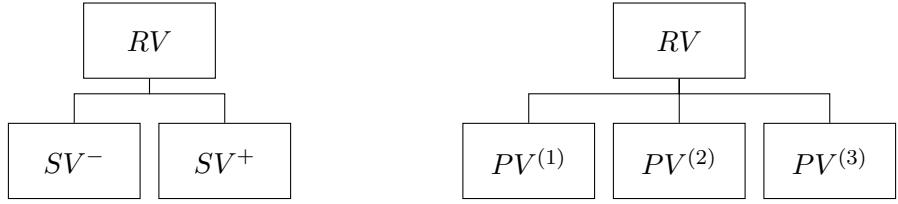
\begin{figure}[tb]
	\centering
	\begin{tikzpicture}[baseline=(current  bounding  box.center),
		rel/.append style={draw=black, font=\small,
			minimum width=1.75cm,
			minimum height=1cm},
		connection/.style ={inner sep =0, outer sep =0}]
		
		\node[rel] at (0, 0) (svm){$SV^{-}$};
		\node[rel] at (2, 0) (svp){$SV^{+}$};
		
		\node[rel] at (1, 1.6) (RV){$RV$};

		\relation{0.2}{svm}{RV};
		\relation{0.2}{svp}{RV};
	\end{tikzpicture}
	\hspace{2cm}
	\begin{tikzpicture}[baseline=(current  bounding  box.center),
		rel/.append style={draw=black, font=\small,
			minimum width=1.75cm,
			minimum height=1cm},
		connection/.style ={inner sep =0, outer sep =0}]
		
		\node[rel] at (0, 0) (pv1){$PV^{(1)}$};
		\node[rel] at (2, 0) (pv2){$PV^{(2)}$};
		\node[rel] at (4, 0) (pv3){$PV^{(3)}$};
		
		\node[rel] at (2, 1.6) (RV){$RV$};

		\relation{0.2}{pv1}{RV};
		\relation{0.2}{pv2}{RV};
		\relation{0.2}{pv3}{RV};
	\end{tikzpicture}
	\caption{Hierarchical representations of the Bad and Good (left, \citealp{PattonSheppard2015}) and $PV(3)$ (right, \citealp{Bollerslev2022}) decompositions of daily RV.}
	\label{fig:B&G-PV(3)}
\end{figure}

The concept of hierarchical time series in the framework of forecasting daily $RV$ using its intraday decompositions may be illustrated by considering the two simple hierarchies deriving from the intraday $RV$ decompositions reported above, i.e. \cite{PattonSheppard2015} and \cite{Bollerslev2022}, the last in the simple case with $p=2$. Figure \ref{fig:B&G-PV(3)} provides a graphical representation of the two hierarchies.

To illustrate the advantage and flexibility of forecast reconciliation approaches in the $RV$ context, we provide a more general form of temporal and threshold-based decomposition of daily $RV$. 
The use of quantiles computed from intraday returns is, in practice, a special case of a grouping of returns according to pre-defined, possibly time-varying thresholds, with general representation given by:
\begin{equation}
	z_{l,t} = \displaystyle\sum_{t=1}^N r_{i,t}^2 I\left(c_{t,l-1} < r_{i,t} \le c_{t,l}\right), \quad l=1,\ldots,p ,
\end{equation}
where $c_{t,0}=-\infty$ and $c_{t,p}=+\infty$.

Differently, by exploiting the availability of information distributed over a range of minutes within a given day, we might group the intraday returns according to a temporal scheme
\begin{equation}
w_{k,t} = \displaystyle\sum_{i=m(k-1)+1}^{mk} r_{i,t}^2 , \quad k=1,\ldots,\frac{N}{m},
\end{equation}
which is equivalent to
\begin{equation}
	w_{k,t} = \displaystyle\sum_{t=1}^N r_{i,t}^2 I\left(m(k-1) < i \le mk\right), \quad k=1,\ldots,\frac{N}{m}.
\end{equation}

The threshold- and time-based decompositions might be combined giving rise to the most disaggregated (bottom level) time series, defined as
\begin{equation}
x_{l,k,t} = \displaystyle\sum_{t=1}^N r_{i,t}^2
I\left[\left(c_{t,l-1} < r_{i,t} \le c_{t,l}\right)
\cap
\left(m(k-1) < i \le mk\right)\right],
\quad 
\begin{array}{l}
	l=1,\ldots,p \\ 
	k=1,\ldots,\frac{N}{m}
\end{array},
\end{equation}
or
\begin{equation}
x_{l,k,t} = \displaystyle\sum_{i=m(k-1)+1}^{mk} r_{i,t}^2
I\left(c_{t,l-1} < r_{i,t} \le c_{t,l}\right),\quad 
\begin{array}{l}
	l=1,\ldots,p \\ 
	k=1,\ldots,\frac{N}{m}
\end{array},
\end{equation}
For example, assuming a day consisting of 6.5 hours, with data available at the 1-minute frequency ($N=390$), setting $p=3$ and $m=78$, we have the following decompositions of $RV$:
\begin{center}
	\begin{tabular}{c|ccccc}
		$RV$ & $w_1$ & $w_2$ & $w_3$ & $w_4$ & $w_5$ \\
		\midrule
		$z_1$ & $x_{1,1}$ & $x_{1,2}$ & $x_{1,3}$ & $x_{1,4}$ & $x_{1,5}$ \\
		$z_2$ & $x_{2,1}$ & $x_{2,2}$ & $x_{2,3}$ & $x_{2,4}$ & $x_{2,5}$ \\
		$z_3$ & $x_{3,1}$ & $x_{3,2}$ & $x_{3,3}$ & $x_{3,4}$ & $x_{3,5}$ \\
	\end{tabular}
\end{center}
where the threshold-based $z_l = \displaystyle\sum_{k=1}^{5} x_{l,k}$, $l=1, \ldots, 3$, and temporal-based $w_k = \displaystyle\sum_{l=1}^{3} x_{l,k}$, $k=1,\ldots,5$, decompositions represent the marginals of a combined and richer decomposition. We also note that $RV = \displaystyle\sum_{k=1}^{5} w_k = \displaystyle\sum_{l=1}^{3} z_l =
\displaystyle\sum_{l=1}^{3}\displaystyle\sum_{k=1}^{5} x_{l,k}$.

\begin{table}[t]
	\centering
	\footnotesize
	\begin{tabular}[t]{l|cc}
		\toprule
		& \multicolumn{2}{c}{Semi-Variances}\\
		Temporal & $r_{t,i} < 0$ & $r_{t,i} \ge 0$  \\
		decomposition & $SV^{-}$          & $SV^{+}$ \\
		\midrule
		T1: minutes    1-78 & T1$SV^{-}$ & T1$SV^{+}$ \\
		T2: minutes  79-156 & T2$SV^{-}$ & T2$SV^{+}$ \\
		T3: minutes 157-234 & T3$SV^{-}$ & T3$SV^{+}$ \\
		T4: minutes 235-312 & T4$SV^{-}$ & T4$SV^{+}$ \\
		T5: minutes 313-390 & T5$SV^{-}$ & T5$SV^{+}$ \\
		\bottomrule
	\end{tabular}
	\caption{The ten bottom variables from the time-by-`Good \& Bad' volatility decompositions.}
	\label{tab:TSV}
\end{table}

In the following empirical analyses, we will make use of the hierarchies/groupings generated by crossing a temporal decomposition in five non-overlapping intervals of 78 minutes each, with either (ii) the dichotomous intraday decomposition in `Good and Bad' volatility (\citealp{PattonSheppard2015}, see Table \ref{tab:TSV} and Appendices 2 and 3), or (ii) a quantile-based decomposition with $p=3$,  where, on the basis of the results found by \cite{Bollerslevetal2022}, the quantile thresholds are exogenously fixed at 10\% and 75\%, respectively (see Table \ref{tab:TPV} and Appendices 2 and 3). The Appendices report the graphical and structural representations of the hierarchies.

\begin{table}[t]
	\centering
	\footnotesize
	\begin{tabular}[t]{l|ccc}
		\toprule
		& \multicolumn{3}{c}{Quantile-based decomposition}\\
		Temporal & $r_{t,i} \le Q(10\%)$ & $Q(10\%) < r_{t,i} \le Q(75\%)$ & $r_{t,i} > Q(75\%)$ \\
		decomposition & $PV^{(1)}$          & $PV^{(2)}$ & $PV^{(3)}$ \\
		\midrule
		T1: minutes    1-78 & T1$PV^{(1)}$ & T1$PV^{(2)}$ & T1$PV^{(3)}$ \\
		T2: minutes  79-156 & T2$PV^{(1)}$ & T2$PV^{(2)}$ & T2$PV^{(3)}$ \\
		T3: minutes 157-234 & T3$PV^{(1)}$ & T3$PV^{(2)}$ & T3$PV^{(3)}$ \\
		T4: minutes 235-312 & T4$PV^{(1)}$ & T4$PV^{(2)}$ & T4$PV^{(3)}$ \\
		T5: minutes 313-390 & T5$PV^{(1)}$ & T5$PV^{(2)}$ & T5$PV^{(3)}$ \\
		\bottomrule
	\end{tabular}
	\caption{The fifteen bottom variables from the time-by-quantile daily decompositions according to $PV(3)$.}
	\label{tab:TPV}
\end{table}

%\clearpage

\section{The empirical setup}
\label{sec:forecexperiment}

\subsection{Data description and analysis}
\label{sec:data}

We evaluate the impact of forecast reconciliation in forecasting daily $RV$ of individual stocks included in the Dow Jones Industrial Average (DJIA) index, from the beginning of January 2003 to the end of June 2022. We use price data at the 1-minute frequency, adjusted for splits and dividends. We consider prices recorded from 9:00 AM to 3:59 PM (time identifies the start of each intraday interval), obtaining 390 observations per day. Our dataset includes 4,908 full days excluding weekends, holidays and closed market days. The data have been recovered from Kibot.com.\footnote{The quality of data from Kibot.com is comparable to that of NYSE TAQ data. A comparison on a selected equity is available from authors upon request.} We consider the 26 stocks whose data are available to us for the entire sample, denoted by the following tickers:\footnote{Appendix A1 contains a summary description of the data used in the forecasting experiment.}
AAPL, AMGN, AXP, BA, CAT, CSCO, CVX, DIS, GS, HD, HON, IBM, INTC, JNJ, JPM, KO, MCD, MMM, MRK, MSFT, NKE, PG, UNH, VZ, WBA, and WMT.
The use of DJIA constituents is in line with the choice made by \cite{Bollerslevetal2022} and allows dealing with the possible presence of large amount of zeros at the intraday level. In fact, for those highly liquid stocks, the presence of zeros is extremely limited. 

Starting from the 1-minute data, we estimate the daily $RV$, and the decompositions (hierarchies) we previously mentioned. First of all, we decompose the $RV$ into the Good and the Bad components, following \cite{PattonSheppard2015}; this gives a hierarchy with two bottom series. Second, following \cite{Bollerslevetal2022}, we decompose $RV$ into the Partial Variances $PV^{(g)}$. Differently from the authors, we do not optimally select the quantiles used for the decomposition, nor we allow for a time-change in the quantile. On the contrary, building on the evidence in \cite{Bollerslevetal2022}, and in particular on the values in Table 2 of their paper, we select the $PV(3)$ decomposition with fixed quantiles set at the 10\% and 75\% thresholds. This gives a hierarchy with three bottom series, and allows us to simplify the treatment and the following analyses.
Third, we apply both the Good and Bad and $PV(3)$ decompositions on sub-samples of the day. We first divide the entire day in 5 sub-intervals of length equal to 78 observations (minutes), and in each sub-sample we apply either the Good and Bad or the $PV(3)$ decompositions. We note that this gives, overall, 10 bottom series in the former case, and 15 in the latter. In addition, we do have two possible intermediate aggregations, by temporal sub-sample, or by volatility (either $SV$ or $PV(3)$) components. The hierarchies/groupings in this case are thus much richer than if only volatility-based decompositions are used (see the Appendix for additional details and results).

\subsection{Base forecasts: direct forecasts from benchmark models and intraday components' forecasts}

%In this paper, we investigate alternative ways of leveraging intraday $RV$ decompositions in forecasting daily $RV$. The key question of this study is whether it is beneficial to model and forecast daily $RV$ at the sub-component level, thus exploiting the informative content (from a forecasting point of view) of bottom time series, or whether a direct strategy should be preferred. The latter refers to the prediction of $RV$ by directly modeling it, even when the explanatory variables include a decomposition of the $RV$ itself. Differently, indirect (bottom-up) forecasting is based on the aggregation of models fit on the bottom series, whose forecasts are then aggregated to recover the prediction of the top series. Forecast reconciliation adds a further element, by restoring the aggregation constraints linking the bottom, possibly the intermediate, and the top-level time series. The idea is that an appropriate `imposition' to the forecasts of the same constraints valid for the observed data should improve the overall forecasting accuracy.

In the past years, the subject of the comparison of the forecast accuracy of aggregating disaggregate forecasts versus forecasts based on aggregated data has received attention in different fields, as macro-economic (\citealp{Marcellino2003}, \citealp{Frale2011}, \citealp{Poncela2014}, \citealp{Grassi2015}), demand (\citealp{Petropoulos2014}, \citealp{Mircetic2022}), and energy (\citealp{Silva2018}, \citealp{Wang2021}) forecasting. However, as far as we know, a detailed comparison of direct, indirect (bottom-up) and combination (forecast reconciliation) procedures for daily $RV$ forecasting has not been provided yet.

At this end,
%To answer this question, 
we conduct an out-of-sample forecasting experiment where we compare direct daily-$RV$ forecasts with three reference models proposed, respectively, by \cite{Corsi2009}, \cite{PattonSheppard2015}, and \cite{Bollerslevetal2022}, two indirect forecasts obtained through simple bottom-up of the HAR forecasts of either semi-variances (\citealp{Barndorffetal2010}) or partial-variances (\citealp {Bollerslevetal2022}) components, and finally the daily forecasts of $RV$ obtained through forecast reconciliation of both the aggregate (daily $RV$) and disaggregate (corresponding components of the daily $RV$) forecasts. %Anticipating our findings, we will show that both bottom-up and regression-based reconciliation procedures (\citealp{Wickramasuriya2019}) perform relatively well, mostly in the $PV(3)$ model by \cite{Bollerslevetal2022} framework.

The modeling strategy we adopt for each of the bottom and top time series, and that we will use to produce direct and base forecasts, is very simple. As our purpose is to introduce the use of forecast reconciliation tools in the prediction of $RV$, and not to identify the best forecasting univariate model, we fit on all (possibly disaggregate) time series the HAR model of \cite{Corsi2009}. Let $x_t$ be a generic time series, that is, either $RV_t$ or one of the bottom series according to one of the hierarchies previously introduced. We model $x_t$ as follows:
\begin{equation}
	x_t = \beta_0 + \beta_D x_{t-1} + \beta_W x_{t-1:t-5} +\beta_M x_{t-1:t-22} + \varepsilon_t ,
	\label{eq:HAR}
\end{equation}
where $x_{t-1:t-m}=\displaystyle\frac{1}{m} \sum_{j=1}^m x_{t-m}$ and $\varepsilon_t$ is an innovation term. Parameters refer to the intercept ($\beta_0$) and to the daily, weekly and monthly effects ($\beta_D$, $\beta_W$, and $\beta_M$, respectively). Parameter estimation is based on least squares and we adopt robust standard errors to to be coherent with the \textit{volatility-of-volatility} effect \citep{Corsietal2008}. Further, we do not consider the modeling of logarithms of $RV_t$ leaving to future researches the generalization of our approach along this line.\footnote{We stress the use of the logarithmic transformation of $RV$ sensibly impacts on the aggregation constraints, with the need of moving toward probabilistic hierarchical forecasting approaches.}

For the top-level variable (i.e., daily $RV$) forecasts, we consider other two reference models proposed by \cite{PattonSheppard2015} and \cite{Bollerslevetal2022} to account for, respectively, Good and Bad volatility and Partial Variances.
%, respectively:
%\begin{eqnarray}
%	&& RV_t = \beta_0 + \beta_{D^+} SV_{t-1}^{+} + \beta_{D^-} SV_{t-1}^{-} + \beta_W RV_{t-1:t-5} +\beta_M RV_{t-1:t-22} + \varepsilon_t, \nonumber\\
%	&& RV_t = \beta_0 + \beta_{D^{(1)}} PV_{t-1}^{(1)} + \beta_{D^{(2)}} PV_{t-1}^{(2)} + \beta_{D^{(3)}} PV_{t-1}^{(3)} + \beta_W RV_{t-1:t-5} +\beta_M RV_{t-1:t-22} + \varepsilon_t, \nonumber
%\end{eqnarray} 
%where, coherently with the above-cited contributions, we introduce the $RV$ decomposition only in the daily lag,  $RV_{t-1:t-m}=\displaystyle\frac{1}{m} \sum_{j=1}^m RV_{t-m}$, and $\varepsilon_t$ is an innovation term.
%Before moving to the empirical evidence, the following Section provides additional details on our empirical framework.
%In our forecast setting, we take into account three different modeling strategies. First, we obtain direct forecasts for the daily $RV$ according to the benchmark model by \cite{Corsi2009}, and its developments by \cite{PattonSheppard2015} and \cite{Bollerslevetal2022}.
In this last case, for simplicity, we consider an \textit{ex-ante} choice of the two quantiles defining the three partial-variances decomposition (see Table \ref{tab:mod}).

\begin{table}[!t]
\centering
	\setlength{\tabcolsep}{1pt}
	\begin{tabular}{rcl}
	\toprule
	\addlinespace
	\multicolumn{3}{c}{\textbf{Models for daily $RV$ (direct and base forecasts)}}\\
	\addlinespace[0.25cm]
	\multicolumn{3}{l}{$HAR$: \textit{Heterogeneous AutoRegressive model}}\\
	$RV_t$ & $=$ & $\alpha_0 + \alpha_D RV_{t-1} + \alpha_W RV_{t-1:t-5} + \alpha_M RV_{t-1:t-22} + \varepsilon_t^{HAR}$ \\
	\addlinespace[0.25cm]
	\multicolumn{3}{l}{$SV$: \textit{Semi-Variances Heterogeneous AutoRegressive model}}\\
	$RV_t$ & $=$ & $\beta_0 + \beta_D^+ SV_{t-1}^+ + \beta_D^- SV_{t-1}^- + \beta_W RV_{t-1:t-5} + \beta_M RV_{t-1:t-22} + \varepsilon_t^{SHAR}$ \\
	\addlinespace[0.25cm]
	\multicolumn{3}{l}{$PV(3)$: \textit{Partial-Variances Heterogeneous AutoRegressive model}}\\
	$RV_t$ & $=$ & $\gamma_0 + \displaystyle\sum_{l=1}^{3}\gamma_D^{(j)} PV^{(l)}_{t-1} + \gamma_W RV_{t-1:t-5} + \gamma_M RV_{t-1:t-22} + \varepsilon_t^{PV_3}$ \\
		\addlinespace[0.25cm]
 \midrule
 	\addlinespace
	\multicolumn{3}{c}{\textbf{$HAR-$type models for intraday $RV$ decompositions (base forecasts)}}\\
		\addlinespace[0.25cm]
	\multicolumn{3}{l}{\textit{Semi-variances decomposition}}\\
	$SV_t^+$ & $=$ & $\delta_0^+ + \delta_D^+ SV_{t-1}^+ + \delta_W^+ SV_{t-1:t-5}^+ + \delta_M^+ SV_{t-1:t-22}^+ + \varepsilon_t^{SHAR^+}$ \\
 	\addlinespace
	$SV_t^-$ & $=$ & $\delta_0^- + \delta_D^- SV_{t-1}^- + \delta_W^- SV_{t-1:t-5}^- + \delta_M^- SV_{t-1:t-22}^- + \varepsilon_t^{SHAR^-}$ \\
	\addlinespace[0.25cm]
	\multicolumn{3}{l}{\textit{Partial-variances decomposition} ($l=1,2,3$)}\\
	$PV^{(l)}_t$ & $=$ & $\theta_0^{(l)} + \theta_D^{(l)} PV^{(1)}_{t-1} + \theta_W^{(l)} PV^{(l)}_{t-1:t-5} + \theta_M^{(l)} PV^{(l)}_{t-1:t-22} + \varepsilon_t^{PV(l)}$\\
	\addlinespace[0.25cm]
	\multicolumn{3}{l}{\textit{Time-variances decomposition} ($j=1,2,3,4,5$)}\\
	$T_{j,t}$ & $=$ & $\eta_0^j + \eta_1^j T_{j,t-1} + \eta_2^j T_{j,t-1:t-5} + \eta_3^j T_{j,t-1:t-22} + \varepsilon_t^{T(j)}$\\
	\addlinespace[0.25cm]
	\multicolumn{3}{l}{\textit{Time- and semi-variances decomposition} ($j=1,2,3,4,5$)}\\
	$T_jSV^{-}_t$ & $=$ & $\lambda_0^{j-} + \lambda_1^{j-} T_jSV^{-}_{t-1} + \lambda_2^{j-} T_jSV^{-}_{t-1:t-5} + \lambda_3^{j-} T_jSV^{-}_{t-1:t-22} + \varepsilon_t^{T(j)SV^{-}}$ \\
	\addlinespace
	$T_jSV^{+}_t$ & $=$ & $\lambda_0^{j+} + \lambda_1^{j+} T_jSV^{+}_{t-1} + \lambda_2^{j+} T_jSV^{-}_{t-1:t-5} + \lambda_3^{j+} T_jSV^{+}_{t-1:t-22} + \varepsilon_t^{T(j)SV^{+}}$ \\
	\addlinespace[0.25cm]
	\multicolumn{3}{l}{\textit{Time- and partial-variances decomposition} ($j=1,2,3,4,5$, $l=1,2,3$)}\\
	$\qquad T_jPV^{(l)}_t$ & $=$ & $\eta_0^{j,l} + \eta_1^{j,l} T_jPV^{(l)}_{t-1} + \eta_2^{j,l} T_jPV^{(l)}_{t-1:t-5} + \eta_3^{j,l} T_jPV^{(l)}_{t-1:t-22} + \varepsilon_t^{T(j)PV^{(l)}}\qquad $ \\
		\addlinespace[0.25cm]
 \bottomrule
\end{tabular}
\caption{Models used to produce daily (direct and base) $RV$ forecasts, and base forecasts of intraday $RV$ decompositions according to either semi-variances or partial-variances, alone or with time-groupings of non-overlapping 78 consecutive minutes intervals.}
\label{tab:mod}
\end{table}

We obtain daily forecasts for the time series of the daily semi-variances according to the `Good \& Bad' (\citealp{Barndorffetal2010}, \citealp{PattonSheppard2015}) and to the $PV(3)$ decompositions (\citealp{Bollerslevetal2022}), then we apply a simple bottom-up procedure to compute indirect forecasts of the daily $RV$.
Finally, the forecasts obtained in the two previous steps are combined through the forecast reconciliation approach proposed by \cite{Wickramasuriya2019} (see also \citealp{Hyndman2011}), which is a regression-based forecast combination approach exploiting the simple hierarchical structure of the two considered decomposition settings.
%\noindent In this work, we start by considering the very simple hierarchies induced by (i) the `Bad \& Good' decomposition (SSV), and (ii) the Power Variances with two quantile thresholds fixed at 10\% and 75\%, respectively (SPV(3))
The competing forecasting approaches, and the corresponding acronyms, are the following ones (reported for a one-period forecast horizon):
\begin{itemize}[nosep, label = {}, leftmargin = 0cm]
	\item \textit{Direct forecasting procedures}
	\begin{itemize}
		\item[-] $HAR$: $\widehat{RV}_{t+1}^{HAR}$ \citep{Corsi2009};
		\item[-] $SV$: $\widehat{RV}_{t+1}^{SV}$ \citep{PattonSheppard2015};
		\item[-] $PV(3)$: $\widehat{RV}_{t+1}^{PV(3)}$ \citep{Bollerslevetal2022};
	\end{itemize}
	\item \textit{Indirect forecasting procedures (bottom-up)}
	\begin{itemize}
		\item[-] $SV_{bu}$: $\widehat{RV}_{t+1}^{SV_{bu}} = \widehat{SV}_{t+1}^+ + \widehat{SV}_{t+1}^-$; 
		\item[-] $PV(3)_{bu}$: $\widehat{RV}_{t+1}^{PV(3)_{bu}} = \widehat{PV}^{(1)}_{t+1} + \widehat{PV}^{(2)}_{t+1} + \widehat{PV}^{(3)}_{t+1}$;
	\end{itemize}
	\item \textit{Forecast reconciliation procedures}
	\begin{itemize}
		\item[-] $SV_{shr}$: $\widehat{RV}_{t+1}^{SV_{shr}} = f(\widehat{RV}_{t+1}^{SV}, \widehat{SV}_{t+1}^+, \widehat{SV}_{t+1}^-)$;
		\item[-] $PV(3)_{shr}$: $\widehat{RV}_{t+1}^{PV(3)_{shr}} = f(\widehat{RV}_{t+1}^{PV_3}, \widehat{PV^{(1)}}_{t+1}, \widehat{PV^{(2)}}_{t+1}, \widehat{PV^{(3)}}_{t+1})$.
	\end{itemize}
\end{itemize}

We will always use the above-reported acronyms independently from the forecast horizon we consider. Coherently with the common practice, see, for instance, \citep{PattonSheppard2015}, when the forecast horizon differs from 1 and become $h$, we set the dependent variable of our model to the $h-$period average cumulative value.\footnote{In this case, equation \ref{eq:HAR} becomes $x_{t+h-1:t} = \beta_0 + \beta_D x_{t-1} + \beta_W x_{t-1:t-5} +\beta_M x_{t-1:t-22} + \varepsilon_t$, with $x_{t+h-1:t}$ being the average of $x_{t+i}$ for $i=0,1,\ldots h-1.$}

\subsection{Out-of-sample forecast evaluation}

We perform a fixed length rolling window forecasting experiment on the DJIA series and 26 individual stocks previously mentioned. The first training set spans the period January 2, 2003 - December 29, 2006 (1,007 days). From each training set three direct multistep forecasts
%\footnote{The direct method involves developing a separate model for each forecast time step. In the case of predicting the $RV$ for the next one, five, and twenty-two  days (i.e., $h=1,5,22$), three separate models for predicting the $RV$ on day 1, the sum of five consecutive $RV$'s on week 1, and the sum of twenty-two consecutive $RV$'s on month 1, are used.} 
for, respectively, one-, five- and twenty-two-steps (day) ahead are computed, and this is done for all the time series components of the various hierarchies defined by the time-and/or-quantile-based $RV_t$ intraday decompositions.

The base forecasts of the top-level series in each hierarchy ($RV_t$) are obtained according to the $HAR$, $SV$ and $PV(3)$ models, respectively. The base forecasts of either the semi- or partial-variances series forming each hierarchy are obtained using appropriately adapted $HAR$ models. The base forecasts are then reconciled through the MinT-shr approach (\citealp{Wickramasuriya2019}) using the \texttt{R} package \texttt{FoReco} (\citealp{FoReco2023}). The point forecast accuracy of daily $RV_t$ is evaluated using the Mean Square Error ($MSE$), and the $QLIKE$ index (\citealp{Patton2011a}):\footnote{The $QLIKE$ index is computed as average of a simple modification of the familiar Gaussian log-likelihood loss function, which belongs to the family of robust and homogeneous loss functions defined by \cite{Patton2011a}, with parameter $b=-2$. The modification is such that the index amounts to zero when $\widehat{RV}_t = RV_t$, that is, the daily observed $RV$ is forecast without error (\citealp{PattonSheppard2009}, \citealp{Patton2011b}). It gives asymmetric weights to the forecast errors, so that underestimating the $RV$ is more important than overestimating.}
\begin{equation}
	\label{MSE_QLIKE}
	\begin{array}{rcl}
		MSE & = & 
		\displaystyle\frac{1}{|\mathcal{S}|}
		\sum_{t=1}^{|\mathcal{S}|} \left(\widehat{RV}_t - RV_t\right)^2 \\
		%MAE  & = & \displaystyle\frac{1}{|\mathcal{S}|}
		%	\sum_{t=1}^{|\mathcal{S}|}\left|\widehat{RV}_t - RV _t\right| \\
		QLIKE & = &
		\displaystyle\frac{1}{|\mathcal{S}|}\sum_{t=1}^{|\mathcal{S}|}
		\left(\frac{RV_t}{\widehat{RV}_t} - \displaystyle\frac{\log RV_t}{\log\widehat{RV}_t} - 1\right)
		,
	\end{array}
\end{equation}
where $\widehat{RV}_t$ and $|\mathcal{S}|$ denote the one-step-ahead forecast and the number of days in the test set, respectively.
Both $MSE$ and $QLIKE$ belong to the family of loss functions of \cite{Patton2011a}, that are robust to the noise in the volatility proxy. We consider the $MSE$ and $QLIKE$ ratios, defined respectively as
\begin{equation}
	\label{MSEratio}
	rMSE = \displaystyle\frac{MSE_i}{MSE_{HAR}} \qquad
	rQLIKE = \displaystyle\frac{QLIKE_i}{QLIKE_{HAR}} ,
\end{equation}
where $MSE_i$ ($QLIKE_i$) is defined as the forecast $MSE$ ($QLIKE$) over the out-of-sample period of any our competing models, and $MSE_{HAR}$ ($QLIKE_{HAR}$) is the respective value of the $HAR$ benchmark model. The values less than 1 are associated with the superior forecast ability of the proposed model, and vice versa.

In order to examine the advantages of the individual $HAR$ models and the $HAR$ models with forecast reconciliation methods over the $HAR$ benchmark model, we then employ the \cite{DieboldMariano1995} test (DM test) to investigate the null hypothesis of equal predict accuracy (EPA) where the $HAR$ model is used as a benchmark.

Finally, we utilize the Model Confidence Set (MCS) approach developed by Hansen et al. (2011) to compare the point forecast accuracy between the direct daily forecasts and the reconciliation-based forecasts using intraday decompositions of $RV$. Given a set of candidate forecast models, $\mathcal{M}_0$, the goal of the MCS procedure is to identify the MCS $\hat{\mathcal{M}}^*_{1-\alpha} \subset \mathcal{M}_0$, which is the set of the models that contains the ``best'' forecast model given a level of conﬁdence $\alpha$. The MCS procedures start with the full set of models
$\mathcal{M} = \mathcal{M}_0 = \left\{1, \ldots, m_0\right\}$
and repeatedly test the null hypothesis of EPA:
\begin{equation}
	H_{0,\mathcal{M}}: E\left(d_{ij,t}\right) = 0 , \quad \forall i,j \in \mathcal{M}
\end{equation}
where $d_{ij,t} = \mathcal{L}_{i,t} - \mathcal{L}_{j,t}$ is the loss diﬀerential between models $i$ and $j$ in the set\footnote{The loss function $\mathcal{L}$ is either $MSE$ or $QLIKE$.}.
The MCS procedure sequentially eliminates the worst performance model from $\mathcal{M}$, as long as the null is rejected at the significance level of $\alpha$. This trimming of models is repeated until the null is not rejected any longer, and the surviving set of models form the MCS, $\widehat{\mathcal{M}}^*_{1 - \alpha}$. If a fixed significance level of $\alpha$ is used at each step, $\widehat{\mathcal{M}}^*_{1 - \alpha}$ contains the best model from $\widehat{\mathcal{M}}$ with $(1 - \alpha)$ confidence.

\cite{Hansenetal2011} present three different types of statistics for testing the EPA hypothesis. We employ the MCS procedure based on the $t$-statistics
\begin{equation}
	t_{ij} = \displaystyle\frac{\overline{d}_{ij}}{\sqrt{\widehat{var}\left(\overline{d}_{ij}\right)}} \quad \text{for } i,j \in \mathcal{M} ,
\end{equation}
where $\overline{d}_{ij} = \displaystyle\frac{1}{T_h}\sum_{t=N}^{T-h}d_{ij,t}$.
The quantity
%$t$-statistics,
$t_{ij}$, provides scaled information on the average difference in the point forecast quality of models $i$ and $j$.
$\widehat{var}\left(\overline{d}_{ij}\right)$ is an estimate of $var\left(\overline{d}_{ij}\right)$, obtained by using the stationary block bootstrap of \cite{PolitisRomano1994} following \cite{Hansenetal2011}. The range statistics $T_R$ %and the semi-quadratic statistics, $T_{SQ}$ are 
is given by
\begin{equation}
	\begin{array}{rclcl}
		T_R & = & \max_{i,j \in \mathcal{M}} |t_{ij}| & = & \max_{i,j \in \mathcal{M}} \displaystyle\frac{|\overline{d}_{ij}|}{\sqrt{\widehat{var}\left(\overline{d}_{ij}\right)}}\\
%		T_{SQ} & = & \displaystyle\sum_{i,j \in \mathcal{M}}t_{ij}^2 & = & \sum_{i,j \in \mathcal{M}} \frac{\left(\overline{d}_{ij}\right)^2}{\widehat{var}\left(\overline{d}_{ij}\right)}
	\end{array} .
\end{equation}
The MCS procedure assigns $p$-values to each model in the initial set. For a given model $i \in \mathcal{M}$, the MCS $p$-value, $\hat{p}_i$, is the threshold confidence level that determines whether the model belongs to the MCS. It holds that $\widehat{\mathcal{M}}^*_{1 - \alpha}$ if and only if $\hat{p}_i \ge \alpha$.

\section{Does forecast reconciliation help in $RV$ forecasting?}
\label{sec:results}

Our final purpose is to answer the following question: when `volatility-based' decompositions of the daily realized volatility are available, does considering forecast reconciliation significantly improve the forecast accuracy of daily $RV$ compared to the benchmark $HAR$-type models by \cite{Corsi2009}, \cite{PattonSheppard2015} and \cite{Bollerslevetal2022}?

We start by evaluating the direct forecasts accuracy for the DJIA using the $MSE$ and $QLIKE$ ratios (Table \ref{tab:geomean_2007-2022_new}, Panel A). First of all, it appears that the benchmark $HAR$ model is almost always outperformed by both $SV$ and $PV(3)$ in terms of $QLIKE$, the only exception being for $h=1$. On the contrary, $h=1$ and $PV(3)$ model is the only combination forecast horizon/model at which the $HAR$ model is outperformed in terms of $MSE$. Second, the $QLIKE$ indices of the forecast reconciliation-based approaches, either indirect (bu) or regression-based (shr), improve on both the $HAR$ benchmark model (apart $SV_{bu}$ at $h=1$) and their direct counterparts. Again, this picture is not confirmed by the $MSE$ indices, because of the different view at the forecasting accuracy offered by these two indices.
%$MSE$ at $h=22$. Third, the $CTPV_{shr}$ forecasts of DJIA are always better than the HAR-based direct forecasts; this is the only case where $MSE$ and $QLIKE$ ratios are less than 1 for all forecast horizons.

The individual stocks' forecast performance analysis (Table \ref{tab:geomean_2007-2022_new}, Panel B) provides more compelling findings. We note that considering very simple intraday decompositions of $RV_t$ in a forecast-reconciliation framework, either indirect (i.e., $SV_{bu}$ and $PV(3)_{bu}$), or regression-based (i.e., $SV_{shr}$ and $PV(3)_{shr}$), always improves on the forecasting accuracy of the $HAR$ benchmark model: both $MSE$ and $QLIKE$ are less than one at any forecast horizon. In addition, $SV_{shr}$ and $PV(3)_{shr}$ always improve on their direct forecasting approaches counterparts, at any forecast horizon and in terms of both $MSE$ and $QLIKE$ indices, with the most notable results being offered by $PV(3)_{shr}$, which stably gives the best accuracy indices (highlighted in bold in Table \ref{tab:geomean_2007-2022_new}, Panel B). %, with the only exception of $CTSV_{shr}$ for $h=5$, where the $MSE$ ratio is greater than one.

\begin{table}
	\centering
	\footnotesize
	\setlength{\tabcolsep}{5pt}
	
\begin{tabular}[t]{>{}l|cc>{}c|ccc}
\toprule
\multicolumn{1}{c}{ } & \multicolumn{3}{c}{MSE} & \multicolumn{3}{c}{QLIKE} \\
 & $h=1$ & $h=5$ & $h=22$ & $h=1$ & $h=5$ & $h=22$\\
\midrule
\addlinespace[0.3em]
\multicolumn{7}{c}{\textit{Panel A: DJIA index}}\\
$SV$ & \textcolor{red}{1.028} & \textcolor{red}{1.017} & \textcolor{red}{1.003} & \textcolor{black}{0.973} & \textcolor{black}{0.997} & \textcolor{black}{0.928}\\
$SV_{bu}$ & \textcolor{black}{0.960} & \textcolor{black}{0.972} & \textcolor{red}{1.005} & \textcolor{red}{1.010} & \textcolor{black}{0.237} & \textcolor{black}{0.748}\\
$SV_{shr}$ & \textcolor{black}{0.976} & \textcolor{black}{0.991} & \textcolor{red}{\textbf{1.002}} & \textcolor{black}{0.983} & \textcolor{black}{0.233} & \textcolor{black}{0.728}\\
$PV(3)$ & \textcolor{black}{\textbf{0.816}} & \textcolor{red}{1.077} & \textcolor{red}{1.025} & \textcolor{red}{2.272} & \textcolor{black}{0.977} & \textcolor{black}{0.962}\\
$PV(3)_{bu}$ & \textcolor{black}{0.924} & \textcolor{black}{\textbf{0.957}} & \textcolor{red}{1.009} & \textcolor{red}{1.001} & \textcolor{black}{0.235} & \textcolor{black}{\textbf{0.596}}\\
$PV(3)_{shr}$ & \textcolor{black}{0.833} & \textcolor{red}{1.015} & \textcolor{red}{1.015} & \textcolor{black}{\textbf{0.945}} & \textcolor{black}{\textbf{0.225}} & \textcolor{black}{0.596}\\
\addlinespace[0.3em]
\multicolumn{7}{c}{\textit{Panel B: Individual stocks}}\\
$SV$ & \textcolor{red}{1.016} & \textcolor{red}{1.008} & \textcolor{red}{1.001} & \textcolor{red}{1.092} & \textcolor{black}{0.998} & \textcolor{black}{0.993}\\
$SV_{bu}$ & \textcolor{black}{0.964} & \textcolor{black}{0.992} & \textcolor{black}{0.998} & \textcolor{black}{0.921} & \textcolor{black}{0.977} & \textcolor{black}{0.971}\\
$SV_{shr}$ & \textcolor{black}{0.980} & \textcolor{black}{0.996} & \textcolor{black}{0.998} & \textcolor{black}{0.899} & \textcolor{black}{0.947} & \textcolor{black}{0.960}\\
$PV(3)$ & \textcolor{black}{0.899} & \textcolor{black}{0.976} & \textcolor{red}{1.010} & \textcolor{red}{1.364} & \textcolor{red}{1.025} & \textcolor{red}{1.055}\\
$PV(3)_{bu}$ & \textcolor{black}{0.943} & \textcolor{black}{0.982} & \textcolor{black}{0.995} & \textcolor{black}{0.900} & \textcolor{black}{0.895} & \textcolor{black}{0.945}\\
$PV(3)_{shr}$ & \textcolor{black}{\textbf{0.896}} & \textcolor{black}{\textbf{0.964}} & \textcolor{black}{\textbf{0.994}} & \textcolor{black}{\textbf{0.833}} & \textcolor{black}{\textbf{0.824}} & \textcolor{black}{\textbf{0.916}}\\
\bottomrule
\end{tabular}

	\caption{Forecast accuracy at forecast horizons $h=1, 5, 22$. $MSE$ and $QLIKE$ ratios over the benchmark $HAR$ model for the DJIA index (panel A), and geometric means of the $MSE$ and $QLIKE$ ratios for individual stocks (panel B). Values larger than one are highlighted in red. The best index value in each column is highlighted in bold.
	}
	\label{tab:geomean_2007-2022_new}
\end{table}

Following \cite{Hansenetal2011}, we implement the MCS procedure using the block bootstrap of \cite{PolitisRomano1994} (see \citealp{Hansen2005}), in which blocks have length of 22 days, and results are based on 10,000 resamples. We choose both the $MSE$ and the $QLIKE$ loss functions, and use the test statistic $T_{max}$ to test the null hypothesis of no difference between the forecast accuracy of the considered model. The results for the forecast horizons $h=1, 5, 22$ are shown in Tables \ref{Table_2007_2022_h1_red}, \ref{Table_2007_2022_h5_red} and \ref{Table_2007_2022_h22_red}, respectively.

For the DJIA index, results sensibly differ between $MSE$ and $QLIKE$, coherently with Table \ref{tab:geomean_2007-2022_new}: while for $MSE$ the MCS includes all models (apart $SV_{bu}$ for $h=5$), in the case of $QLIKE$, $PV(3)_{shr}$ is the best model at $h=1$ and $h=5$ (for $h=22$ even under $QLIKE$ most models are equivalent). For the Diebold-Mariano test, under $MSE$ only few cases lead to a rejection of the null hypothesis, while with $QLIKE$ for $h=5$ and $h=22$ the models with forecast reconciliation improves over the benchmark models in a statistically significant way in most cases.
Moving to the individual stocks, we stress that Tables \ref{Table_2007_2022_h1_red}, \ref{Table_2007_2022_h5_red} and \ref{Table_2007_2022_h22_red} report aggregated results, thus providing an overall evaluation in the cross-section of the 26 stocks. We highlight that even in this case performances differ between $MSE$ and $QLIKE$: for the former, improvements are limited and only in few cases we do have rejections of the null for the Diebold-Mariano test, or models excluded from the confidence set; for the latter, the use of forecast reconciliation leads to a clear improvement, with $PV(3)_{shr}$ providing, overall, better performances.

\begin{table}[p]
\begin{center}\footnotesize

\begin{tabular}[t]{>{}l|cc>{}c|cccc}
\toprule
& $RV$ & $SV$ & $PV(3)$ & $SV_{bu}$ & $PV(3)_{bu}$ & $SV_{shr}$ & $PV(3)_{shr}$\\
\midrule
\addlinespace[0.3em]
\multicolumn{8}{l}{\textit{Panel A: DJIA index}}\\
$MSE$ & 6.352 & 6.527 & \textbf{5.184} & 6.099 & 5.867 & 6.202 & 5.293\\
$p$-value $dm_{RV}$ & $-$ & 0.785 & 0.168 & 0.142 & 0.075 & 0.119 & 0.066\\
$p$-value $dm_{SV}$ & $-$ & $-$ & 0.131 & 0.148 & 0.079 & 0.115 & \textbf{0.044}\\
$p$-value $dm_{PV}$ & $-$ & $-$ & $-$ & 0.800 & 0.759 & 0.821 & 0.579\\
$p$-value MCS & \textbf{0.264} & \textbf{0.224} & \textbf{1.000} & \textbf{0.358} & \textbf{0.431} & \textbf{0.292} & \textbf{0.713}\\
\addlinespace[0.3em]
$QLIKE$ & 0.216 & 0.210 & 0.491 & 0.218 & 0.217 & 0.212 & \textbf{0.204}\\
$p$-value $dm_{RV}$ & $-$ & \textbf{0.004} & 0.908 & 0.692 & 0.528 & 0.176 & \textbf{0.008}\\
$p$-value $dm_{SV}$ & $-$ & $-$ & 0.911 & 0.988 & 0.956 & 0.796 & \textbf{0.031}\\
$p$-value $dm_{PV}$ & $-$ & $-$ & $-$ & 0.098 & 0.097 & 0.093 & 0.087\\
$p$-value MCS & 0.155 & 0.122 & 0.192 & 0.030 & 0.162 & 0.028 & \textbf{1.000}\\
\midrule
\addlinespace[0.3em]
\multicolumn{8}{l}{\textit{Panel B: Individual stocks}}\\
$\overline{MSE}$ & 39.595 & 40.993 & 36.374 & 37.805 & 36.947 & 38.970 & \textbf{35.553}\\
$p$-value $dm_{RV}$ & $-$ & 0 & 0 & 1 & 2 & 1 & 2\\
$p$-value $dm_{SV}$ & $-$ & $-$ & 0 & 0 & 1 & 5 & 2\\
$p$-value $dm_{PV}$ & $-$ & $-$ & $-$ & 1 & 1 & 2 & 1\\
$p$-value MCS & 23 & 19 & 26 & 20 & 25 & 22 & 26\\
\addlinespace[0.3em]
$\overline{QLIKE}$ & 0.185 & 0.214 & 0.269 & 0.170 & 0.165 & 0.165 & \textbf{0.151}\\
$p$-value $dm_{RV}$ & $-$ & 2 & 1 & 2 & 4 & 7 & 19\\
$p$-value $dm_{SV}$ & $-$ & $-$ & 1 & 6 & 7 & 10 & 20\\
$p$-value $dm_{PV}$ & $-$ & $-$ & $-$ & 7 & 8 & 8 & 10\\
$p$-value MCS & 10 & 8 & 19 & 6 & 4 & 6 & 26\\
\bottomrule
\end{tabular}
\\[0.5cm]
\caption{One-day-ahead forecasting performance: 2007-2022 (3,880 days)}
\label{Table_2007_2022_h1_red}
\end{center}
\begin{footnotesize}
	\textit{Note}: The table reports the \textbf{1-step ahead} forecasting performance of the different models. The top panel shows the results for the DJIA index, while the bottom panel refers to individual stocks. $MSE$ and $QLIKE$ refer to the loss function value for a given model (top panel) or average loss function across individual stocks (bottom panel). The one-sided tests between each forecasting model against $HAR$, $SV$, and $PV(3)$ are denoted by $dm_{HAR}$, $dm_{SV}$, and $dm_{PV(3)}$, respectively. The top panel includes p-values while the bottom panel reports 5\% rejection frequencies. MCS denotes the $p$-value of that model being in the Model Confidence Set (top panel), or the number of times that model is in the 80\% Model Confidence Set (lower panel). $PV(3)$ uses three intraday decompositions defined by two thresholds at 10\% and 75\%. In the upper panel we highlight in bold $p$-values $< 0.05$ for Diebold-Mariano and $p$-values $ > 0.2$ for MCS. In the both panel we highlight in bold the minimum (average) loss function.
\end{footnotesize}
\end{table}	

\begin{table}[p]
\begin{center}\footnotesize

\begin{tabular}[t]{>{}l|cc>{}c|cccc}
\toprule
& $RV$ & $SV$ & $PV(3)$ & $SV_{bu}$ & $PV(3)_{bu}$ & $SV_{shr}$ & $PV(3)_{shr}$\\
\midrule
\addlinespace[0.3em]
\multicolumn{8}{l}{\textit{Panel A: DJIA index}}\\
$MSE$ & 5.109 & 5.195 & 5.503 & 4.966 & \textbf{4.888} & 5.064 & 5.185\\
$p$-value $dm_{RV}$ & $-$ & 0.774 & 0.833 & \textbf{0.021} & \textbf{0.007} & 0.235 & 0.613\\
$p$-value $dm_{SV}$ & $-$ & $-$ & 0.783 & 0.054 & \textbf{0.023} & \textbf{0.049} & 0.485\\
$p$-value $dm_{PV}$ & $-$ & $-$ & $-$ & 0.105 & 0.077 & 0.139 & \textbf{0.027}\\
$p$-value MCS & \textbf{0.408} & \textbf{0.369} & \textbf{0.369} & 0.090 & \textbf{1.000} & \textbf{0.288} & \textbf{0.443}\\
\addlinespace[0.3em]
$QLIKE$ & 0.930 & 0.927 & 0.908 & 0.220 & 0.219 & 0.217 & \textbf{0.209}\\
$p$-value $dm_{RV}$ & $-$ & \textbf{0.000} & 0.441 & \textbf{0.008} & \textbf{0.008} & \textbf{0.008} & \textbf{0.007}\\
$p$-value $dm_{SV}$ & $-$ & $-$ & 0.449 & \textbf{0.008} & \textbf{0.008} & \textbf{0.008} & \textbf{0.007}\\
$p$-value $dm_{PV}$ & $-$ & $-$ & $-$ & \textbf{0.007} & \textbf{0.007} & \textbf{0.007} & \textbf{0.006}\\
$p$-value MCS & \textbf{0.462} & \textbf{0.486} & 0.140 & 0.006 & 0.104 & 0.016 & \textbf{1.000}\\
\midrule
\addlinespace[0.3em]
\multicolumn{8}{l}{\textit{Panel B: Individual stocks}}\\
$\overline{MSE}$ & 25.109 & 25.371 & 22.936 & 24.804 & 24.433 & 24.923 & \textbf{22.753}\\
$p$-value $dm_{RV}$ & $-$ & 0 & 1 & 2 & 9 & 4 & 2\\
$p$-value $dm_{SV}$ & $-$ & $-$ & 1 & 2 & 4 & 4 & 3\\
$p$-value $dm_{PV}$ & $-$ & $-$ & $-$ & 0 & 0 & 0 & 0\\
$p$-value MCS & 26 & 22 & 26 & 25 & 26 & 24 & 26\\
\addlinespace[0.3em]
$\overline{QLIKE}$ & 0.186 & 0.187 & 0.201 & 0.181 & 0.161 & 0.174 & \textbf{0.147}\\
$p$-value $dm_{RV}$ & $-$ & 6 & 7 & 4 & 8 & 9 & 24\\
$p$-value $dm_{SV}$ & $-$ & $-$ & 7 & 2 & 4 & 3 & 21\\
$p$-value $dm_{PV}$ & $-$ & $-$ & $-$ & 2 & 4 & 2 & 8\\
$p$-value MCS & 11 & 13 & 22 & 9 & 6 & 9 & 24\\
\bottomrule
\end{tabular}
\\[0.5cm]
\caption{Five-day-ahead forecasting performance: 2007-2022 (3,880 days)}
\label{Table_2007_2022_h5_red}
\end{center}
\begin{footnotesize}
\textit{Note}: The table reports the \textbf{1-step ahead} forecasting performance of the different models. The top panel shows the results for the DJIA index, while the bottom panel refers to individual stocks. $MSE$ and $QLIKE$ refer to the loss function value for a given model (top panel) or average loss function across individual stocks (bottom panel). The one-sided tests between each forecasting model against $HAR$, $SV$, and $PV(3)$ are denoted by $dm_{HAR}$, $dm_{SV}$, and $dm_{PV(3)}$, respectively. The top panel includes p-values while the bottom panel reports 5\% rejection frequencies. MCS denotes the $p$-value of that model being in the Model Confidence Set (top panel), or the number of times that model is in the 80\% Model Confidence Set (lower panel). $PV(3)$ uses three intraday decompositions defined by two thresholds at 10\% and 75\%. In the upper panel we highlight in bold $p$-values $< 0.05$ for Diebold-Mariano and $p$-values $ > 0.2$ for MCS. In the lower panel we highlight in bold the minimum (average) loss function.
\end{footnotesize}
\end{table}	

\begin{table}[p]
\begin{center}\footnotesize

\begin{tabular}[t]{>{}l|cc>{}c|cccc}
\toprule
& $RV$ & $SV$ & $PV(3)$ & $SV_{bu}$ & $PV(3)_{bu}$ & $SV_{shr}$ & $PV(3)_{shr}$\\
\midrule
\addlinespace[0.3em]
\multicolumn{8}{l}{\textit{Panel A: DJIA index}}\\
$MSE$ & \textbf{4.424} & 4.436 & 4.533 & 4.448 & 4.463 & 4.434 & 4.488\\
$p$-value $dm_{RV}$ & $-$ & 0.613 & 0.851 & 0.631 & 0.683 & 0.569 & 0.780\\
$p$-value $dm_{SV}$ & $-$ & $-$ & 0.843 & 0.575 & 0.646 & 0.478 & 0.765\\
$p$-value $dm_{PV}$ & $-$ & $-$ & $-$ & 0.209 & 0.257 & 0.158 & 0.225\\
$p$-value MCS & \textbf{1.000} & \textbf{0.869} & \textbf{0.727} & \textbf{0.841} & \textbf{0.841} & \textbf{0.761} & \textbf{0.811}\\
\addlinespace[0.3em]
$QLIKE$ & 0.973 & 0.902 & 0.936 & 0.727 & \textbf{0.580} & 0.708 & 0.580\\
$p$-value $dm_{RV}$ & $-$ & 0.160 & 0.150 & \textbf{0.027} & \textbf{0.002} & \textbf{0.016} & \textbf{0.002}\\
$p$-value $dm_{SV}$ & $-$ & $-$ & 0.830 & \textbf{0.049} & \textbf{0.004} & \textbf{0.028} & \textbf{0.003}\\
$p$-value $dm_{PV}$ & $-$ & $-$ & $-$ & \textbf{0.031} & \textbf{0.002} & \textbf{0.017} & \textbf{0.002}\\
$p$-value MCS & \textbf{0.314} & \textbf{0.437} & \textbf{0.453} & 0.135 & \textbf{1.000} & \textbf{0.323} & \textbf{0.951}\\
\midrule
\addlinespace[0.3em]
\multicolumn{8}{l}{\textit{Panel B: Individual stocks}}\\
$\overline{MSE}$ & 14.243 & 14.277 & 14.316 & 14.205 & 14.231 & 14.207 & \textbf{14.122}\\
$p$-value $dm_{RV}$ & $-$ & 1 & 2 & 2 & 4 & 3 & 3\\
$p$-value $dm_{SV}$ & $-$ & $-$ & 2 & 1 & 2 & 2 & 2\\
$p$-value $dm_{PV}$ & $-$ & $-$ & $-$ & 3 & 2 & 3 & 6\\
$p$-value MCS & 24 & 24 & 24 & 24 & 25 & 23 & 24\\
\addlinespace[0.3em]
$\overline{QLIKE}$ & 0.269 & 0.268 & 0.292 & 0.262 & 0.255 & 0.259 & \textbf{0.247}\\
$p$-value $dm_{RV}$ & $-$ & 5 & 6 & 4 & 10 & 6 & 22\\
$p$-value $dm_{SV}$ & $-$ & $-$ & 5 & 3 & 6 & 4 & 17\\
$p$-value $dm_{PV}$ & $-$ & $-$ & $-$ & 8 & 10 & 9 & 14\\
$p$-value MCS & 18 & 19 & 24 & 13 & 15 & 13 & 24\\
\bottomrule
\end{tabular}
\\[0.5cm]
\caption{Twenty-two-day-ahead forecasting performance: 2007-2022 (3,880 days)}
\label{Table_2007_2022_h22_red}
\end{center}
\begin{footnotesize}
\textit{Note}: The table reports the \textbf{1-step ahead} forecasting performance of the different models. The top panel shows the results for the DJIA index, while the bottom panel refers to individual stocks. $MSE$ and $QLIKE$ refer to the loss function value for a given model (top panel) or average loss function across individual stocks (bottom panel). The one-sided tests between each forecasting model against $HAR$, $SV$, and $PV(3)$ are denoted by $dm_{HAR}$, $dm_{SV}$, and $dm_{PV(3)}$, respectively. The top panel includes p-values while the bottom panel reports 5\% rejection frequencies. MCS denotes the $p$-value of that model being in the Model Confidence Set (top panel), or the number of times that model is in the 80\% Model Confidence Set (lower panel). $PV(3)$ uses three intraday decompositions defined by two thresholds at 10\% and 75\%. In the upper panel we highlight in bold $p$-values $< 0.05$ for Diebold-Mariano and $p$-values $ > 0.2$ for MCS. In the lower panel we highlight in bold the minimum (average) loss function.
\end{footnotesize}
\end{table}	

\begin{figure}[t]
\centering
\includegraphics[width = .495\linewidth]{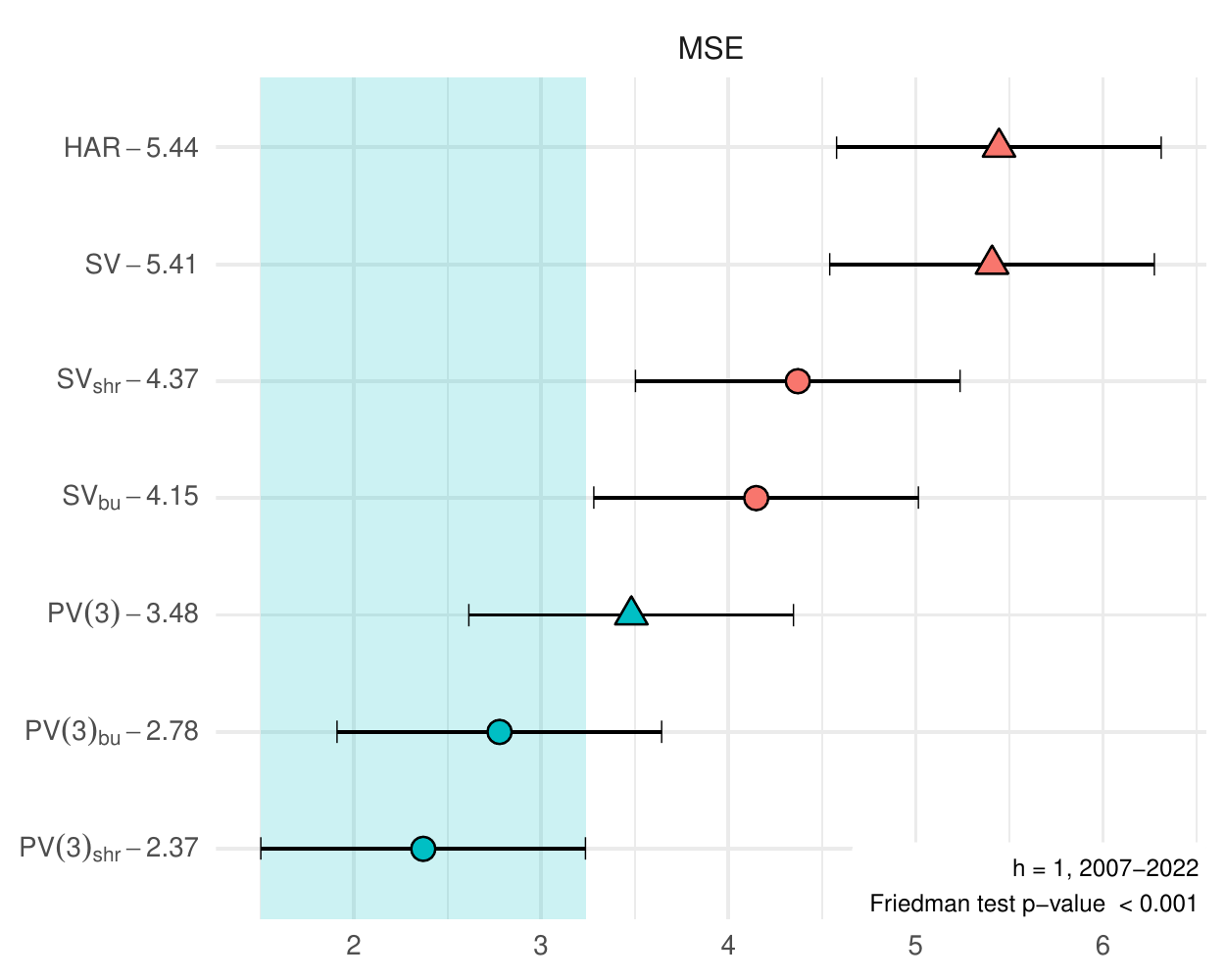}
\includegraphics[width = .495\linewidth]{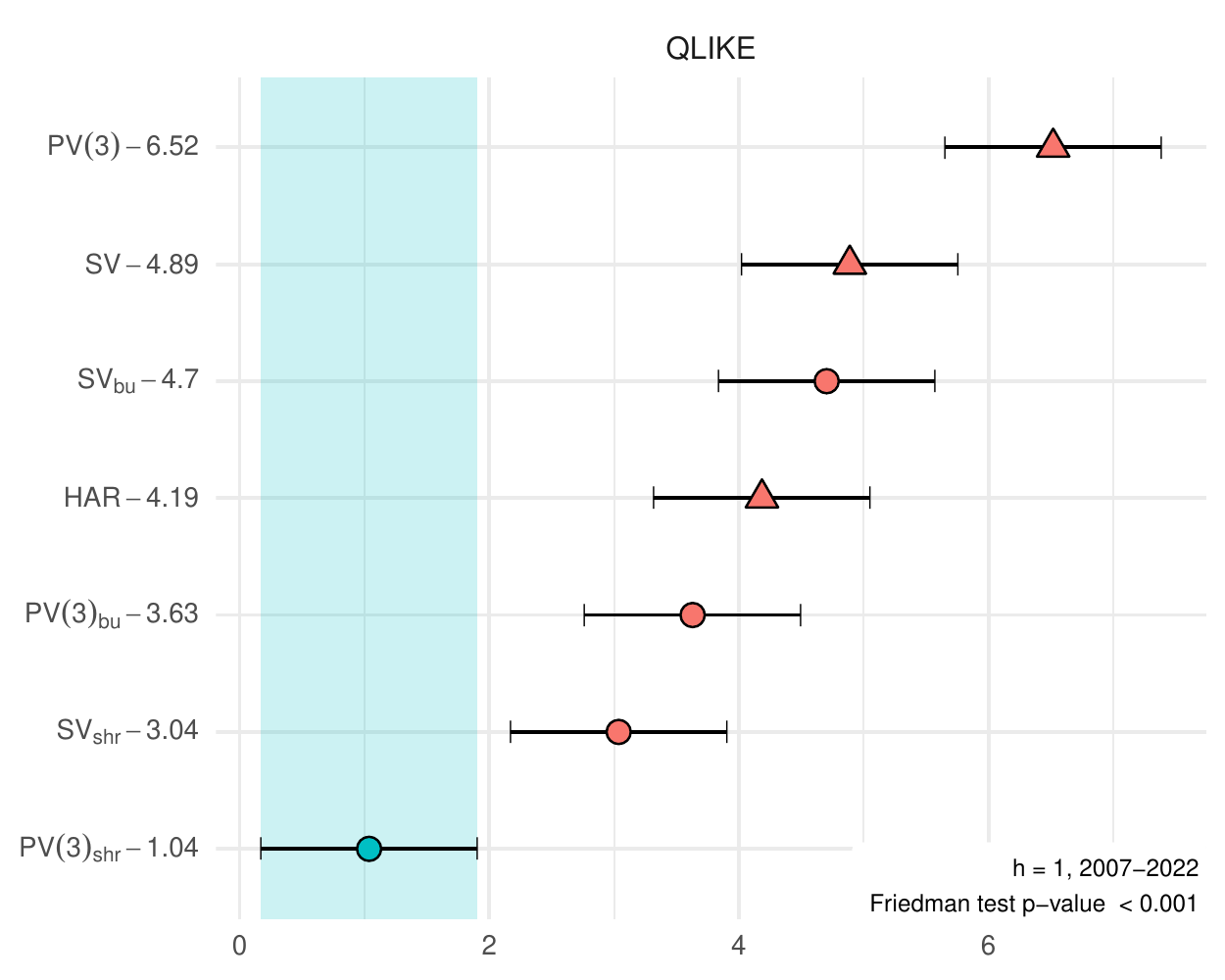}
\caption{{MCB Nemenyi test results: average ranks and 95\% confidence intervals for the \textbf{one-step ahead} $RV$ forecasts of the DJIA index and 26 individual stocks. Direct daily $RV$ forecasts from $HAR$, $SV$ and $PV(3)$ models, and from their extensions with the bottom-up (bu) and the MinT-shr (shr) forecast reconciliation-based approaches according to the corresponding intraday $RV$ decomposition. The forecasting approaches are sorted vertically according to the $MSE$ mean rank (left panel) and the $QLIKE$ mean rank (right panel). The mean rank of each method is displayed to the right of their names. If the intervals of two forecasting models do not overlap, this indicates a statistically different performance. Thus, methods that do not overlap with the light blue interval are considered significantly worse than the best and vice-versa.}}
\label{fig:mcb}
\end{figure}

These evidences are confirmed and enriched by Figure \ref{fig:mcb}, which shows the results of the Multiple Comparison with the Best (MCB) Nemenyi test, a non-parametric multiple comparison procedure frequently adopted in the forecasting literature (see \citealp{Koning2005}, \citealp{Kourentzes2019}, and \citealp{Makridakis2022}, among others). In particular, the single model $PV(3)$ does not significantly improve on the benchmark $HAR$ (the corresponding lines in the `Multiple Comparison with the Best' graphs for both $MSE$ and $QLIKE$ are overlapping). Further, the forecasts produced by the $PV(3)_{shr}$ approach are significantly better than the benchmarks $HAR$ and $SV$, both in terms of $MSE$ and $QLIKE$, while direct $PV(3)$ forecasts appear significantly worse if the $QLIKE$ loss function is used to evaluate the forecast accuracy. Finally, overall $PV(3)_{shr}$ shows the best forecast accuracy: it is not significantly worse than $PV(3)_{bu}$, the most performing approach in terms of $MSE$, and ranks first in terms of $QLIKE$, while $SV_{shr}$ is the only approach with a statistically equivalent forecasting accuracy.

\begin{figure}[t]
\centering
\includegraphics[width=.495\linewidth]{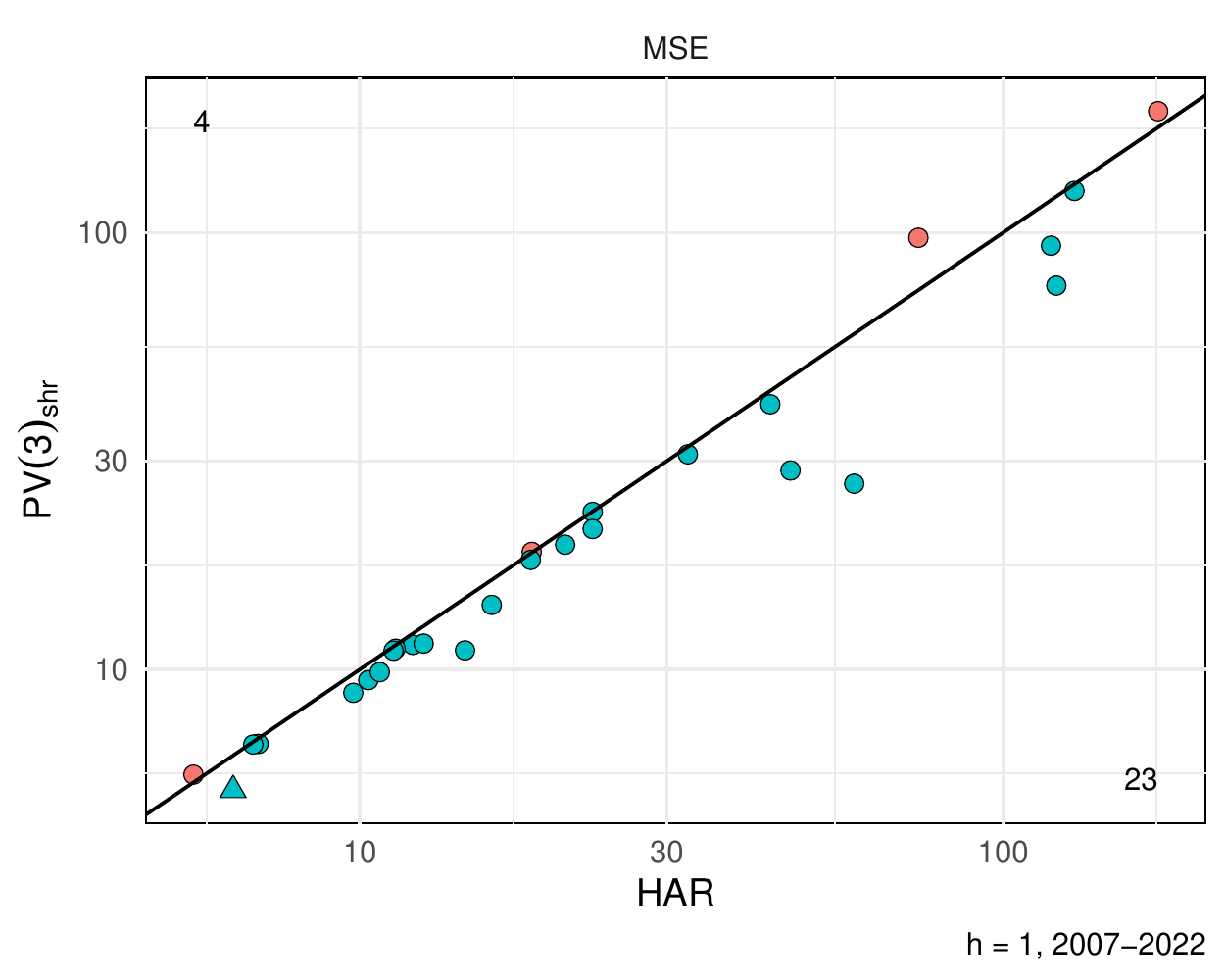}
\includegraphics[width=.495\linewidth]{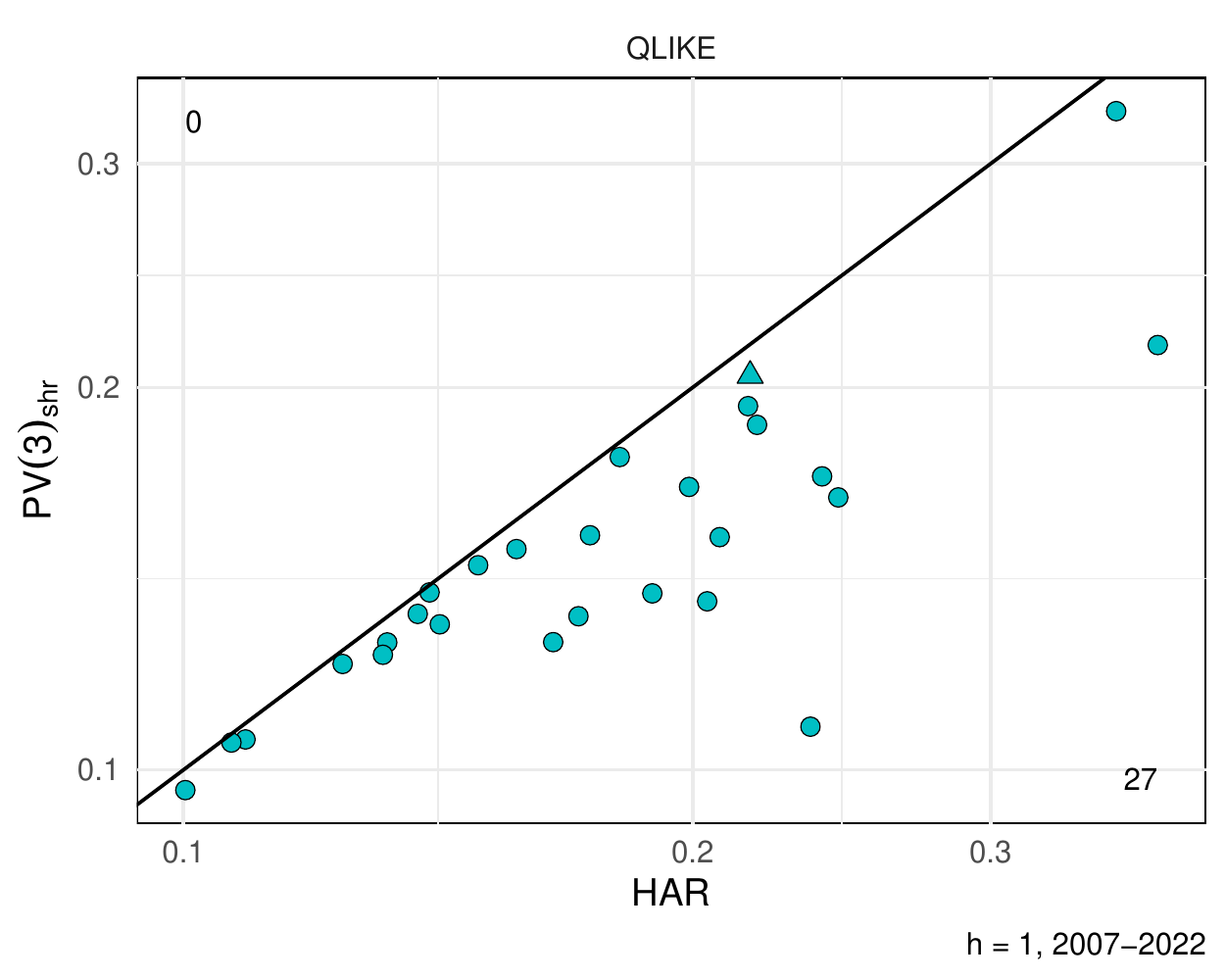}
\caption{{Accuracy of the \textbf{one-step ahead} daily $RV$ forecasts for the DJIA index (triangle) and 26 individual stocks (circle) in terms of $MSE$ (left panel) and $QLIKE$ (right panel) indices. Comparison between $HAR$ direct and $PV(3)_{shr}$ reconciliation-based forecasts. The black line represents the bisector, where either MSE's or QLIKE's for both approaches are equal. On the top-left (bottom-right) corner of each graph, the number of points above (below) the bisector is reported.}}
\label{fig:point_HAR_PV3shr}
\end{figure}

\begin{figure}[htb]
\centering
\includegraphics[width=.495\linewidth]{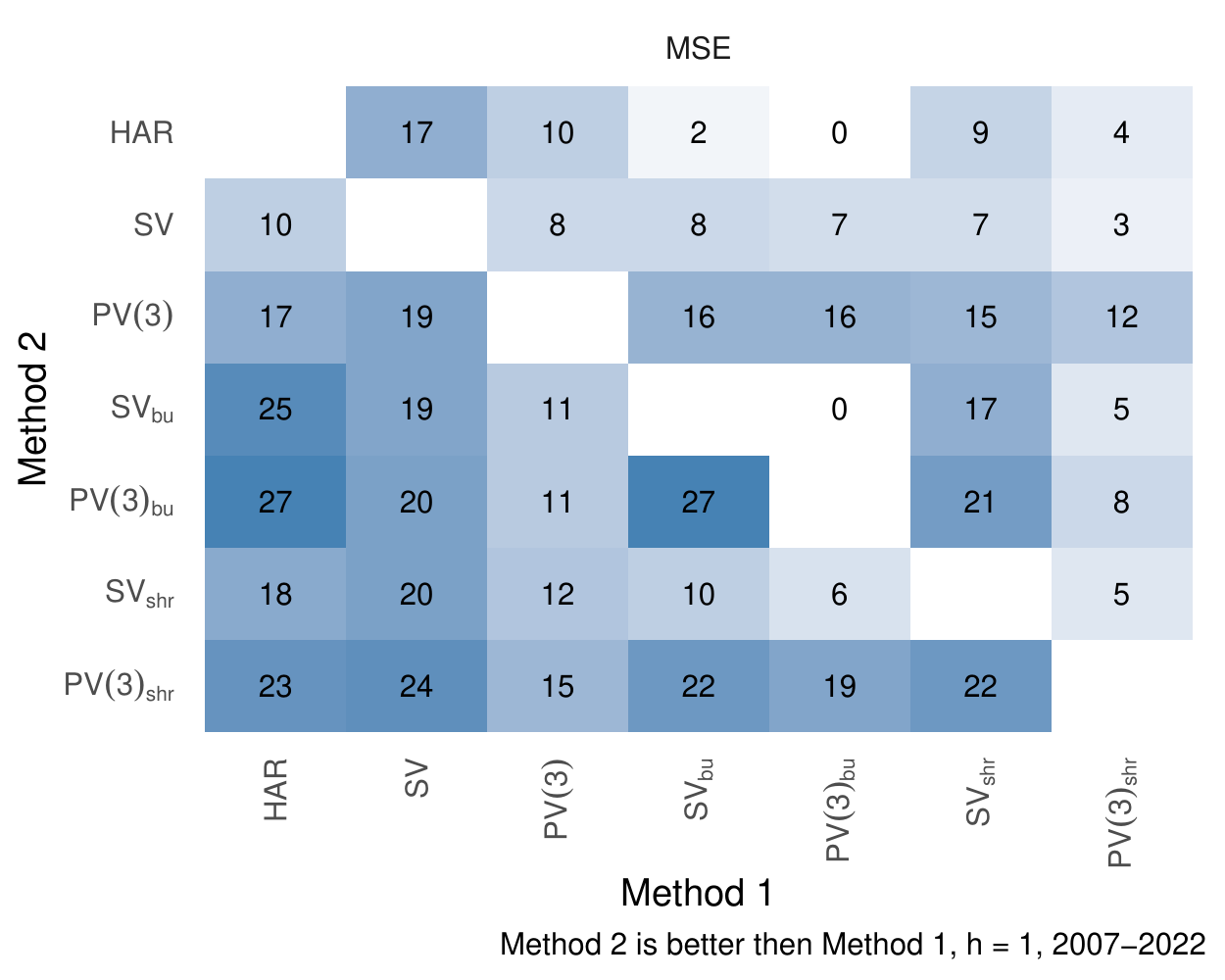}
\includegraphics[width=.495\linewidth]{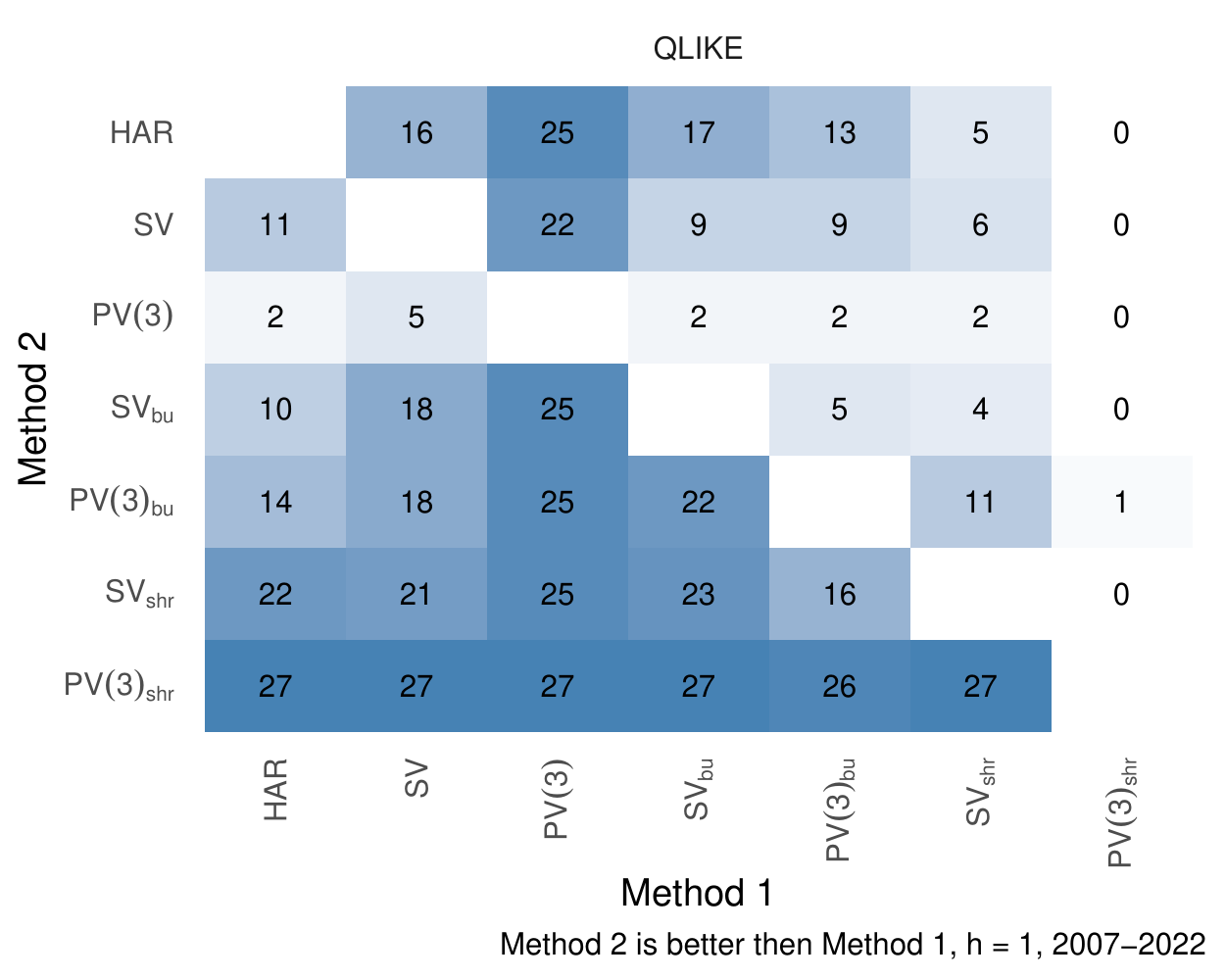}
\caption{Qualitative evaluation of the \textbf{one-step ahead} forecasting accuracy. Each cell reports the number of times the forecasting model in the row outperforms the model in the column.}
\label{fig:qualeval}
\end{figure}

Further insights are offered in a simple, but effective descriptive view, by Figure	\ref{fig:point_HAR_PV3shr}, where the scatter plots of the 27 couples of, respectively, $MSE$ and $QLIKE$ indices obtained using the benchmark $HAR$ model and $PV(3)_{shr}$ are represented. It emerges that $PV(3)_{shr}$ outperforms the benchmark $HAR$  in the majority of cases (23 out of 27) in terms of $MSE$, and always if $QLIKE$ is used to evaluate the forecasting accuracy. This consideration is somehow extended, and further supported, by the results shown in Figure \ref{fig:qualeval}, which contains a summary view of the number of times each forecasting approach provided better forecasting accuracy than the other procedures considered in the comparison. Summarizing, $PV(3)_{shr}$ registers a better prediction performance than all other approaches, with success rates in terms of $QLIKE$ ranging from 96.3\% (26 of 27) to 100\% (27 of 27), and from 55.6\% (15 of 27) to 88.9\% (24 of 27) if $MSE$ is used. 

\section{Robustness of the results}
\label{sec:robust}

\subsection{Sub-sample analysis}
The interpretation of the results so far can be further detailed and specified by considering 4 different time windows of the complete 2007-2022 interval previously analysed, namely: 2006-2010, 2011-2014, 2015-2019, 2020-2022 (see Appendix 2). We observe that the predictive performance of models that make use of intraday decompositions in a reconciliation framework is constantly better in periods of high market volatility (2007-2010 and 2020-2022), while they do not worsen the predictive accuracy of the benchmark in the period 2015-2019, which is characterized by lower variances. The remaining period 2011-2014 offers different indications depending on whether one considers $QLIKE$ or MSE. In the latter case, the $PV(3)$ model, both in the single version and in the version that makes use of forecast reconciliation, shows a better overall accuracy, although not significantly to the $HAR$ benchmark.

\subsection{Grouped series}
Here we consider the forecasting performance of reconciliation approaches applied to grouped time series defined by intraday $RV$ decompositions based on time and returns' characteristics:
\begin{itemize}[nosep]
\item $TSV_{bu}$: indirect (bottom-up) daily $RV$ forecasts from ten bottom time series cross-classified by time interval and semi-variances;
\item $TSV_{shr}$: two hierarchies sharing the ten bottom variables above, with eight upper variables;
\item $TPV(3)_{bu}$: indirect (bottom-up) daily $RV$ forecasts from fifteen bottom time series cross-classified by time interval and partial variances;
\item $TPV_{shr}$: two hierarchies sharing the fifteen bottom variables above, with nine upper variables.
\end{itemize}

\begin{figure}[t]
	\centering
	\includegraphics[width=.495\linewidth]{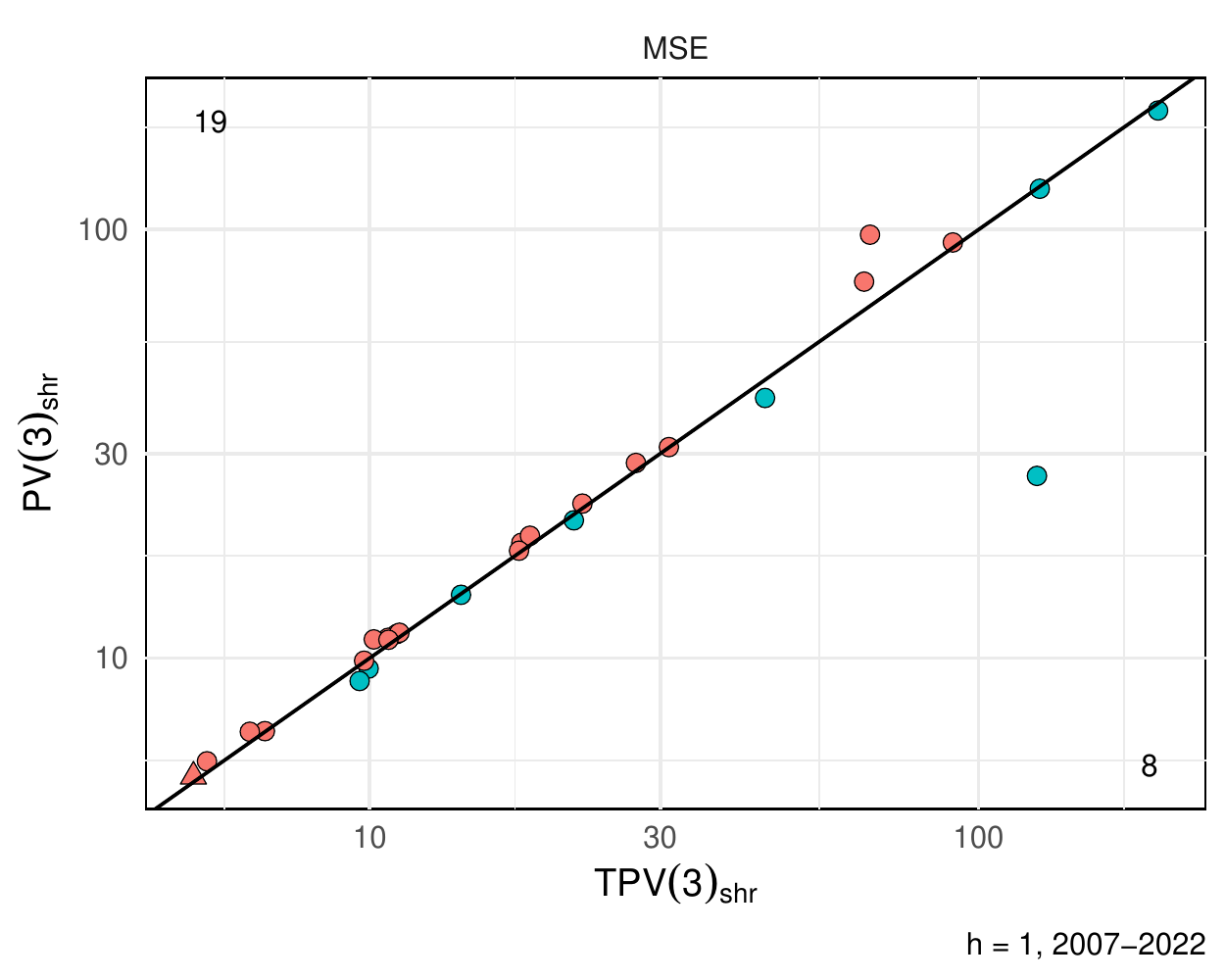}
	\includegraphics[width=.495\linewidth]{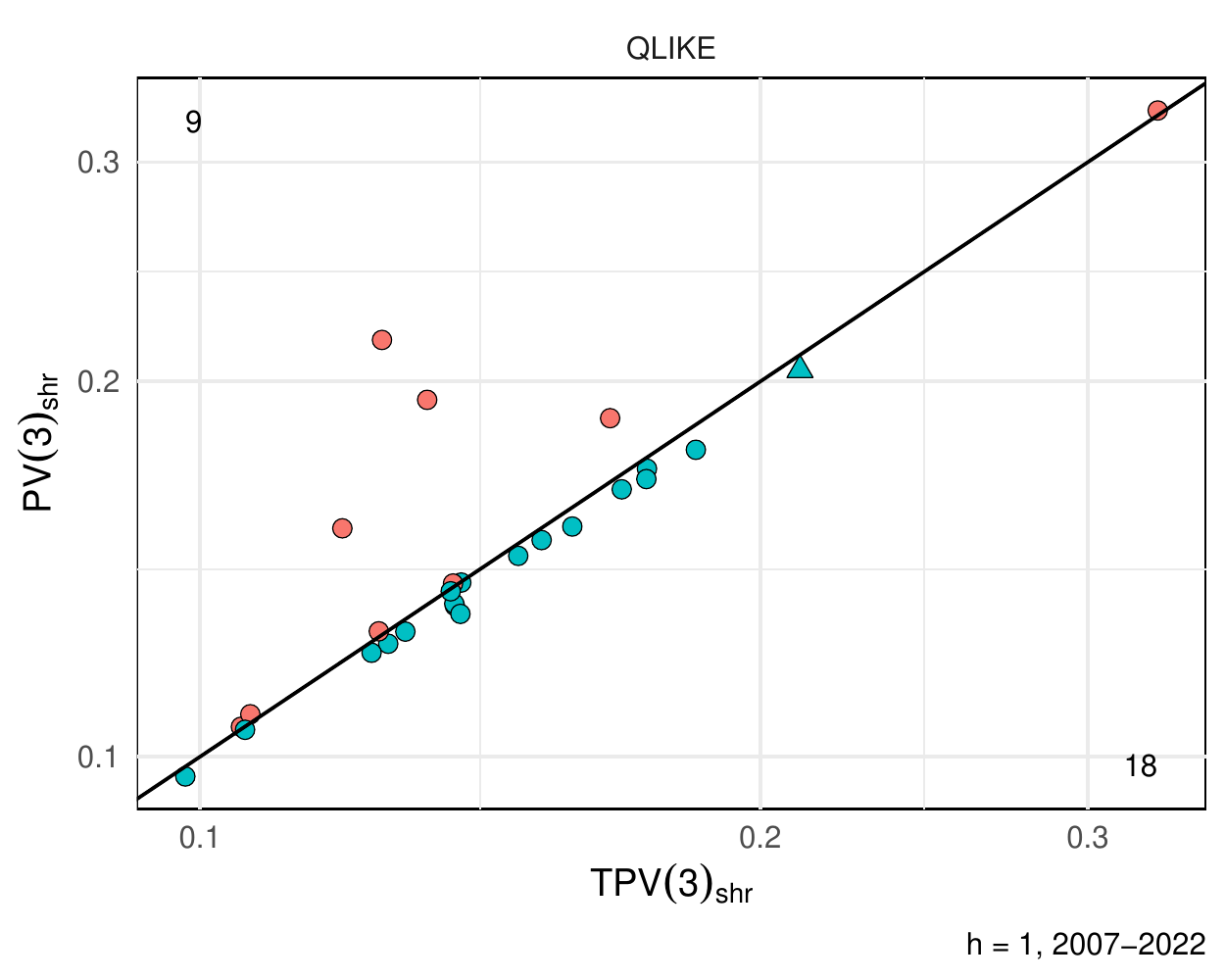}
	\caption{{Accuracy of the \textbf{one-step ahead} daily $RV$ forecasts for the DJIA index (triangle) and 26 individual stocks (circle) in terms of $MSE$ (left panel) and $QLIKE$ (right panel) indices. Comparison between $HAR$ direct and $PV(3)_{shr}$ reconciliation-based forecasts. The black line represents the bisector, where either MSE's or QLIKE's for both approaches are equal. On the top-left (bottom-right) corner of each graph, the number of points above (below) the bisector is reported.}}
	\label{fig:point_TPV3_PV3shr}
\end{figure}

From \autoref{tab:geomean_2007-2022_grouped} it appears that the regression-based reconciliation approach $TPV(3)_{shr}$ largely benefits from the adoption of a time decomposition (all indices are less than one).
However, looking at \autoref{fig:point_TPV3_PV3shr}, $TPV(3)_{shr}$ gives results largely similar to those of the simpler $PV(3)_{shr}$ approach, that does not make use of any temporal decomposition.

\begin{table}[t]
	\centering
	\footnotesize
	\setlength{\tabcolsep}{5pt}
	
\begin{tabular}[t]{>{}l|cc>{}c|ccc}
\toprule
\multicolumn{1}{c}{ } & \multicolumn{3}{c}{MSE} & \multicolumn{3}{c}{QLIKE} \\
 & $h=1$ & $h=5$ & $h=22$ & $h=1$ & $h=5$ & $h=22$\\
\midrule
\addlinespace[0.3em]
\multicolumn{7}{c}{\textit{Panel A: DJIA index}}\\
$TSV_{bu}$ & \textcolor{black}{0.824} & \textcolor{black}{0.801} & \textcolor{red}{1.099} & \textcolor{red}{1.085} & \textcolor{black}{0.249} & \textcolor{black}{0.385}\\
$TSV_{shr}$ & \textcolor{black}{0.869} & \textcolor{black}{0.878} & \textcolor{black}{0.965} & \textcolor{red}{1.011} & \textcolor{black}{0.236} & \textcolor{black}{0.379}\\
$TPV(3)_{bu}$ & \textcolor{black}{0.822} & \textcolor{black}{\textbf{0.795}} & \textcolor{red}{1.103} & \textcolor{red}{1.090} & \textcolor{black}{0.250} & \textcolor{black}{0.386}\\
$TPV(3)_{shr}$ & \textcolor{black}{\textbf{0.808}} & \textcolor{black}{0.925} & \textcolor{black}{\textbf{0.939}} & \textcolor{black}{\textbf{0.971}} & \textcolor{black}{\textbf{0.227}} & \textcolor{black}{\textbf{0.374}}\\
\addlinespace[0.3em]
\multicolumn{7}{c}{\textit{Panel B: Individual stocks}}\\
$TSV_{bu}$ & \textcolor{black}{0.991} & \textcolor{red}{1.054} & \textcolor{red}{1.019} & \textcolor{black}{0.893} & \textcolor{black}{0.908} & \textcolor{black}{0.875}\\
$TSV_{shr}$ & \textcolor{black}{0.987} & \textcolor{red}{1.007} & \textcolor{black}{0.984} & \textcolor{black}{0.850} & \textcolor{black}{0.861} & \textcolor{black}{0.853}\\
$TPV(3)_{bu}$ & \textcolor{black}{0.982} & \textcolor{red}{1.039} & \textcolor{red}{1.012} & \textcolor{black}{0.894} & \textcolor{black}{0.902} & \textcolor{black}{0.868}\\
$TPV(3)_{shr}$ & \textcolor{black}{\textbf{0.924}} & \textcolor{black}{\textbf{0.967}} & \textcolor{black}{\textbf{0.965}} & \textcolor{black}{\textbf{0.805}} & \textcolor{black}{\textbf{0.793}} & \textcolor{black}{\textbf{0.824}}\\
\bottomrule
\end{tabular}

	\caption{Forecast accuracy at forecast horizons $h=1, 5, 22$. $MSE$ and $QLIKE$ ratios over the benchmark $HAR$ model for the DJIA index (panel A), and geometric means of the $MSE$ and $QLIKE$ ratios for individual stocks (panel B). Values larger than one are highlighted in red. The best index value in each column is highlighted in bold.
	}
	\label{tab:geomean_2007-2022_grouped}
\end{table}

\subsection{An alternative $PV$ decomposition}

In the $PV(3)$ decomposition, the use of $RV_t$ in the standardization of intraday returns exposes the estimators of standardized returns quantile to the influence of price jumps (i.e., extreme positive or negative returns) and of the intraday volatility pattern. 

We consider here a slight modification in the construction of the decomposition by employing the contribution of \cite{Boudtetal2011} and exploiting a result of \cite{Andersenetal2007b}. The former provide a methodology for estimating and filtering out the intraday variance pattern while the latter shows that, in large sample, the intraday returns standardized by the Bi-power variation follow an asymptotically normal distribution. We combine these two elements as follows. First, we compute returns filtered from the intraday volatility pattern by adopting the \textit{WDS} approach of \cite{Boudtetal2011}. The filtered returns are defined as  $r_{i,t}^{\star}=\frac{r_{i,t}}{\sigma_i}$ where $\sigma_i$ is the intraday volatility at interval $i$, estimated with the \textit{WDS} approach. Then, we evaluate the Bi-power variation, a jump-robust estimator of the integrated volatility, on the filtered returns: $\widetilde{BPV_t}= \frac{\pi}{2}\frac{N}{N-1} \sum_{i=2}^M \vert r_{i,t}^{\star}\vert\vert r_{i,t}^{\star}\vert$. Finally, we assume the following distribution for the filtered and standardized returns
\begin{equation}
\frac{r_{i,t}}{\sigma_i\sqrt{N^{-1}\widetilde{BPV_t}}}\sim \mathcal{N}\left(0,1\right),
\end{equation}
This quantity has already been used in \cite{Boudtetal2011} for jump detection. Going back to the $PV$ framework, the empirical quantiles of standardized returns might be replaced by theoretical quantiles of the asymptotic distribution leading to
\begin{equation}
\tilde{c}_{i,t,j}=\sigma_i\sqrt{N^{-1}\widetilde{BPV_t}} \Phi^{-1}\left(\tau_j\right), \quad j=1,2,\ldots p-1.
\end{equation}
Finally, we note that thresholds are both day-specific and intra-day-interval-specific. Similarly to the $PV(3)$, we now define a $PV(3)^{\star}$ decomposition where we employ the previously defined quantiles with $\tau_1=0.1$ and $\tau_2=0.75$. 

\begin{figure}[H]
	\centering
	\includegraphics[width=.495\linewidth]{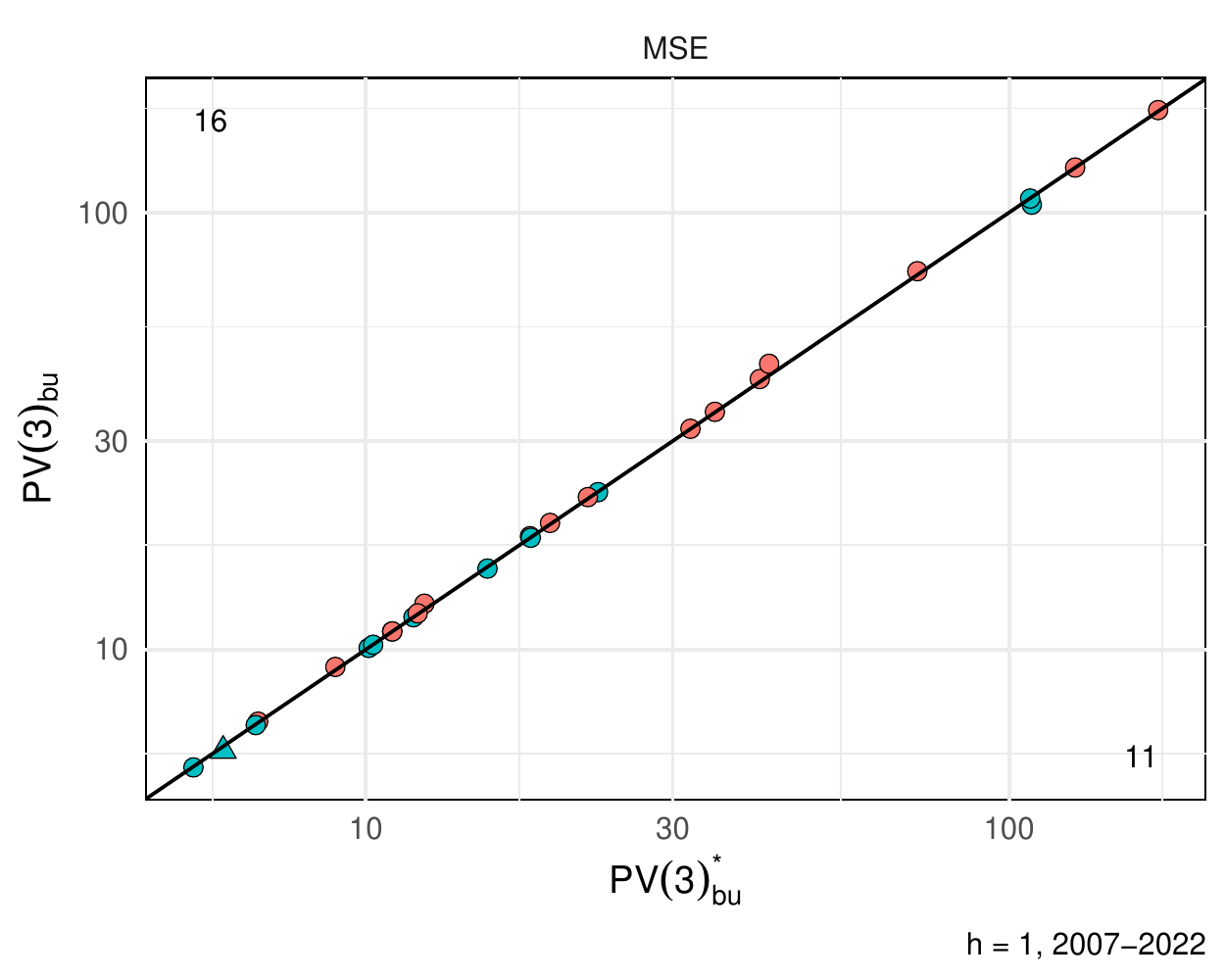}
	\includegraphics[width=.495\linewidth]{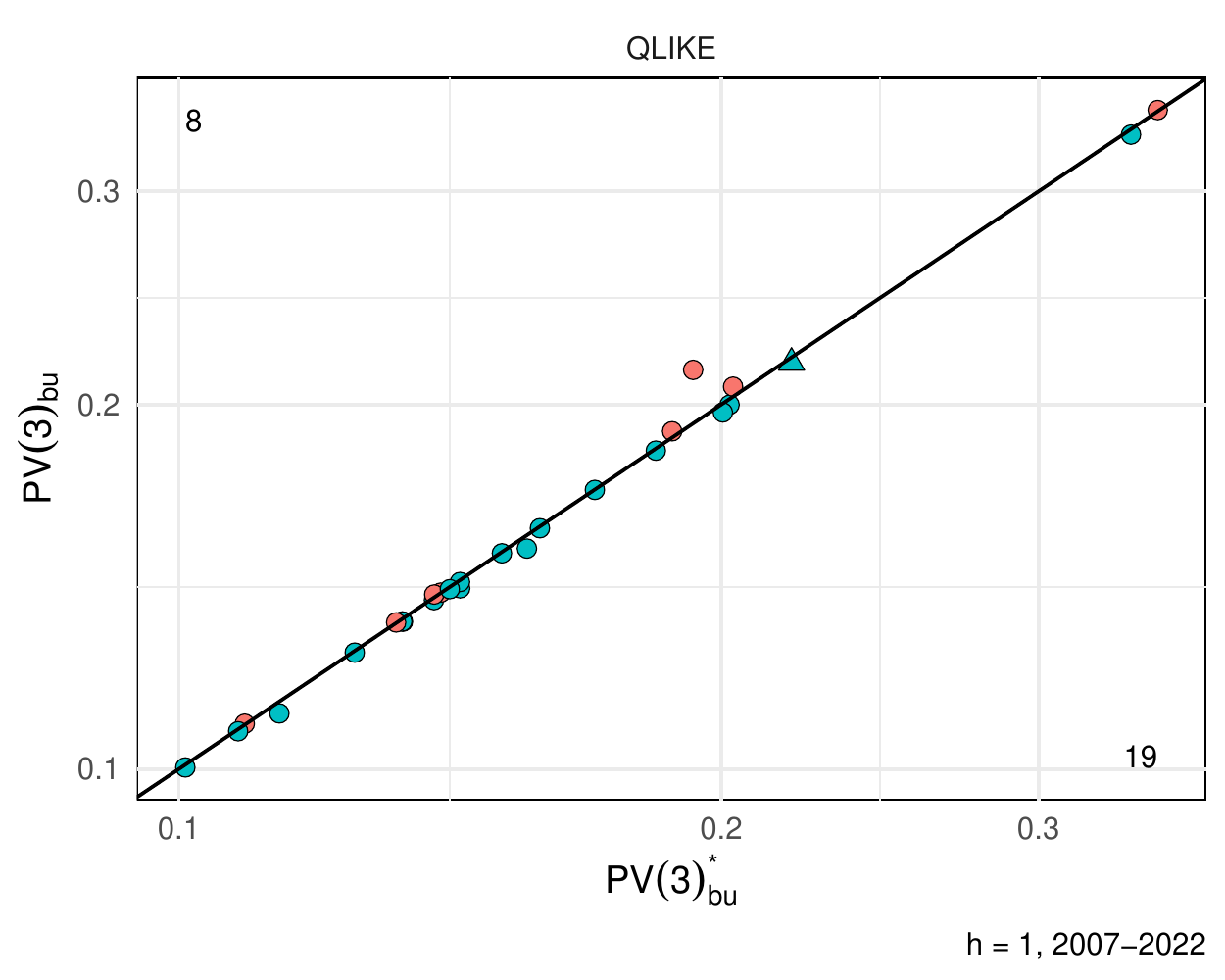}
	\caption{Accuracy of the \textbf{one-step ahead} daily $RV$ forecasts for the DJIA index (triangle) and 26 individual stocks (circle) in terms of $MSE$ (left panel) and $QLIKE$ (right panel) indices. Comparison between $HAR$ direct and $PV(3)_{shr}$ reconciliation-based forecasts. The black line represents the bisector, where either MSE's or QLIKE's for both approaches are equal. On the top-left (bottom-right) corner of each graph, the number of points above (below) the bisector is reported.}
	\label{fig:point_PV3bu}
\end{figure}

\begin{figure}[H]
	\centering
	\includegraphics[width=.495\linewidth]{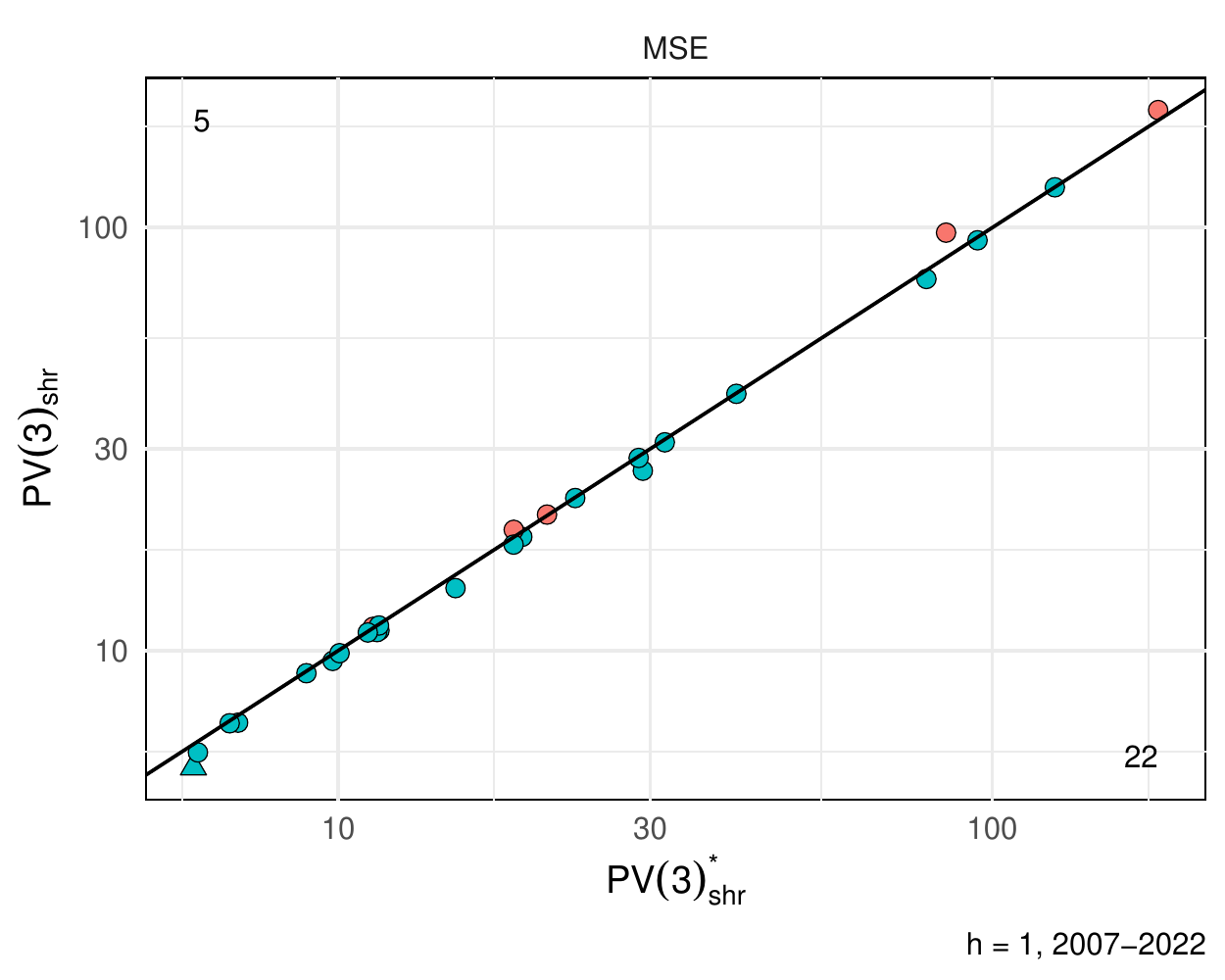}
	\includegraphics[width=.495\linewidth]{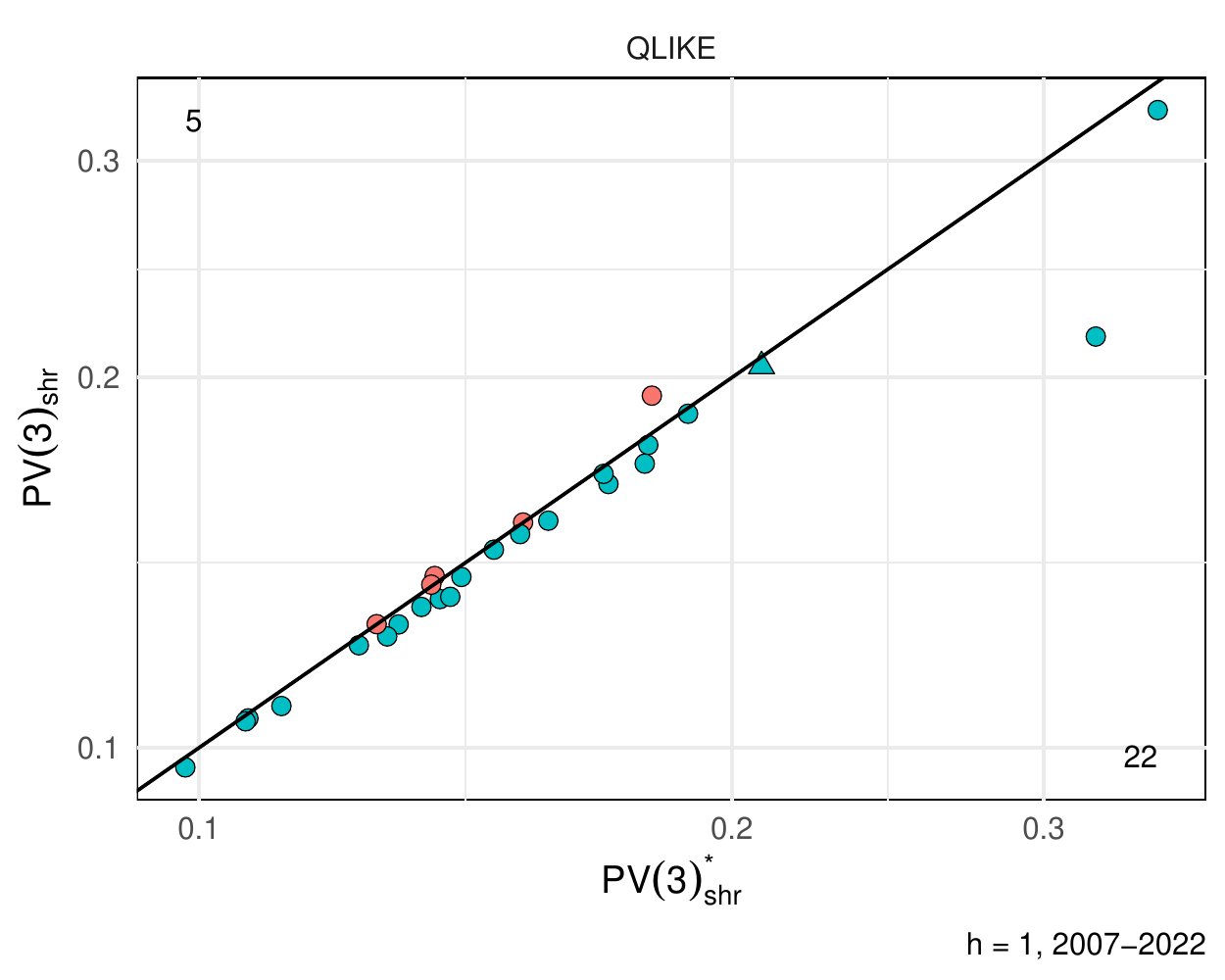}
	\caption{Accuracy of the \textbf{one-step ahead} daily $RV$ forecasts for the DJIA index (triangle) and 26 individual stocks (circle) in terms of $MSE$ (left panel) and $QLIKE$ (right panel) indices. Comparison between $HAR$ direct and $PV(3)_{shr}$ reconciliation-based forecasts. The black line represents the bisector, where either MSE's or QLIKE's for both approaches are equal. On the top-left (bottom-right) corner of each graph, the number of points above (below) the bisector is reported.}
	\label{fig:point_PV3shr}
\end{figure}

In \autoref{fig:point_PV3bu} the $MSE$ and $QLIKE$ indices for the bottom-up reconciliations in the two cases are shown. It clearly appears that no meaningful  accuracy improvement is obtained for the considered assets, and this turns out to be confirmed by \autoref{fig:point_PV3shr}, where the regression-based reconciliation approaches in the two cases are considered. 


%\clearpage

\section{Concluding remarks}
\label{sec:conclusion}

In this paper, we address whether using the disaggregate components of the daily realized volatility or combining them with the daily realized volatility forecasts improves the forecasting accuracy compared to using the daily realized volatility series alone.
To this end,
we investigate alternative ways of leveraging intraday $RV$ decompositions in forecasting daily $RV$. The key question of this study is whether it is beneficial to model and forecast daily $RV$ at the sub-component level, thus exploiting the informative content (from a forecasting point of view) of bottom time series, or whether a direct strategy should be preferred. The latter refers to the prediction of $RV$ by directly modeling it, even when the explanatory variables include a decomposition of the $RV$ itself. Differently, indirect (bottom-up) forecasting is based on the aggregation of models fit on the bottom series, whose forecasts are then aggregated to recover the prediction of the top series. Forecast reconciliation adds a further element, by restoring the aggregation constraints linking the bottom, possibly the intermediate, and the top-level time series. The idea is that an appropriate `imposition' to the forecasts of the same constraints valid for the observed data should improve the overall forecasting accuracy.
%Disaggregated data refer to the decomposition of daily $RV$ into several sub-components.
We obtain a forecast of different sub-components individually, and then combine them to estimate the forecast of the aggregated series in an indirect (bottom-up) and a reconciliation forecasting framework. %This alternative could increase the accuracy of the forecast; we modeled the sub-components by taking their characteristics into account. We used this alternative in the present work to understand if there was a reduction in the forecast error of the daily realized volatility by estimating a model for each sub-components.

Our main results can be summarized as follows.
Through a simple out-of-sample forecasting experiment, we show that both bottom-up and regression-based reconciliation procedures (\citealp{Wickramasuriya2019}) perform relatively well compared to the benchmark direct forecasting models by \cite{Corsi2009}, \cite{PattonSheppard2015}, and \cite{Bollerslevetal2022}, mostly when the $PV(3)$ model by \cite{Bollerslevetal2022} is used to produce base forecasts of the daily $RV$ top-level series.
We find substantial and significant reductions in forecast errors when using the new proposed indirect/reconciliation approaches. The disaggregated procedures do quite well in forecasting $RV$, and the reconciliation approach is generally more promising. For the HAR model, the differences between the bottom-up and the direct approach are clearly visible, and the regression-based reconciliation approach offers some additional improvements. This is somehow confirmed for both $SV$ and $PV(3)$ forecasts, where generally the combination scheme of single component models provides smaller forecast errors than both the indirect and direct forecasting approaches.

\singlespacing
\bibliographystyle{elsarticle-harv} 
\bibliography{biblio_CDFG}

\end{document}

\maketitle
\blfootnote{Email: \emailfoot{massimiliano.caporin@unipd.it} (M.C.), \emailfoot{tommaso.difonzo@unipd.it} (T.D.F.), \\ \phantom{Email: }\emailfoot{daniele.girolimetto@phd.unipd.it} (D.G.)}
%\clearpage
\doublespacing
\section*{A1: Data description}
\begin{figure}[H]
	\centering
	\includegraphics[width=.9\linewidth]{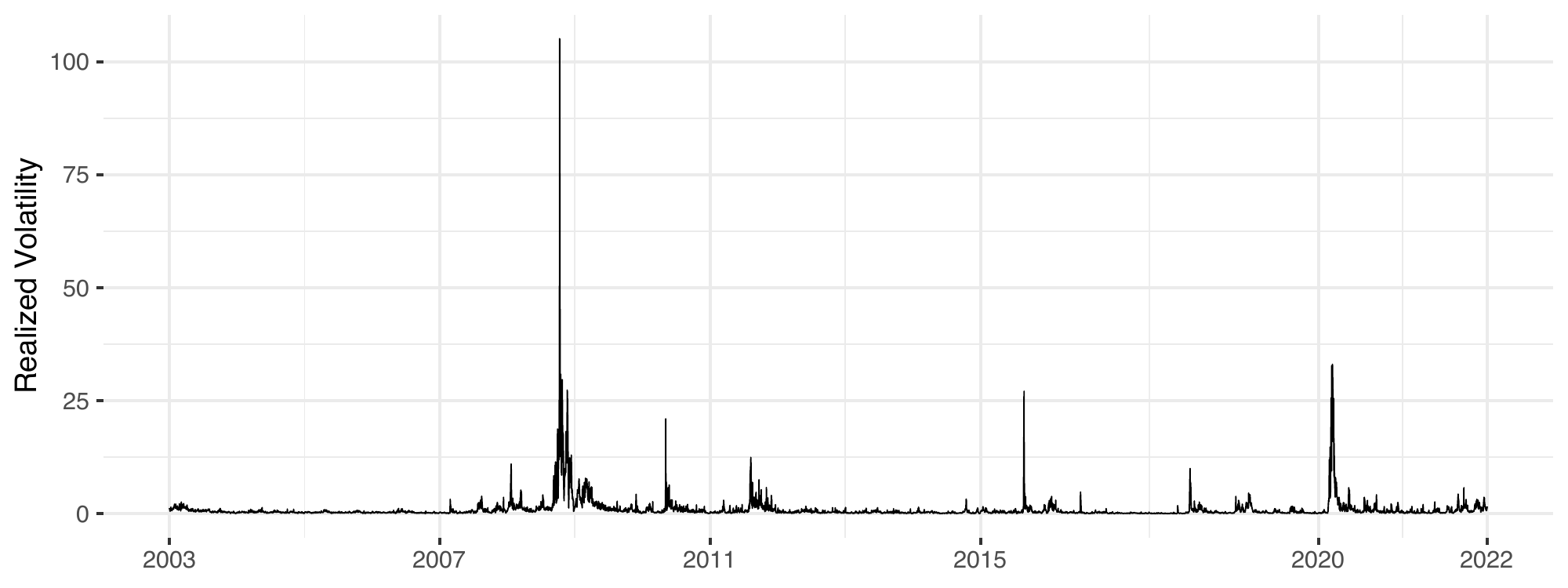}
	\caption{DJIA index: daily Realized Variance, January 2, 2003 - June 30, 2022 (4,908 days).}
	\label{fig:RVDJIA}
\end{figure}

\begin{figure}[H]
	\centering
	\includegraphics[width=.9\linewidth]{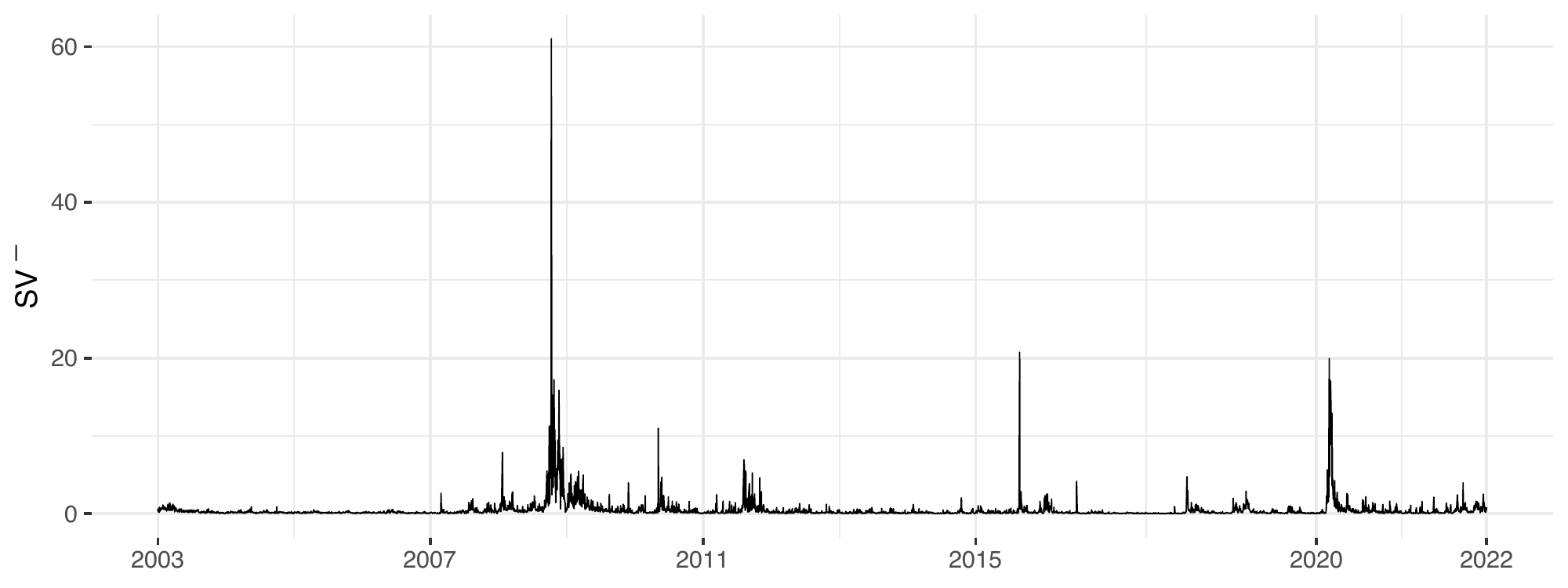}
	\includegraphics[width=.9\linewidth]{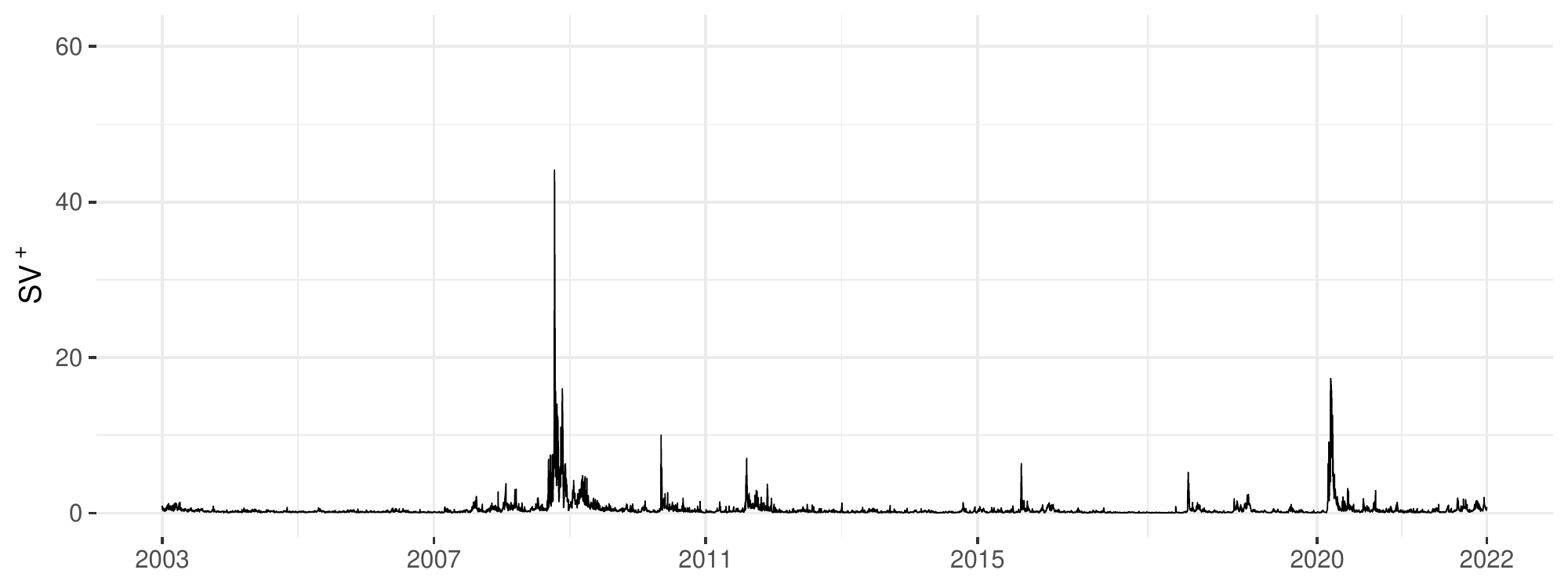}
	\caption{DJIA index: daily Realized Semi-Variances, January 2, 2003 - June 30, 2022 (4,908 days).}
	\label{fig:SVDJIA}
\end{figure}

\begin{figure}[H]
	\centering
	\includegraphics[width=.9\linewidth]{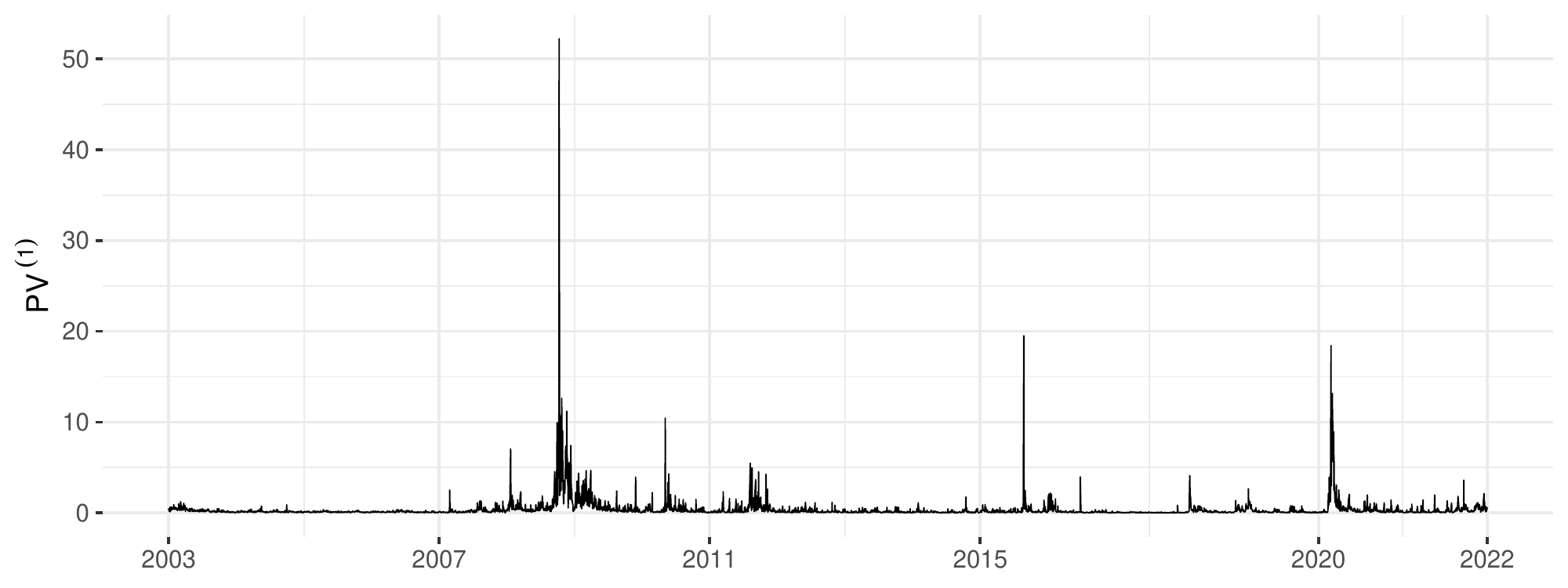}
	\includegraphics[width=.9\linewidth]{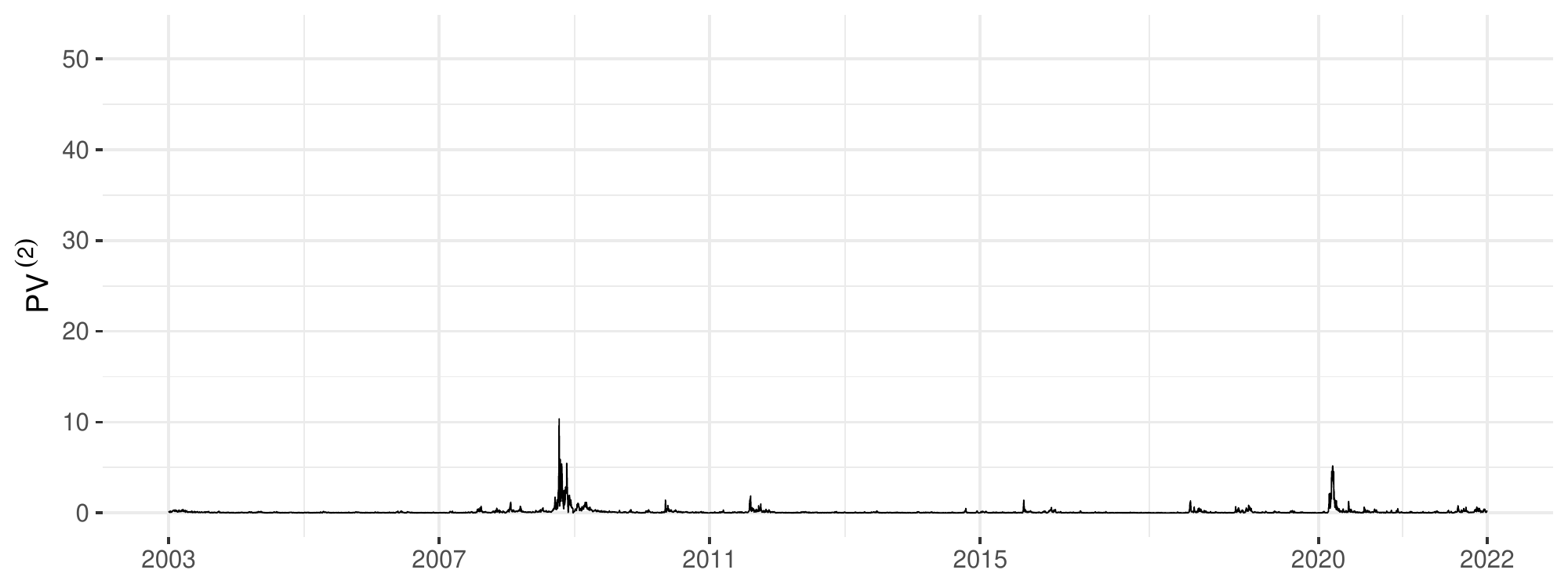}
	\includegraphics[width=.9\linewidth]{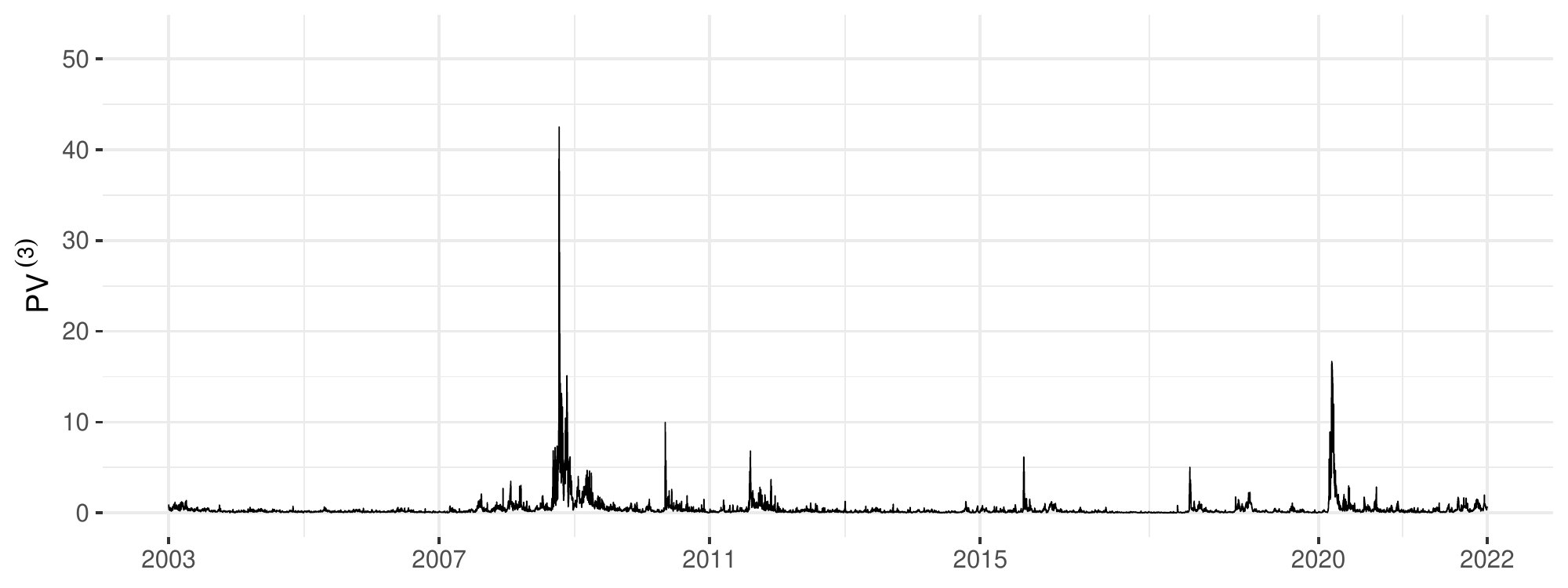}
	\caption{DJIA index: daily Realized Power-Variances, January 2, 2003 - June 30, 2022 (4,908 days).}
	\label{fig:PV3DJIA}
\end{figure}

\begin{figure}[H]
	\centering
	\includegraphics[width=.9\linewidth]{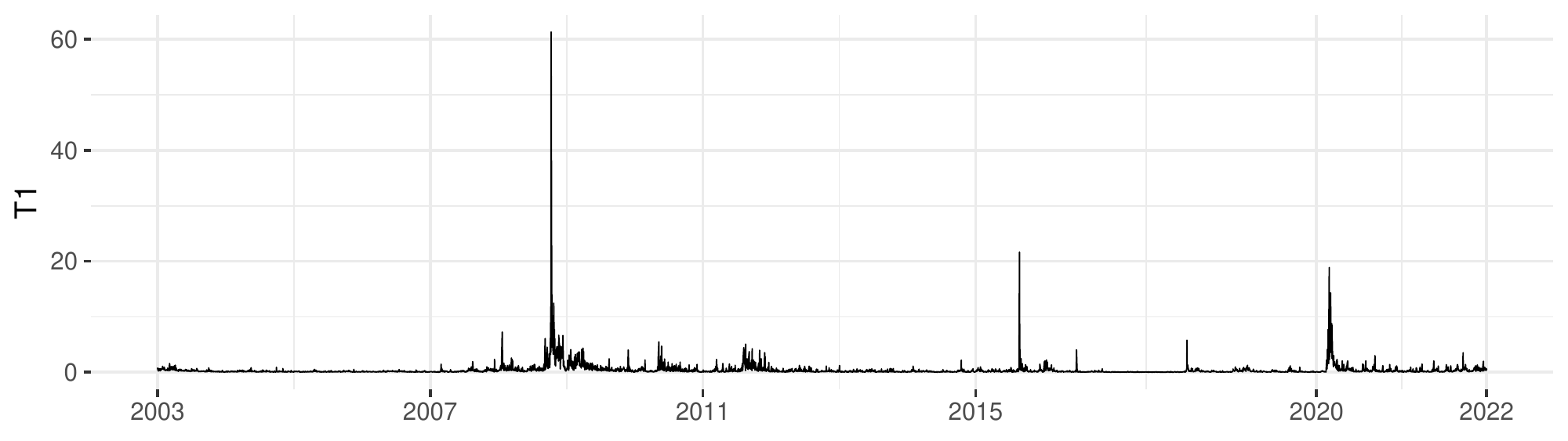}
	\includegraphics[width=.9\linewidth]{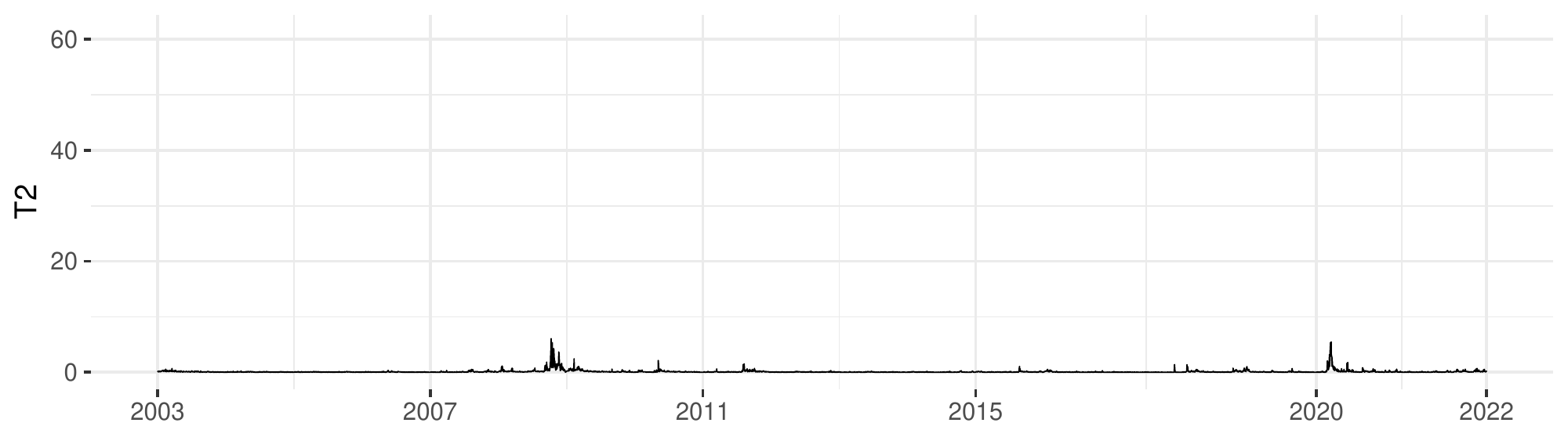}
	\includegraphics[width=.9\linewidth]{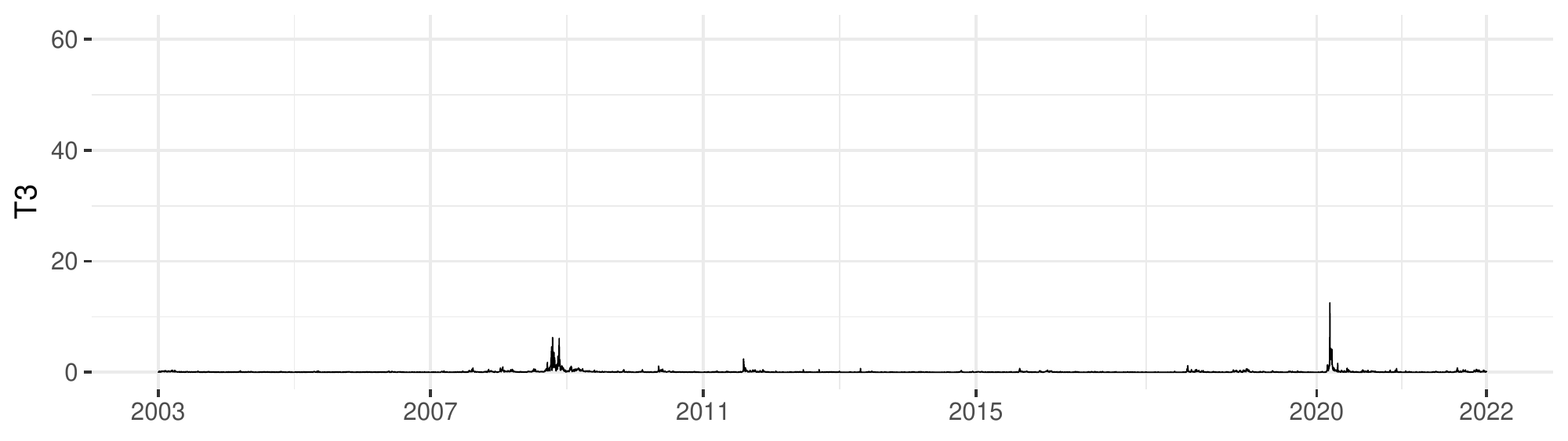}
	\includegraphics[width=.9\linewidth]{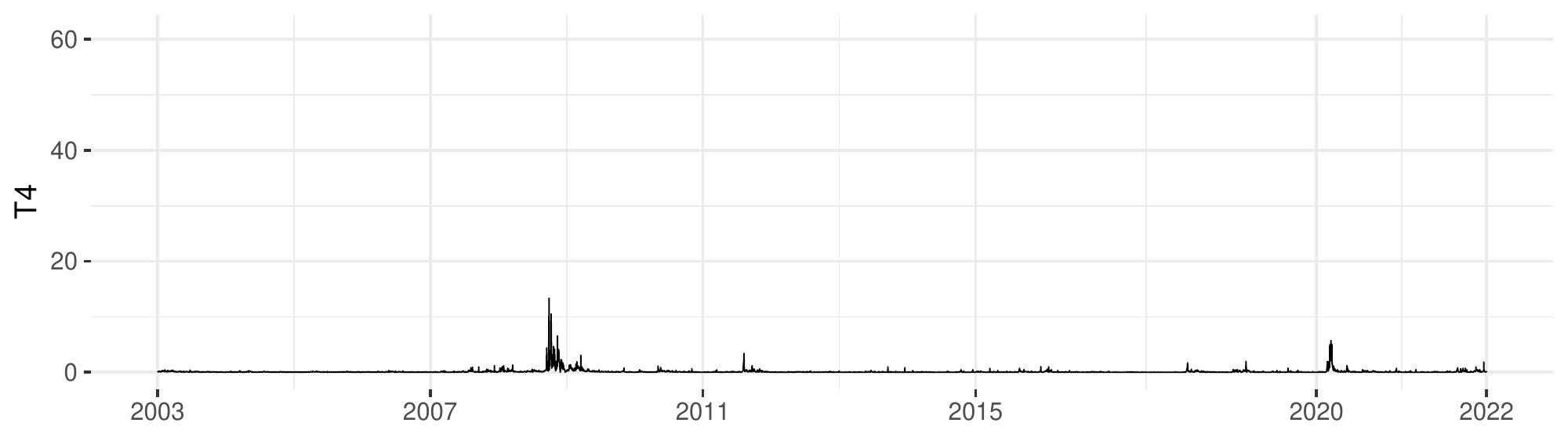}
	\includegraphics[width=.9\linewidth]{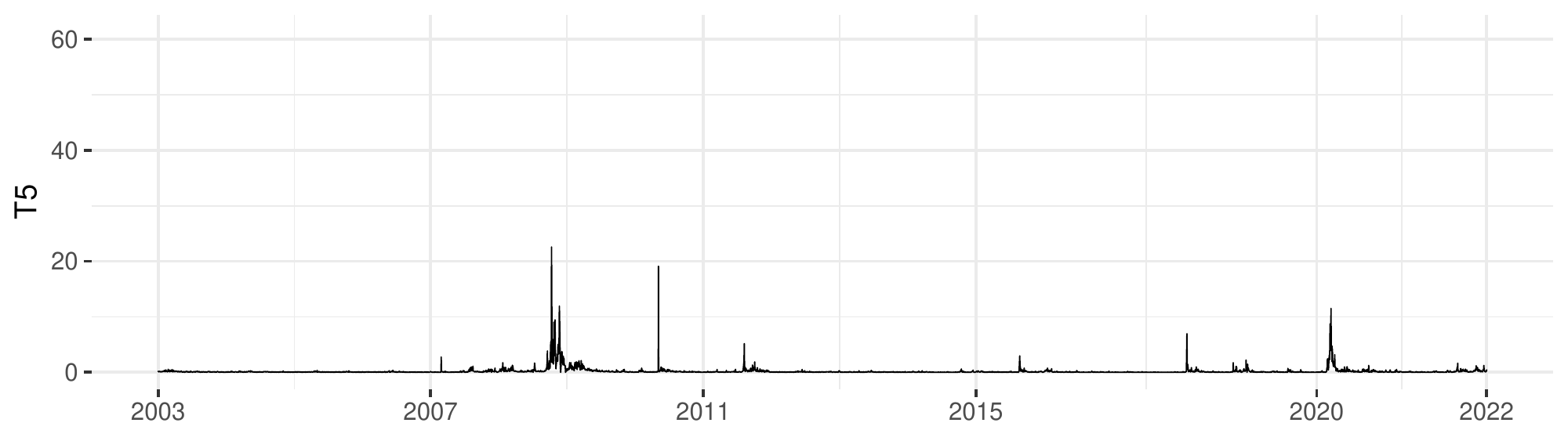}
	\caption{DJIA index: Realized Variances in non-overlapping 78-minutes segments of the day, January 2, 2003 - June 30, 2022 (4,908 days).}
	\label{fig:RVDJIAtime}
\end{figure}

\begin{figure}[H]
	\centering
	\includegraphics[width=.85\linewidth]{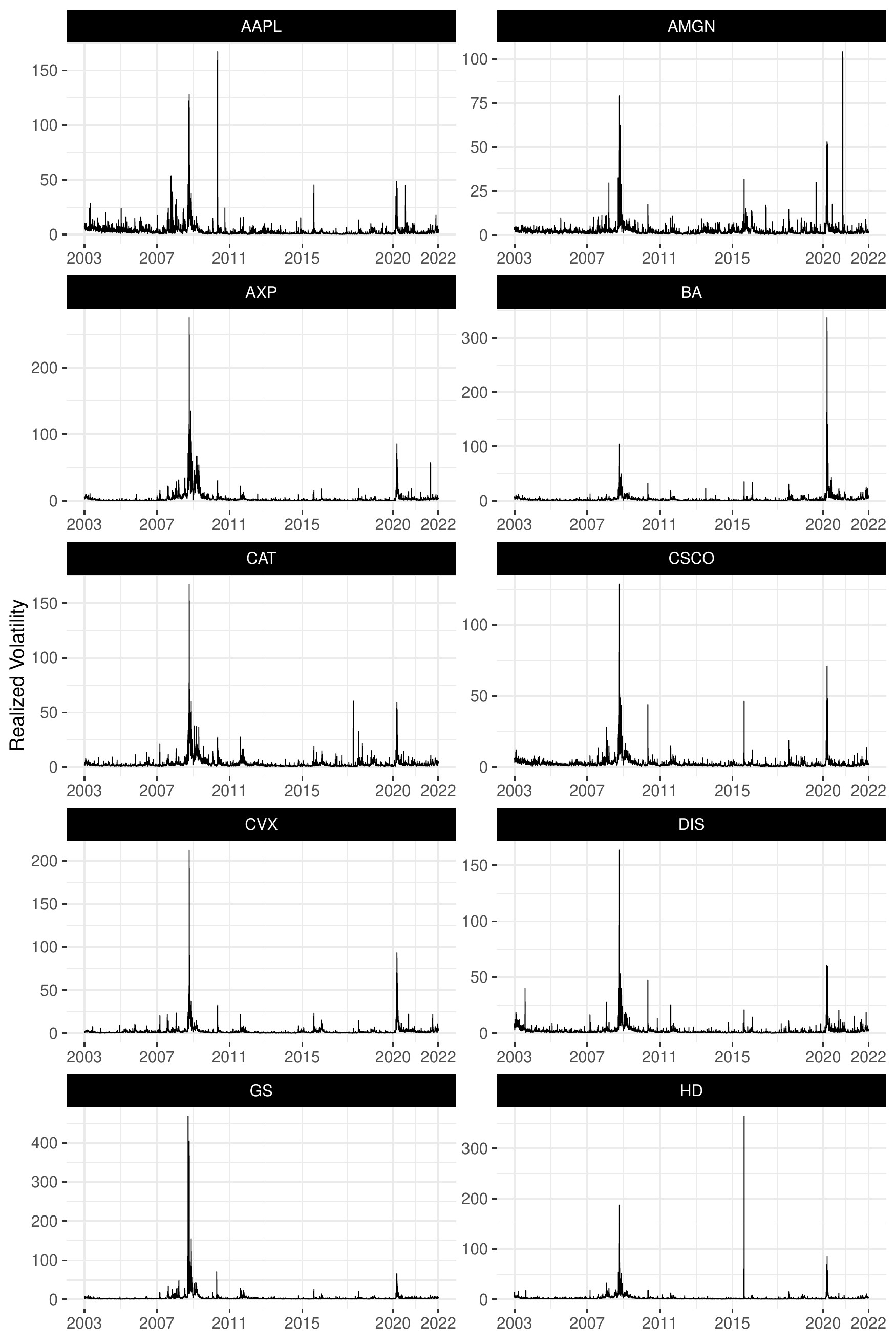}
	\caption{Daily Realized Variance of individual stocks, January 2, 2003 - June 30, 2022 (4,908 days). Tickers are described in Table \ref{Table_RV_descriptive_stats}.}
	\label{fig:RVstock1}
\end{figure}

\begin{figure}[H]
	\centering
	\includegraphics[width=.85\linewidth]{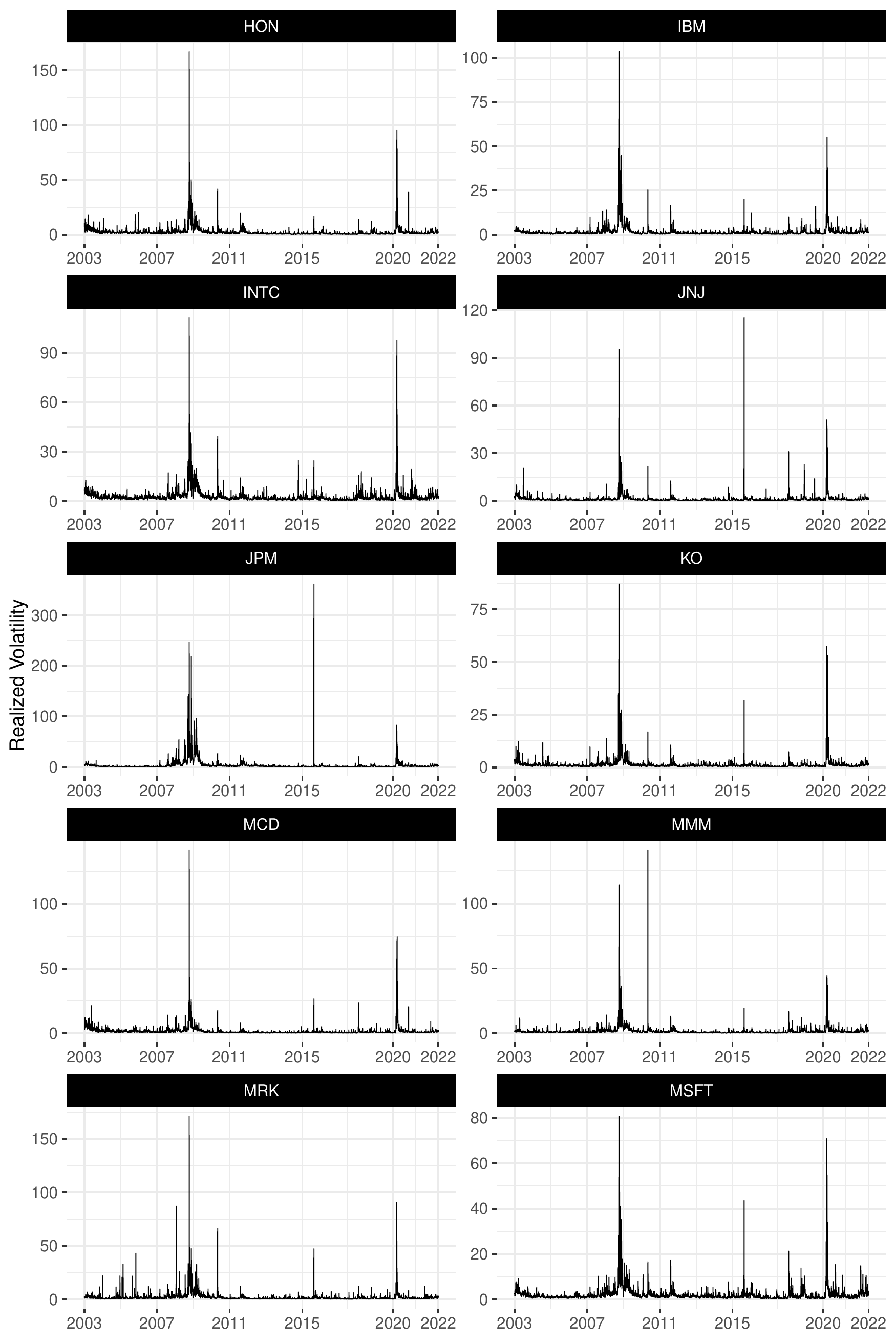}
	\caption{Daily Realized Variance of individual stocks, January 2, 2003 - June 30, 2022 (4,908 days). Tickers are described in Table \ref{Table_RV_descriptive_stats}.}
	\label{fig:RVstock2}
\end{figure}

\begin{figure}[H]
	\centering
	\includegraphics[width=.85\linewidth]{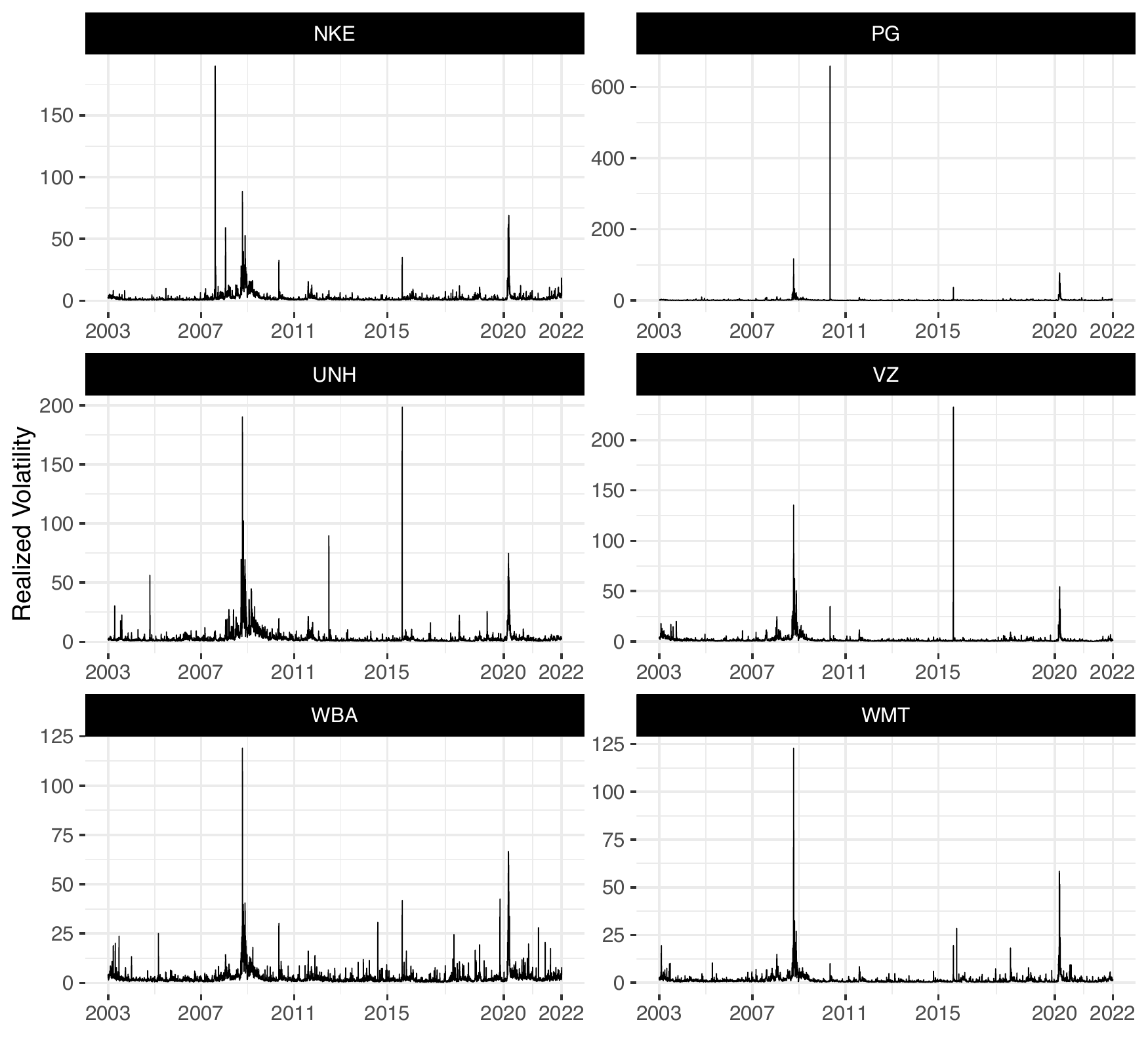}
	\caption{Daily Realized Variance of individual stocks, January 2, 2003 - June 30, 2022 (4,908 days). Tickers are described in Table \ref{Table_RV_descriptive_stats}.}
	\label{fig:RVstock3}
\end{figure}

A summary description of the daily $RV$s for the DJIA index and 26 individual stocks
is in Table \ref{Table_RV_descriptive_stats}.
Figures \ref{fig:RVDJIA}, \ref{fig:SVDJIA}, \ref{fig:PV3DJIA} show the time series of $RV$ and its $SV$ and $PV(3)$ components for the DJIA index, while the $RV$ of the DJIA index for the 5 time intervals of the day are shown in Figure \ref{fig:RVDJIAtime}.
The $RV$ graphs for the individual stocks are shown in Figures \ref{fig:RVstock1}, \ref{fig:RVstock2}, \ref{fig:RVstock3}.

%\vspace{.5cm}

%\textbf{[To be completed with a short preliminary descriptive data analysis]}

\begin{landscape}
	\begin{table}[htb]
		\centering
		\begin{footnotesize}
			\begin{tabular}[t]{lllccccccc}
\toprule
Ticker & Company & GICS$^*$ Sector & Min & Mean & Median & Max & St.dv. & Skew. & Kurt.\\
\midrule
DJIA & \multicolumn{2}{l}{\textit{Dow Jones Industrial Average index}} & 0.020 & 0.937 & 0.382 & 105.107 & 2.655 & 17.016 & 525.675\\
\midrule
AAPL & Apple Inc. & Information technology & 0.122 & 3.231 & 1.899 & 167.230 & 5.729 & 12.440 & 256.828\\
AMGN & Amgen & Health Care & 0.311 & 2.408 & 1.673 & 104.363 & 3.708 & 12.014 & 222.839\\
AXP & American Express & Financials & 0.131 & 3.450 & 1.210 & 275.174 & 8.992 & 10.565 & 212.313\\
BA & Boeing & Industrials & 0.187 & 3.147 & 1.566 & 336.957 & 8.591 & 20.090 & 598.972\\
CAT & Caterpillar Inc. & Industrials & 0.217 & 3.017 & 1.704 & 167.521 & 5.463 & 10.647 & 212.560\\
CSCO & Cisco & Information technology & 0.237 & 2.575 & 1.681 & 128.759 & 4.264 & 11.737 & 230.317\\
CVX & Chevron Corporation & Energy & 0.219 & 2.316 & 1.264 & 212.251 & 5.488 & 16.986 & 498.805\\
DIS & Disney & Communication Services & 0.168 & 2.399 & 1.349 & 163.579 & 4.587 & 13.627 & 355.131\\
GS & Goldman Sachs & Financials & 0.204 & 3.622 & 1.614 & 468.029 & 12.412 & 22.562 & 709.009\\
HD & Home Depot (The) & Consumer Discretionary & 0.163 & 2.581 & 1.344 & 364.037 & 7.350 & 29.328 & 1287.173\\
HON & Honeywell & Industrials & 0.146 & 2.365 & 1.349 & 166.977 & 4.922 & 13.936 & 334.939\\
IBM & IBM & Information technology & 0.140 & 1.604 & 0.916 & 103.611 & 3.403 & 12.103 & 236.646\\
INTC & Intel & Information technology & 0.337 & 2.851 & 1.904 & 111.192 & 4.407 & 11.059 & 192.990\\
JNJ & Johnson \& Johnson & Health Care & 0.118 & 1.219 & 0.700 & 115.272 & 3.132 & 19.937 & 578.804\\
JPM & JPMorgan Chase & Financials & 0.186 & 3.738 & 1.410 & 362.245 & 11.103 & 14.267 & 331.107\\
KO & The Coca-Cola Company & Consumer Staples & 0.188 & 1.311 & 0.796 & 87.093 & 2.786 & 14.112 & 298.911\\
MCD & McDonald’s & Consumer Discretionary & 0.125 & 1.638 & 0.918 & 141.630 & 3.742 & 17.477 & 495.913\\
MMM & 3M & Industrials & 0.135 & 1.680 & 0.979 & 141.197 & 3.784 & 18.862 & 565.079\\
MRK & Merck & Health Care & 0.242 & 2.154 & 1.215 & 171.104 & 4.636 & 16.048 & 440.157\\
MSFT & Microsoft & Information technology & 0.089 & 2.162 & 1.353 & 80.692 & 3.495 & 9.640 & 143.703\\
NKE & Nike Inc. & Consumer Discretionary & 0.208 & 2.321 & 1.249 & 189.811 & 4.891 & 16.803 & 500.581\\
PG & Procter \& Gamble & Consumer Staples & 0.152 & 1.404 & 0.733 & 658.769 & 10.002 & 58.625 & 3811.064\\
UNH & UnitedHealth Group & Health Care & 0.227 & 3.115 & 1.592 & 198.469 & 6.953 & 12.739 & 271.191\\
VZ & Verizon & Communication Services & 0.156 & 1.864 & 1.034 & 232.582 & 4.999 & 26.244 & 1046.776\\
WBA & Walgreens Boots Alliance & Consumer Staples & 0.215 & 2.581 & 1.666 & 119.073 & 3.959 & 11.147 & 219.696\\
WMT & Walmart & Consumer Staples & 0.127 & 1.464 & 0.890 & 122.954 & 2.974 & 19.248 & 635.463\\
\bottomrule\\[-.2cm]
%\multicolumn{10}{l}{\textit{Notes}: 
%	The table reports descriptive statistics of the (open-to-close) daily Realized Variances of the DJIA index and 26}\\
%\multicolumn{10}{l}{individual stocks for the period January 2, 2003 - June 30, 2022 (4,908 daily observations).}\\
\multicolumn{10}{l}{$^*$ Global Index Classification Standard.}\\
\end{tabular}\\[0.2cm]

		\end{footnotesize}
		\caption{Descriptive statistics of the (open-to-close) daily Realized Variances of the DJIA index and 26 individual stocks for the period January 2, 2003 - June 30, 2022 (4,908 daily observations).}
%		\caption{Descriptive statistics of daily Realized Variances of the DJIA index and individual stocks, 2003-2022 (4,908 days).}
		\label{Table_RV_descriptive_stats}
	\end{table}
\end{landscape}

% Grafici a colori con evidenziati i 4 diversi periodi 2007-2010, 2011-2014, 2015-2019, 2020-2022
%\begin{figure}[H]
%		\centering
%	\includegraphics[width=.9\linewidth]{img/INDU.pdf}
%	\caption{DJIA index: daily Realized Variance 2003-2022 (4,908 days)}
%\end{figure}
%\begin{figure}[H]
%	\centering
%	\includegraphics[width=.9\linewidth]{img/GBPV.pdf}
%	\caption{}
%\end{figure}
%\begin{figure}[H]
%	\centering
%	\includegraphics[width=.9\linewidth]{img/TIME.pdf}
%	\caption{}
%\end{figure}

\section*{A2: Alternative hierarchical/grouped representations of different intraday decompositions of daily $RV$}

In Table \ref{tab:hierarchylabels} are reported the eleven hierarchies/groupings deriving from different Temporal-and-Volatility based decompositions. A complete hierarchy/grouping involves the bottom level series produced by a decomposition according to time and/or volatility, and the series forming at least one upper level (by time, quantile or both). A simple hierarchy consists of the top-level series (daily $RV$) and a single level of bottom time series disaggregated according time, quantile, or both. The largest decomposition, CTPV(3), involves 24 series, i.e. daily $RV$, 3 series distinguished only by quantile thresholds, 5 temporally aggregated series over nonoverlapping, consecutive 78 minutes intervals, 15 series cross-classified by quantile and temporal intervals.
This is a grouped series, formed by hierarchies PV(3)-T (see figure \ref{fig:PV(3)-T}) and T-PV(3) (see figure \ref{fig:T-PV(3)}), whose matrix structural representation is given by figure \ref{fig:eqCTPV(3)}).
Instead, the simplest hierarchy, SSV, considers the daily $RV$ as top-level series and two bottom series corresponding to `Bad' ($r_{i,t} <0$) and `Good' ($r_{i,t} > 0$) volatility, respectively.

\begin{table}[htb]
	\centering
	\caption{Temporal-and-Volatility based intraday RV decompositions}
	\begin{tabular}{llcccccc}
		\toprule
		Name & Intraday decomposition & H/G & $n_b$ & $n_a$ & $n$ & Fig. \# & Eq. \#\\
		\midrule
		ST   & Simple Temporal & H & 5 & 1 & 6 & \ref{fig:ST} & \ref{fig:eqST}\\
		\midrule
		SSV  & Simple SV (Good \& Bad) & H & 2 & 1 & 3 & \ref{fig:SSV} & \ref{fig:eqSSV}\\
		STSV & Simple Temporal-and-SV & H & 10 & 1 & 11 & \ref{fig:STSV} & \ref{fig:eqSTSV}\\
		SV-T & SV with Temporal & H & 10 & 3 & 13 & \ref{fig:SV-T} & \ref{fig:eqSV-T}\\
		T-SV & Temporal with SV & H & 10 & 6 & 16 & \ref{fig:T-SV} & \ref{fig:eqT-SV}\\
		CTSV & Complete Temporal-and-SV & G & 10 & 8 & 18 & $-$ & \ref{fig:eqCTSV}\\
		\midrule
		SPV(3)  & Simple PV(3) & H & 3 & 1 & 4 & \ref{fig:SPV(3)} & \ref{fig:eqSPV(3)}\\
		STPV(3) & Simple Temporal-and-PV(3) & H & 15 & 1 & 16 & \ref{fig:STPV(3)} & \ref{fig:eqSTPV(3)}\\
		PV(3)-T & PV(3) with Temporal & H & 15 & 4 & 19 & \ref{fig:PV(3)-T} & \ref{fig:eqPV(3)-T}\\	
		T-PV(3) & Temporal with PV(3) & H & 15 & 6 & 21 & \ref{fig:T-PV(3)} & \ref{fig:eqT-PV(3)}\\
		CTPV(3) & Complete Temporal-and-PV(3) & G & 15 & 9 & 24 & $-$ & \ref{fig:eqCTPV(3)}\\
		\bottomrule
		\multicolumn{7}{l}{H: hierarchy; G: grouping.}\\
		\multicolumn{7}{l}{$n_b$: n. of bottom variables; $n_a$: n. of upper variables; $n = n_b + n_a$.}\\
	\end{tabular}
	\label{tab:hierarchylabels}
\end{table}

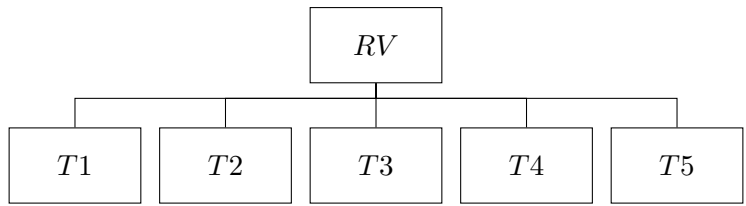
\begin{figure}[H]
	\centering
	\begin{tikzpicture}[baseline=(current  bounding  box.center),
		rel/.append style={draw=black, font=\small,
			minimum width=1.75cm,
			minimum height=1cm},
		connection/.style ={inner sep =0, outer sep =0}]
		
		\node[rel] at (0, 0) (t1){$T1$};
		\node[rel] at (2, 0) (t2){$T2$};
		\node[rel] at (4, 0) (t3){$T3$};
		\node[rel] at (6, 0) (t4){$T4$};
		\node[rel] at (8, 0) (t5){$T5$};
		
		\node[rel] at (4, 1.6) (RV){$RV$};
		
		\relation{0.2}{t1}{RV};
		\relation{0.2}{t2}{RV};
		\relation{0.2}{t3}{RV};
		\relation{0.2}{t4}{RV};
		\relation{0.2}{t5}{RV};
	\end{tikzpicture}
	\caption{Hierarchical representation of the Simple Temporal (ST) decomposition of daily $RV$ using five intraday intervals. $n_b=5, n_a=1, n=6$. More details are in Table \ref{tab:hierarchylabels}.}
	\label{fig:ST}
\end{figure}

%\clearpage

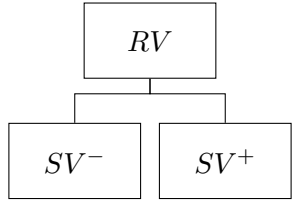
\begin{figure}[H]
	\centering
	\begin{tikzpicture}[baseline=(current  bounding  box.center),
		rel/.append style={draw=black, font=\small,
			minimum width=1.75cm,
			minimum height=1cm},
		connection/.style ={inner sep =0, outer sep =0}]
		
		\node[rel] at (0, 0) (svm){$SV^{-}$};
		\node[rel] at (2, 0) (svp){$SV^{+}$};
		
		\node[rel] at (1, 1.6) (RV){$RV$};

		\relation{0.2}{svm}{RV};
		\relation{0.2}{svp}{RV};
	\end{tikzpicture}
	\caption{Hierarchical representation of the Simple SV (SSV) decomposition of daily RV. $n_b=2$, $n_a=1$, $n=3$. More details are in Table \ref{tab:hierarchylabels}.}
	\label{fig:SSV}
\end{figure}

\begin{figure}[H]
	\centering
	\resizebox{\linewidth}{!}{
		\begin{tikzpicture}[baseline=(current  bounding  box.center),
			rel/.append style={draw=black, font=\small,
				minimum width=1.75cm,
				minimum height=1cm},
			connection/.style ={inner sep =0, outer sep =0}]
			
			\node[rel] at (0, 0) (t1svm){$T1SV^{-}$};
			\node[rel] at (2, 0) (t2svm){$T2SV^{-}$};
			\node[rel] at (4, 0) (t3svm){$T3SV^{-}$};
			\node[rel] at (6, 0) (t4svm){$T4SV^{-}$};
			\node[rel] at (8, 0) (t5svm){$T5SV^{-}$};
			
			\node[rel] at (10, 0) (t1svp){$T1SV^{+}$};
			\node[rel] at (12, 0) (t2svp){$T2SV^{+}$};
			\node[rel] at (14, 0) (t3svp){$T3SV^{+}$};
			\node[rel] at (16, 0) (t4svp){$T4SV^{+}$};
			\node[rel] at (18, 0) (t5svp){$T5SV^{+}$};
			
			\node[rel] at (9, 1.6) (RV){$RV$};
			
			\relation{0.2}{t1svm}{RV};
			\relation{0.2}{t1svp}{RV};
			
			\relation{0.2}{t2svm}{RV};
			\relation{0.2}{t2svp}{RV};
			
			\relation{0.2}{t3svm}{RV};
			\relation{0.2}{t3svp}{RV};
			
			\relation{0.2}{t4svm}{RV};
			\relation{0.2}{t4svp}{RV};
			
			\relation{0.2}{t5svm}{RV};
			\relation{0.2}{t5svp}{RV};
	\end{tikzpicture}}
	\caption{Hierarchical representation of the Simple Temporal-and-SV (STSV) decomposition of daily RV. $n_b=10$, $n_a=1$, $n=11$. More details are in Table \ref{tab:hierarchylabels}.}
	\label{fig:STSV}
\end{figure}

\begin{figure}[H]
	\centering
	\resizebox{\linewidth}{!}{
		\begin{tikzpicture}[baseline=(current  bounding  box.center),
			rel/.append style={draw=black, font=\small,
				minimum width=1.75cm,
				minimum height=1cm},
			connection/.style ={inner sep =0, outer sep =0}]
			
			\node[rel] at (0, 0) (t1svm){$T1SV^{-}$};
			\node[rel] at (2, 0) (t2svm){$T2SV^{-}$};
			\node[rel] at (4, 0) (t3svm){$T3SV^{-}$};
			\node[rel] at (6, 0) (t4svm){$T4SV^{-}$};
			\node[rel] at (8, 0) (t5svm){$T5SV^{-}$};
			
			\node[rel] at (10, 0) (t1svp){$T1SV^{+}$};
			\node[rel] at (12, 0) (t2svp){$T2SV^{+}$};
			\node[rel] at (14, 0) (t3svp){$T3SV^{+}$};
			\node[rel] at (16, 0) (t4svp){$T4SV^{+}$};
			\node[rel] at (18, 0) (t5svp){$T5SV^{+}$};
			
			\node[rel] at (4, 1.6) (svm){$SV^{-}$};
			\node[rel] at (14, 1.6) (svp){$SV^{+}$};
			
			\node[rel] at (9, 3.2) (RV){$RV$};
			
			\relation{0.2}{t1svm}{svm};
			\relation{0.2}{t1svp}{svp};
			
			\relation{0.2}{t2svm}{svm};
			\relation{0.2}{t2svp}{svp};
			
			\relation{0.2}{t3svm}{svm};
			\relation{0.2}{t3svp}{svp};
			
			\relation{0.2}{t4svm}{svm};
			\relation{0.2}{t4svp}{svp};
			
			\relation{0.2}{t5svm}{svm};
			\relation{0.2}{t5svp}{svp};
			
			\relation{0.2}{svm}{RV};
			\relation{0.2}{svp}{RV};
	\end{tikzpicture}}
	\caption{Hierarchical representation of the SV with Temporal (SV-T) decomposition of daily $RV$. $n_b=10$, $n_a=3$, $n=13$. More details are in Table \ref{tab:hierarchylabels}.}
	\label{fig:SV-T}
\end{figure}

\begin{figure}[H]
	\centering
	\resizebox{\linewidth}{!}{
		\begin{tikzpicture}[baseline=(current  bounding  box.center),
			rel/.append style={draw=black, font=\small,
				minimum width=1.75cm,
				minimum height=1cm},
			connection/.style ={inner sep =0, outer sep =0}]
			
			\node[rel] at (0, 0) (t1svm){$T1SV^{-}$};
			\node[rel] at (2, 0) (t1svp){$T1SV^{+}$};
			
			\node[rel] at (4, 0) (t2svm){$T2SV^{-}$};
			\node[rel] at (6, 0) (t2svp){$T2SV^{+}$};
			
			\node[rel] at (8, 0) (t3svm){$T3SV^{-}$};
			\node[rel] at (10, 0) (t3svp){$T3SV^{+}$};
			
			\node[rel] at (12, 0) (t4svm){$T4SV^{-}$};
			\node[rel] at (14, 0) (t4svp){$T4SV^{+}$};
			
			\node[rel] at (16, 0) (t5svm){$T5SV^{-}$};
			\node[rel] at (18, 0) (t5svp){$T5SV^{+}$};

			\node[rel] at (1, 1.6) (t1){$T1$};
			\node[rel] at (5, 1.6) (t2){$T2$};
			\node[rel] at (9, 1.6) (t3){$T3$};
			\node[rel] at (13, 1.6) (t4){$T4$};
			\node[rel] at (17, 1.6) (t5){$T5$};
			
			\node[rel] at (9, 3.2) (RV){$RV$};
			
			\relation{0.2}{t1svm}{t1};
			\relation{0.2}{t1svp}{t1};
			
			\relation{0.2}{t2svm}{t2};
			\relation{0.2}{t2svp}{t2};
			
			\relation{0.2}{t3svm}{t3};
			\relation{0.2}{t3svp}{t3};
			
			\relation{0.2}{t4svm}{t4};
			\relation{0.2}{t4svp}{t4};
			
			\relation{0.2}{t5svm}{t5};
			\relation{0.2}{t5svp}{t5};
			
			\relation{0.2}{t1}{RV};
			\relation{0.2}{t2}{RV};
			\relation{0.2}{t3}{RV};
			\relation{0.2}{t4}{RV};
			\relation{0.2}{t5}{RV};
	\end{tikzpicture}}
	\caption{Hierarchical representation of the Temporal with SV (T-SV) decomposition of daily $RV$. $n_b=10$, $n_a=6$, $n=16$. More details are in Table \ref{tab:hierarchylabels}.}
	\label{fig:T-SV}
\end{figure}

\begin{figure}[H]
	\centering
	\begin{tikzpicture}[baseline=(current  bounding  box.center),
		rel/.append style={draw=black, font=\small,
			minimum width=1.75cm,
			minimum height=1cm},
		connection/.style ={inner sep =0, outer sep =0}]
		
		\node[rel] at (0, 0) (pv1){$PV^{(1)}$};
		\node[rel] at (2, 0) (pv2){$PV^{(2)}$};
		\node[rel] at (4, 0) (pv3){$PV^{(3)}$};
		
		\node[rel] at (2, 1.6) (RV){$RV$};

		\relation{0.2}{pv1}{RV};
		\relation{0.2}{pv2}{RV};
		\relation{0.2}{pv3}{RV};
	\end{tikzpicture}
	\caption{Hierarchical representation of the Simple PV(3) (SPV(3)) decomposition of daily RV. $n_b=3$, $n_a=1$, $n=4$. More details in Table \ref{tab:hierarchylabels}.}
	\label{fig:SPV(3)}
\end{figure}

\begin{landscape}
	\begin{figure}[H]
		\centering
		\resizebox{\linewidth}{!}{
			\begin{tikzpicture}[baseline=(current  bounding  box.center),
				rel/.append style={draw=black, font=\small,
					minimum width=1.75cm,
					minimum height=1cm},
				connection/.style ={inner sep =0, outer sep =0}]
				
				\node[rel] at (0, 0) (t1pv1){$T1PV^{(1)}$};
				\node[rel] at (2, 0) (t2pv1){$T2PV^{(1)}$};
				\node[rel] at (4, 0) (t3pv1){$T3PV^{(1)}$};
				\node[rel] at (6, 0) (t4pv1){$T4PV^{(1)}$};
				\node[rel] at (8, 0) (t5pv1){$T5PV^{(1)}$};
				\node[rel] at (10, 0) (t1pv2){$T1PV^{(2)}$};
				\node[rel] at (12, 0) (t2pv2){$T2PV^{(2)}$};
				\node[rel] at (14, 0) (t3pv2){$T3PV^{(2)}$};
				\node[rel] at (16, 0) (t4pv2){$T4PV^{(2)}$};
				\node[rel] at (18, 0) (t5pv2){$T5PV^{(2)}$};
				\node[rel] at (20, 0) (t1pv3){$T1PV^{(3)}$};
				\node[rel] at (22, 0) (t2pv3){$T2PV^{(3)}$};
				\node[rel] at (24, 0) (t3pv3){$T3PV^{(3)}$};
				\node[rel] at (26, 0) (t4pv3){$T4PV^{(3)}$};
				\node[rel] at (28, 0) (t5pv3){$T5PV^{(3)}$};
				
				\node[rel] at (14, 1.6) (RV){$RV$};

				\relation{0.2}{t1pv1}{RV};
				\relation{0.2}{t2pv1}{RV};
				\relation{0.2}{t3pv1}{RV};
				\relation{0.2}{t4pv1}{RV};
				\relation{0.2}{t5pv1}{RV};
				\relation{0.2}{t1pv2}{RV};
				\relation{0.2}{t2pv2}{RV};
				\relation{0.2}{t3pv2}{RV};
				\relation{0.2}{t4pv2}{RV};
				\relation{0.2}{t5pv2}{RV};
				\relation{0.2}{t1pv3}{RV};
				\relation{0.2}{t2pv3}{RV};
				\relation{0.2}{t3pv3}{RV};
				\relation{0.2}{t4pv3}{RV};
				\relation{0.2}{t5pv3}{RV};
		\end{tikzpicture}}
		\caption{Hierarchical representation of the Simple Temporal-and-PV(3) (STPV(3)) decomposition of daily $RV$. $n_b=15$, $n_a=1$, $n=16$. More details in Table \ref{tab:hierarchylabels}.}
		\label{fig:STPV(3)}
	\end{figure}
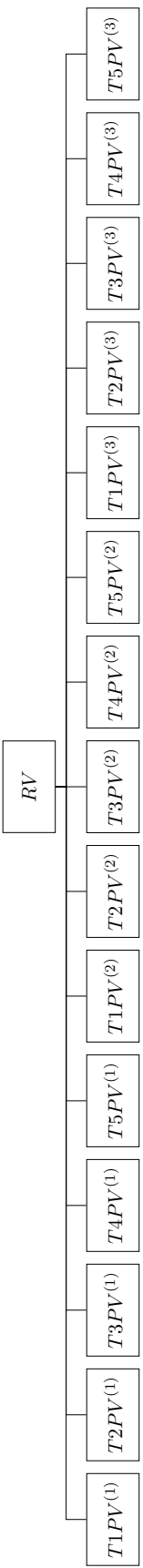
	
	\begin{figure}[H]
		\centering
		\resizebox{\linewidth}{!}{
			\begin{tikzpicture}[baseline=(current  bounding  box.center),
				rel/.append style={draw=black, font=\small,
					minimum width=1.75cm,
					minimum height=1cm},
				connection/.style ={inner sep =0, outer sep =0}]
				
				\node[rel] at (0, 0) (t1pv1){$T1PV^{(1)}$};
				\node[rel] at (2, 0) (t2pv1){$T2PV^{(1)}$};
				\node[rel] at (4, 0) (t3pv1){$T3PV^{(1)}$};
				\node[rel] at (6, 0) (t4pv1){$T4PV^{(1)}$};
				\node[rel] at (8, 0) (t5pv1){$T5PV^{(1)}$};
				\node[rel] at (10, 0) (t1pv2){$T1PV^{(2)}$};
				\node[rel] at (12, 0) (t2pv2){$T2PV^{(2)}$};
				\node[rel] at (14, 0) (t3pv2){$T3PV^{(2)}$};
				\node[rel] at (16, 0) (t4pv2){$T4PV^{(2)}$};
				\node[rel] at (18, 0) (t5pv2){$T5PV^{(2)}$};
				\node[rel] at (20, 0) (t1pv3){$T1PV^{(3)}$};
				\node[rel] at (22, 0) (t2pv3){$T2PV^{(3)}$};
				\node[rel] at (24, 0) (t3pv3){$T3PV^{(3)}$};
				\node[rel] at (26, 0) (t4pv3){$T4PV^{(3)}$};
				\node[rel] at (28, 0) (t5pv3){$T5PV^{(3)}$};

				\node[rel] at (4, 1.6) (pv1){$PV^{(1)}$};
				\node[rel] at (14, 1.6) (pv2){$PV^{(2)}$};
				\node[rel] at (24, 1.6) (pv3){$PV^{(3)}$};
				
				\node[rel] at (14, 3.2) (RV){$RV$};

				\relation{0.2}{t1pv1}{pv1};
				\relation{0.2}{t2pv1}{pv1};
				\relation{0.2}{t3pv1}{pv1};
				\relation{0.2}{t4pv1}{pv1};
				\relation{0.2}{t5pv1}{pv1};
				\relation{0.2}{t1pv2}{pv2};
				\relation{0.2}{t2pv2}{pv2};
				\relation{0.2}{t3pv2}{pv2};
				\relation{0.2}{t4pv2}{pv2};
				\relation{0.2}{t5pv2}{pv2};
				\relation{0.2}{t1pv3}{pv3};
				\relation{0.2}{t2pv3}{pv3};
				\relation{0.2}{t3pv3}{pv3};
				\relation{0.2}{t4pv3}{pv3};
				\relation{0.2}{t5pv3}{pv3};
				\relation{0.2}{pv1}{RV};
				\relation{0.2}{pv2}{RV};
				\relation{0.2}{pv3}{RV};
		\end{tikzpicture}}
		\caption{Hierarchical representation of the PV(3) with Temporal (PV(3)-T) decomposition of daily $RV$. $n_b=15$, $n_a=4$, $n=19$. More details in Table \ref{tab:hierarchylabels}.}
		\label{fig:PV(3)-T}
	\end{figure}
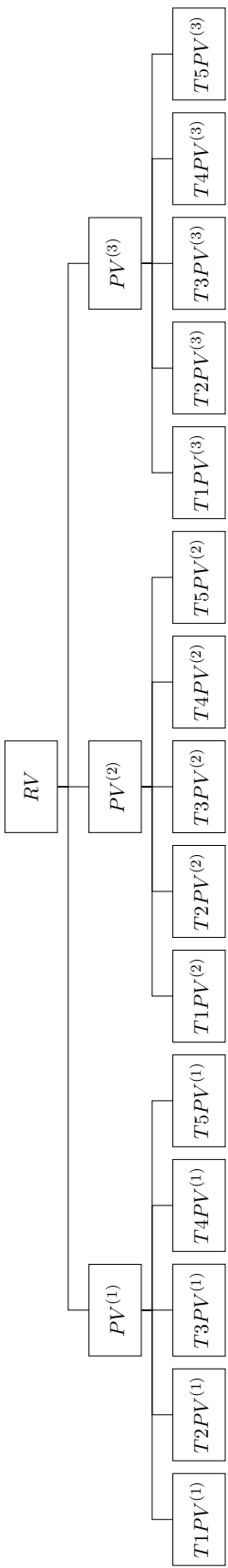	
	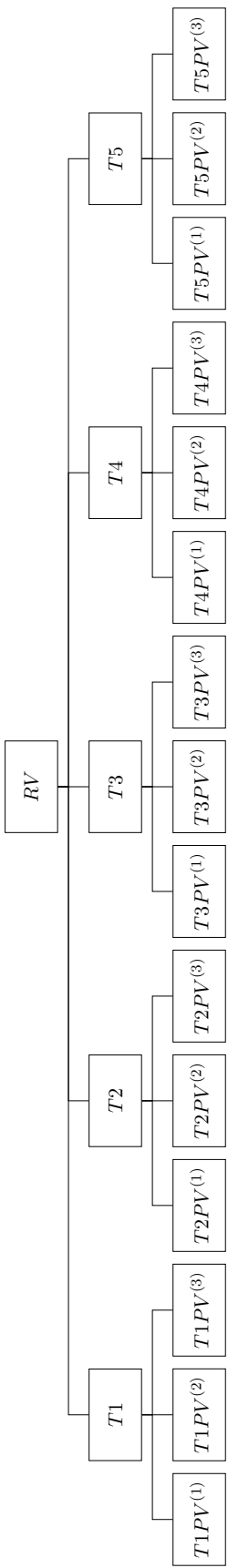
\begin{figure}[H]
		\centering
		\resizebox{\linewidth}{!}{
			\begin{tikzpicture}[baseline=(current  bounding  box.center),
				rel/.append style={draw=black, font=\small,
					minimum width=1.75cm,
					minimum height=1cm},
				connection/.style ={inner sep =0, outer sep =0}]
				
				\node[rel] at (0, 0) (t1pv1){$T1PV^{(1)}$};
				\node[rel] at (2, 0) (t1pv2){$T1PV^{(2)}$};
				\node[rel] at (4, 0) (t1pv3){$T1PV^{(3)}$};
				
				\node[rel] at (6, 0) (t2pv1){$T2PV^{(1)}$};
				\node[rel] at (8, 0) (t2pv2){$T2PV^{(2)}$};
				\node[rel] at (10, 0) (t2pv3){$T2PV^{(3)}$};
				
				\node[rel] at (12, 0) (t3pv1){$T3PV^{(1)}$};
				\node[rel] at (14, 0) (t3pv2){$T3PV^{(2)}$};
				\node[rel] at (16, 0) (t3pv3){$T3PV^{(3)}$};
				
				\node[rel] at (18, 0) (t4pv1){$T4PV^{(1)}$};
				\node[rel] at (20, 0) (t4pv2){$T4PV^{(2)}$};
				\node[rel] at (22, 0) (t4pv3){$T4PV^{(3)}$};
				
				\node[rel] at (24, 0) (t5pv1){$T5PV^{(1)}$};
				\node[rel] at (26, 0) (t5pv2){$T5PV^{(2)}$};
				\node[rel] at (28, 0) (t5pv3){$T5PV^{(3)}$};

				\node[rel] at (2, 1.6) (t1){$T1$};
				\node[rel] at (8, 1.6) (t2){$T2$};
				\node[rel] at (14, 1.6) (t3){$T3$};
				\node[rel] at (20, 1.6) (t4){$T4$};
				\node[rel] at (26, 1.6) (t5){$T5$};
				
				\node[rel] at (14, 3.2) (RV){$RV$};

				\relation{0.2}{t1pv1}{t1};
				\relation{0.2}{t1pv2}{t1};
				\relation{0.2}{t1pv3}{t1};

				\relation{0.2}{t2pv1}{t2};
				\relation{0.2}{t2pv2}{t2};
				\relation{0.2}{t2pv3}{t2};
				
				\relation{0.2}{t3pv1}{t3};
				\relation{0.2}{t3pv2}{t3};
				\relation{0.2}{t3pv3}{t3};
				
				\relation{0.2}{t4pv1}{t4};
				\relation{0.2}{t4pv2}{t4};
				\relation{0.2}{t4pv3}{t4};
				
				\relation{0.2}{t5pv1}{t5};
				\relation{0.2}{t5pv2}{t5};
				\relation{0.2}{t5pv3}{t5};
				
				\relation{0.2}{t1}{RV};
				\relation{0.2}{t2}{RV};
				\relation{0.2}{t3}{RV};
				\relation{0.2}{t4}{RV};
				\relation{0.2}{t5}{RV};
		\end{tikzpicture}}
		\caption{Hierarchical representation of the Temporal with PV(3) (T-PV(3)) decomposition of daily $RV$. $n_b=15$, $n_a=6$, $n=21$. More details in Table \ref{tab:hierarchylabels}.}
		\label{fig:T-PV(3)}
	\end{figure}	
\end{landscape}

\clearpage

\section*{A3. Structural representations of hierarchies and groupings for different intraday decompositions of daily RV}

\begin{figure}[H]
	\centering
	\begin{tikzpicture}[>=stealth,thick,baseline, every node/.style={minimum height=0.55cm}, every matrix/.style={left delimiter=[,right delimiter={]}, nodes in empty cells, row sep=0cm, inner ysep=0pt}]
		\matrix [matrix of math nodes](B){ 
			RV \\
			T1\\
			T2\\
			T3\\
			T4\\
			T5\\
		};
		\node[text centered, right=of B] (Ug) {$=$};
		\matrix [matrix of math nodes, right=of Ug](A){ 
			1 & 1 & 1 & 1 & 1 \\
			\hline
			&   &   &   &   \\
			&   &   &   &   \\
			&   &   &   &   \\
			&   &   &   &   \\
			&   &   &   &   \\
		};
		\matrix [matrix of math nodes, right=of A](C){ 
			T1\\
			T2\\
			T3\\
			T4\\
			T5\\
		};
		\node[text centered] at (A-4-3) (L) {\huge$\mathbf{I}_{5}$};
	\end{tikzpicture}
	\caption{Structural representation of the Simple Temporal (ST) decomposition of daily $RV$ using five intraday intervals. $n_b=5$, $n_a=1$, $n=6$.}
	\label{fig:eqST}
\end{figure}

\begin{figure}[H]
	\centering
	\begin{tikzpicture}[>=stealth,thick,baseline, every node/.style={minimum height=0.55cm}, every matrix/.style={left delimiter=[,right delimiter={]}, nodes in empty cells, row sep=0cm, inner ysep=0pt}]
		\matrix [matrix of math nodes](B){ 
			RV \\
			SV^{-}\\
			SV^{+}\\
		};
		\node[text centered, right=of B] (Ug) {$=$};
		\matrix [matrix of math nodes, right=of Ug](A){ 
			1 & 1 \\
			\hline
			&  \\
			&  \\
		};
		\matrix [matrix of math nodes, right=of A](C){ 
			SV^{-}\\
			SV^{+}\\
		};
		\node[text centered, fit = (A-3-1)(A-3-2)] (L) {\Large$\mathbf{I}_{2}$};
	\end{tikzpicture}
	\caption{Structural representation of the Simple SV (SSV) decomposition of daily RV. $n_b=2$, $n_a=1$, $n=3$.}
	\label{fig:eqSSV}
\end{figure}

\begin{figure}[H]
	\centering
	\begin{tikzpicture}[>=stealth,thick,baseline, every node/.style={minimum height=0.55cm}, every matrix/.style={left delimiter=[,right delimiter={]}, nodes in empty cells, row sep=0cm, inner ysep=0pt}]
		\matrix [matrix of math nodes](B){ 
			RV \\
			T1SV^{-}\\
			T2SV^{-}\\
			T3SV^{-}\\
			T4SV^{-}\\
			T5SV^{-}\\
			T1SV^{+}\\
			T2SV^{+}\\
			T3SV^{+}\\
			T4SV^{+}\\
			T5SV^{+}\\
		};
		\node[text centered, right=of B] (Ug) {$=$};
		\matrix [matrix of math nodes, right=of Ug](A){ 
			1 & 1 & 1 & 1 & 1 & 1 & 1 & 1 & 1 & 1 \\
			\hline
			&   &   &   &   &   &   &   &   &   \\
			&   &   &   &   &   &   &   &   &   \\
			&   &   &   &   &   &   &   &   &   \\
			&   &   &   &   &   &   &   &   &   \\
			&   &   &   &   &   &   &   &   &   \\
			&   &   &   &   &   &   &   &   &   \\
			&   &   &   &   &   &   &   &   &   \\
			&   &   &   &   &   &   &   &   &   \\
			&   &   &   &   &   &   &   &   &   \\
			&   &   &   &   &   &   &   &   &   \\
		};
		\matrix [matrix of math nodes, right=of A](C){ 
			T1SV^{-}\\
			T2SV^{-}\\
			T3SV^{-}\\
			T4SV^{-}\\
			T5SV^{-}\\
			T1SV^{+}\\
			T2SV^{+}\\
			T3SV^{+}\\
			T4SV^{+}\\
			T5SV^{+}\\
		};
		\node[text centered, fit = (A-8-5)(A-7-6)] (L) {\huge$\mathbf{I}_{10}$};
	\end{tikzpicture}
	\caption{Structural representation of the Simple Temporal-and-SV (STSV) decomposition of daily $RV$. $n_b=10$, $n_a=1$, $n=11$.}
	\label{fig:eqSTSV}
\end{figure}

\begin{figure}[H]
	\centering
	\begin{tikzpicture}[>=stealth,thick,baseline, every node/.style={minimum height=0.55cm}, every matrix/.style={left delimiter=[,right delimiter={]}, nodes in empty cells, row sep=0cm, inner ysep=0pt}]
		\matrix [matrix of math nodes](B){ 
			RV \\
			SV^{-}\\
			SV^{+}\\
			T1SV^{-}\\
			T2SV^{-}\\
			T3SV^{-}\\
			T4SV^{-}\\
			T5SV^{-}\\
			T1SV^{+}\\
			T2SV^{+}\\
			T3SV^{+}\\
			T4SV^{+}\\
			T5SV^{+}\\
		};
		\node[text centered, right=of B] (Ug) {$=$};
		\matrix [matrix of math nodes, right=of Ug](A){ 
			1 & 1 & 1 & 1 & 1 & 1 & 1 & 1 & 1 & 1 \\
			1 & 1 & 1 & 1 & 1 & 0 & 0 & 0 & 0 & 0 \\
			0 & 0 & 0 & 0 & 0 & 1 & 1 & 1 & 1 & 1 \\
			\hline
			&   &   &   &   &   &   &   &   &   \\
			&   &   &   &   &   &   &   &   &   \\
			&   &   &   &   &   &   &   &   &   \\
			&   &   &   &   &   &   &   &   &   \\
			&   &   &   &   &   &   &   &   &   \\
			&   &   &   &   &   &   &   &   &   \\
			&   &   &   &   &   &   &   &   &   \\
			&   &   &   &   &   &   &   &   &   \\
			&   &   &   &   &   &   &   &   &   \\
			&   &   &   &   &   &   &   &   &   \\
		};
		\matrix [matrix of math nodes, right=of A](C){ 
			T1SV^{-}\\
			T2SV^{-}\\
			T3SV^{-}\\
			T4SV^{-}\\
			T5SV^{-}\\
			T1SV^{+}\\
			T2SV^{+}\\
			T3SV^{+}\\
			T4SV^{+}\\
			T5SV^{+}\\
		};
		\node[text centered, fit = (A-10-5)(A-9-6)] (L) {\huge$\mathbf{I}_{10}$};
	\end{tikzpicture}
	\caption{Structural representation of the SV with Temporal (SV-T) decomposition of daily $RV$. $n_b=10$, $n_a=3$, $n=13$.}
	\label{fig:eqSV-T}
\end{figure}

\begin{figure}[H]
	\centering
	\begin{tikzpicture}[>=stealth,thick,baseline, every node/.style={minimum height=0.55cm}, every matrix/.style={left delimiter=[,right delimiter={]}, nodes in empty cells, row sep=0cm, inner ysep=0pt}]
		\matrix [matrix of math nodes](B){ 
			RV \\
			T1\\
			T2\\
			T3\\
			T4\\
			T5\\
			T1SV^{-}\\
			T2SV^{-}\\
			T3SV^{-}\\
			T4SV^{-}\\
			T5SV^{-}\\
			T1SV^{+}\\
			T2SV^{+}\\
			T3SV^{+}\\
			T4SV^{+}\\
			T5SV^{+}\\
		};
		\node[text centered, right=of B] (Ug) {$=$};
		\matrix [matrix of math nodes, right=of Ug](A){ 
			1 & 1 & 1 & 1 & 1 & 1 & 1 & 1 & 1 & 1 \\
			1 & 0 & 0 & 0 & 0 & 1 & 0 & 0 & 0 & 0 \\
			0 & 1 & 0 & 0 & 0 & 0 & 1 & 0 & 0 & 0 \\
			0 & 0 & 1 & 0 & 0 & 0 & 0 & 1 & 0 & 0 \\
			0 & 0 & 0 & 1 & 0 & 0 & 0 & 0 & 1 & 0 \\
			0 & 0 & 0 & 0 & 1 & 0 & 0 & 0 & 0 & 1 \\
			\hline
			&   &   &   &   &   &   &   &   &   \\
			&   &   &   &   &   &   &   &   &   \\
			&   &   &   &   &   &   &   &   &   \\
			&   &   &   &   &   &   &   &   &   \\
			&   &   &   &   &   &   &   &   &   \\
			&   &   &   &   &   &   &   &   &   \\
			&   &   &   &   &   &   &   &   &   \\
			&   &   &   &   &   &   &   &   &   \\
			&   &   &   &   &   &   &   &   &   \\
			&   &   &   &   &   &   &   &   &   \\
		};
		\matrix [matrix of math nodes, right=of A](C){ 
			T1SV^{-}\\
			T2SV^{-}\\
			T3SV^{-}\\
			T4SV^{-}\\
			T5SV^{-}\\
			T1SV^{+}\\
			T2SV^{+}\\
			T3SV^{+}\\
			T4SV^{+}\\
			T5SV^{+}\\
		};
		\node[text centered, fit = (A-13-5)(A-12-6)] (L) {\huge$\mathbf{I}_{10}$};
	\end{tikzpicture}
	\caption{Structural representation of the Temporal with SV (T-SV) decomposition of daily $RV$. $n_b=10$, $n_a=6$, $n=16$.}
	\label{fig:eqT-SV}
\end{figure}

\begin{figure}[H]
	\centering
	\begin{tikzpicture}[>=stealth,thick,baseline, every node/.style={minimum height=0.55cm}, every matrix/.style={left delimiter=[,right delimiter={]}, nodes in empty cells, row sep=0cm, inner ysep=0pt}]
		\matrix [matrix of math nodes](B){ 
			RV \\
			SV^{-}\\
			SV^{+}\\
			T1\\
			T2\\
			T3\\
			T4\\
			T5\\
			T1SV^{-}\\
			T2SV^{-}\\
			T3SV^{-}\\
			T4SV^{-}\\
			T5SV^{-}\\
			T1SV^{+}\\
			T2SV^{+}\\
			T3SV^{+}\\
			T4SV^{+}\\
			T5SV^{+}\\
		};
		\node[text centered, right=of B] (Ug) {$=$};
		\matrix [matrix of math nodes, right=of Ug](A){ 
			1 & 1 & 1 & 1 & 1 & 1 & 1 & 1 & 1 & 1 \\
			1 & 1 & 1 & 1 & 1 & 0 & 0 & 0 & 0 & 0 \\
			0 & 0 & 0 & 0 & 0 & 1 & 1 & 1 & 1 & 1 \\
			1 & 0 & 0 & 0 & 0 & 1 & 0 & 0 & 0 & 0 \\
			0 & 1 & 0 & 0 & 0 & 0 & 1 & 0 & 0 & 0 \\
			0 & 0 & 1 & 0 & 0 & 0 & 0 & 1 & 0 & 0 \\
			0 & 0 & 0 & 1 & 0 & 0 & 0 & 0 & 1 & 0 \\
			0 & 0 & 0 & 0 & 1 & 0 & 0 & 0 & 0 & 1 \\
			\hline
			&   &   &   &   &   &   &   &   &   \\
			&   &   &   &   &   &   &   &   &   \\
			&   &   &   &   &   &   &   &   &   \\
			&   &   &   &   &   &   &   &   &   \\
			&   &   &   &   &   &   &   &   &   \\
			&   &   &   &   &   &   &   &   &   \\
			&   &   &   &   &   &   &   &   &   \\
			&   &   &   &   &   &   &   &   &   \\
			&   &   &   &   &   &   &   &   &   \\
			&   &   &   &   &   &   &   &   &   \\
		};
		\matrix [matrix of math nodes, right=of A](C){ 
			T1SV^{-}\\
			T2SV^{-}\\
			T3SV^{-}\\
			T4SV^{-}\\
			T5SV^{-}\\
			T1SV^{+}\\
			T2SV^{+}\\
			T3SV^{+}\\
			T4SV^{+}\\
			T5SV^{+}\\
		};
		\node[text centered, fit = (A-15-5)(A-14-6)] (L) {\huge$\mathbf{I}_{10}$};
	\end{tikzpicture}
	\caption{Structural representation of the grouped time series describing the Complete Temporal-and-SV (CTSV) decomposition of daily $RV$ using five intraday intervals and Semi Variances (Bad \& Good volatility). $n_b=10$, $n_a=8$, $n=18$. The grouped time series is obtained by merging the hierarchies SV-T and T-SV, described in Figures \ref{fig:SV-T} and \ref{fig:T-SV}, respectively.}
	\label{fig:eqCTSV}
\end{figure}

\begin{figure}[H]
	\begin{tikzpicture}[>=stealth,thick,baseline, every node/.style={minimum height=0.55cm}, every matrix/.style={left delimiter=[,right delimiter={]}, nodes in empty cells, row sep=0cm, inner ysep=0pt}]
		\matrix [matrix of math nodes](B){ 
			RV \\
			PV^{(1)}\\
			PV^{(2)}\\
			PV^{(3)}\\
		};
		\node[text centered, right=of B] (Ug) {$=$};
		\matrix [matrix of math nodes, right=of Ug](A){ 
			1 & 1 & 1 \\
			\hline
			&   &   \\
			&   &   \\
			&   &   \\
		};
		\matrix [matrix of math nodes, right=of A](C){ 
			PV^{(1)}\\
			PV^{(2)}\\
			PV^{(3)}\\
		};
		\node[text centered] at (A-3-2) (L) {\huge$\mathbf{I}_{3}$};
		
	\end{tikzpicture}
	\caption{Structural representation of the Simple PV(3) (SPV(3)) decomposition of daily $RV$. $n_b=3$, $n_a=1$, $n=4$.}
	\label{fig:eqSPV(3)}
\end{figure}

\begin{figure}[H]
	\centering
	\begin{tikzpicture}[>=stealth,thick,baseline, every node/.style={minimum height=0.55cm}, every matrix/.style={left delimiter=[,right delimiter={]}, nodes in empty cells, row sep=0cm, inner ysep=0pt}]
		\matrix [matrix of math nodes](B){ 
			RV \\
			T1PV^{(1)}\\
			T2PV^{(1)}\\
			T3PV^{(1)}\\
			T4PV^{(1)}\\
			T5PV^{(1)}\\
			T1PV^{(2)}\\
			T2PV^{(2)}\\
			T3PV^{(2)}\\
			T4PV^{(2)}\\
			T5PV^{(2)}\\
			T1PV^{(3)}\\
			T2PV^{(3)}\\
			T3PV^{(3)}\\
			T4PV^{(3)}\\
			T5PV^{(3)}\\
		};
		\node[text centered, right=of B] (Ug) {$=$};
		\matrix [matrix of math nodes, right=of Ug](A){ 
			1 & 1 & 1 & 1 & 1 & 1 & 1 & 1 & 1 & 1 & 1 & 1 & 1 & 1 & 1 \\
			\hline
			&   &   &   &   &   &   &   &   &   &   &   &   &   &   \\
			&   &   &   &   &   &   &   &   &   &   &   &   &   &   \\
			&   &   &   &   &   &   &   &   &   &   &   &   &   &   \\
			&   &   &   &   &   &   &   &   &   &   &   &   &   &   \\
			&   &   &   &   &   &   &   &   &   &   &   &   &   &   \\
			&   &   &   &   &   &   &   &   &   &   &   &   &   &   \\
			&   &   &   &   &   &   &   &   &   &   &   &   &   &   \\
			&   &   &   &   &   &   &   &   &   &   &   &   &   &   \\
			&   &   &   &   &   &   &   &   &   &   &   &   &   &   \\
			&   &   &   &   &   &   &   &   &   &   &   &   &   &   \\
			&   &   &   &   &   &   &   &   &   &   &   &   &   &   \\
			&   &   &   &   &   &   &   &   &   &   &   &   &   &   \\
			&   &   &   &   &   &   &   &   &   &   &   &   &   &   \\
			&   &   &   &   &   &   &   &   &   &   &   &   &   &   \\
			&   &   &   &   &   &   &   &   &   &   &   &   &   &   \\
		};
		\matrix [matrix of math nodes, right=of A](C){ 
			T1PV^{(1)}\\
			T2PV^{(1)}\\
			T3PV^{(1)}\\
			T4PV^{(1)}\\
			T5PV^{(1)}\\
			T1PV^{(2)}\\
			T2PV^{(2)}\\
			T3PV^{(2)}\\
			T4PV^{(2)}\\
			T5PV^{(2)}\\
			T1PV^{(3)}\\
			T2PV^{(3)}\\
			T3PV^{(3)}\\
			T4PV^{(3)}\\
			T5PV^{(3)}\\
		};
		\node[text centered] at (A-9-8) (L) {\huge$\mathbf{I}_{15}$};
		
	\end{tikzpicture}
	\caption{Structural representation of the Simple Temporal-and-PV(3) (STPV(3)) decomposition of daily $RV$. $n_b=15$, $n_a=1$, $n=16$.}
	\label{fig:eqSTPV(3)}
\end{figure}

\begin{figure}[H]
	\begin{tikzpicture}[>=stealth,thick,baseline, every node/.style={minimum height=0.55cm}, every matrix/.style={left delimiter=[,right delimiter={]}, nodes in empty cells, row sep=0cm, inner ysep=0pt}]
		\matrix [matrix of math nodes](B){ 
			RV \\
			PV^{(1)}\\
			PV^{(2)}\\
			PV^{(3)}\\
			T1PV^{(1)}\\
			T2PV^{(1)}\\
			T3PV^{(1)}\\
			T4PV^{(1)}\\
			T5PV^{(1)}\\
			T1PV^{(2)}\\
			T2PV^{(2)}\\
			T3PV^{(2)}\\
			T4PV^{(2)}\\
			T5PV^{(2)}\\
			T1PV^{(3)}\\
			T2PV^{(3)}\\
			T3PV^{(3)}\\
			T4PV^{(3)}\\
			T5PV^{(3)}\\
		};
		\node[text centered, right=of B] (Ug) {$=$};
		\matrix [matrix of math nodes, right=of Ug](A){ 
			1 & 1 & 1 & 1 & 1 & 1 & 1 & 1 & 1 & 1 & 1 & 1 & 1 & 1 & 1 \\
			1 & 1 & 1 & 1 & 1 & 0 & 0 & 0 & 0 & 0 & 0 & 0 & 0 & 0 & 0 \\
			0 & 0 & 0 & 0 & 0 & 1 & 1 & 1 & 1 & 1 & 0 & 0 & 0 & 0 & 0 \\
			0 & 0 & 0 & 0 & 0 & 0 & 0 & 0 & 0 & 0 & 1 & 1 & 1 & 1 & 1 \\
			\hline
			&   &   &   &   &   &   &   &   &   &   &   &   &   &   \\
			&   &   &   &   &   &   &   &   &   &   &   &   &   &   \\
			&   &   &   &   &   &   &   &   &   &   &   &   &   &   \\
			&   &   &   &   &   &   &   &   &   &   &   &   &   &   \\
			&   &   &   &   &   &   &   &   &   &   &   &   &   &   \\
			&   &   &   &   &   &   &   &   &   &   &   &   &   &   \\
			&   &   &   &   &   &   &   &   &   &   &   &   &   &   \\
			&   &   &   &   &   &   &   &   &   &   &   &   &   &   \\
			&   &   &   &   &   &   &   &   &   &   &   &   &   &   \\
			&   &   &   &   &   &   &   &   &   &   &   &   &   &   \\
			&   &   &   &   &   &   &   &   &   &   &   &   &   &   \\
			&   &   &   &   &   &   &   &   &   &   &   &   &   &   \\
			&   &   &   &   &   &   &   &   &   &   &   &   &   &   \\
			&   &   &   &   &   &   &   &   &   &   &   &   &   &   \\
			&   &   &   &   &   &   &   &   &   &   &   &   &   &   \\
		};
		\matrix [matrix of math nodes, right=of A](C){ 
			T1PV^{(1)}\\
			T2PV^{(1)}\\
			T3PV^{(1)}\\
			T4PV^{(1)}\\
			T5PV^{(1)}\\
			T1PV^{(2)}\\
			T2PV^{(2)}\\
			T3PV^{(2)}\\
			T4PV^{(2)}\\
			T5PV^{(2)}\\
			T1PV^{(3)}\\
			T2PV^{(3)}\\
			T3PV^{(3)}\\
			T4PV^{(3)}\\
			T5PV^{(3)}\\
		};
		\node[text centered
		] at (A-12-8) (L) {\huge$\mathbf{I}_{15}$};
		
	\end{tikzpicture}
	\caption{Structural representation of the PV(3) with Temporal (PV(3)-T) decomposition of daily $RV$. $n_b=15$, $n_a=4$, $n=19$.}
	\label{fig:eqPV(3)-T}
\end{figure}

\begin{figure}[H]
	\begin{tikzpicture}[>=stealth,thick,baseline, every node/.style={minimum height=0.55cm}, every matrix/.style={left delimiter=[,right delimiter={]}, nodes in empty cells, row sep=0cm, inner ysep=0pt}]
		\matrix [matrix of math nodes](B){ 
			RV \\
			T1\\
			T2\\
			T3\\
			T4\\
			T5\\
			T1PV^{(1)}\\
			T2PV^{(1)}\\
			T3PV^{(1)}\\
			T4PV^{(1)}\\
			T5PV^{(1)}\\
			T1PV^{(2)}\\
			T2PV^{(2)}\\
			T3PV^{(2)}\\
			T4PV^{(2)}\\
			T5PV^{(2)}\\
			T1PV^{(3)}\\
			T2PV^{(3)}\\
			T3PV^{(3)}\\
			T4PV^{(3)}\\
			T5PV^{(3)}\\
		};
		\node[text centered, right=of B] (Ug) {$=$};
		\matrix [matrix of math nodes, right=of Ug](A){ 
			1 & 1 & 1 & 1 & 1 & 1 & 1 & 1 & 1 & 1 & 1 & 1 & 1 & 1 & 1 \\
			1 & 0 & 0 & 0 & 0 & 1 & 0 & 0 & 0 & 0 & 1 & 0 & 0 & 0 & 0 \\
			0 & 1 & 0 & 0 & 0 & 0 & 1 & 0 & 0 & 0 & 0 & 1 & 0 & 0 & 0 \\
			0 & 0 & 1 & 0 & 0 & 0 & 0 & 1 & 0 & 0 & 0 & 0 & 1 & 0 & 0 \\
			0 & 0 & 0 & 1 & 0 & 0 & 0 & 0 & 1 & 0 & 0 & 0 & 0 & 1 & 0 \\
			0 & 0 & 0 & 0 & 1 & 0 & 0 & 0 & 0 & 1 & 0 & 0 & 0 & 0 & 1 \\
			\hline
			&   &   &   &   &   &   &   &   &   &   &   &   &   &   \\
			&   &   &   &   &   &   &   &   &   &   &   &   &   &   \\
			&   &   &   &   &   &   &   &   &   &   &   &   &   &   \\
			&   &   &   &   &   &   &   &   &   &   &   &   &   &   \\
			&   &   &   &   &   &   &   &   &   &   &   &   &   &   \\
			&   &   &   &   &   &   &   &   &   &   &   &   &   &   \\
			&   &   &   &   &   &   &   &   &   &   &   &   &   &   \\
			&   &   &   &   &   &   &   &   &   &   &   &   &   &   \\
			&   &   &   &   &   &   &   &   &   &   &   &   &   &   \\
			&   &   &   &   &   &   &   &   &   &   &   &   &   &   \\
			&   &   &   &   &   &   &   &   &   &   &   &   &   &   \\
			&   &   &   &   &   &   &   &   &   &   &   &   &   &   \\
			&   &   &   &   &   &   &   &   &   &   &   &   &   &   \\
			&   &   &   &   &   &   &   &   &   &   &   &   &   &   \\
			&   &   &   &   &   &   &   &   &   &   &   &   &   &   \\
		};
		\matrix [matrix of math nodes, right=of A](C){ 
			T1PV^{(1)}\\
			T2PV^{(1)}\\
			T3PV^{(1)}\\
			T4PV^{(1)}\\
			T5PV^{(1)}\\
			T1PV^{(2)}\\
			T2PV^{(2)}\\
			T3PV^{(2)}\\
			T4PV^{(2)}\\
			T5PV^{(2)}\\
			T1PV^{(3)}\\
			T2PV^{(3)}\\
			T3PV^{(3)}\\
			T4PV^{(3)}\\
			T5PV^{(3)}\\
		};
		\node[text centered] at (A-14-8) (L) {\huge$\mathbf{I}_{15}$};
		
	\end{tikzpicture}
	\caption{Structural representation of the Temporal with PV(3) (T-PV(3)) decomposition of daily $RV$. $n_b=15$, $n_a=6$, $n=21$.}
	\label{fig:eqT-PV(3)}
\end{figure}

\begin{figure}[H]
	\centering
	\begin{tikzpicture}[>=stealth,thick,baseline, every node/.style={minimum height=0.55cm}, every matrix/.style={left delimiter=[,right delimiter={]}, nodes in empty cells, row sep=0cm, inner ysep=0pt}]
		\matrix [matrix of math nodes](B){ 
			RV \\
			PV^{(1)}\\
			PV^{(2)}\\
			PV^{(3)}\\
			T1\\
			T2\\
			T3\\
			T4\\
			T5\\
			T1PV^{(1)}\\
			T2PV^{(1)}\\
			T3PV^{(1)}\\
			T4PV^{(1)}\\
			T5PV^{(1)}\\
			T1PV^{(2)}\\
			T2PV^{(2)}\\
			T3PV^{(2)}\\
			T4PV^{(2)}\\
			T5PV^{(2)}\\
			T1PV^{(3)}\\
			T2PV^{(3)}\\
			T3PV^{(3)}\\
			T4PV^{(3)}\\
			T5PV^{(3)}\\
		};
		\node[text centered, right=of B] (Ug) {$=$};
		\matrix [matrix of math nodes, right=of Ug](A){ 
			1 & 1 & 1 & 1 & 1 & 1 & 1 & 1 & 1 & 1 & 1 & 1 & 1 & 1 & 1 \\
			1 & 1 & 1 & 1 & 1 & 0 & 0 & 0 & 0 & 0 & 0 & 0 & 0 & 0 & 0 \\
			0 & 0 & 0 & 0 & 0 & 1 & 1 & 1 & 1 & 1 & 0 & 0 & 0 & 0 & 0 \\
			0 & 0 & 0 & 0 & 0 & 0 & 0 & 0 & 0 & 0 & 1 & 1 & 1 & 1 & 1 \\
			1 & 0 & 0 & 0 & 0 & 1 & 0 & 0 & 0 & 0 & 1 & 0 & 0 & 0 & 0 \\
			0 & 1 & 0 & 0 & 0 & 0 & 1 & 0 & 0 & 0 & 0 & 1 & 0 & 0 & 0 \\
			0 & 0 & 1 & 0 & 0 & 0 & 0 & 1 & 0 & 0 & 0 & 0 & 1 & 0 & 0 \\
			0 & 0 & 0 & 1 & 0 & 0 & 0 & 0 & 1 & 0 & 0 & 0 & 0 & 1 & 0 \\
			0 & 0 & 0 & 0 & 1 & 0 & 0 & 0 & 0 & 1 & 0 & 0 & 0 & 0 & 1 \\
			\hline
			&   &   &   &   &   &   &   &   &   &   &   &   &   &   \\
			&   &   &   &   &   &   &   &   &   &   &   &   &   &   \\
			&   &   &   &   &   &   &   &   &   &   &   &   &   &   \\
			&   &   &   &   &   &   &   &   &   &   &   &   &   &   \\
			&   &   &   &   &   &   &   &   &   &   &   &   &   &   \\
			&   &   &   &   &   &   &   &   &   &   &   &   &   &   \\
			&   &   &   &   &   &   &   &   &   &   &   &   &   &   \\
			&   &   &   &   &   &   &   &   &   &   &   &   &   &   \\
			&   &   &   &   &   &   &   &   &   &   &   &   &   &   \\
			&   &   &   &   &   &   &   &   &   &   &   &   &   &   \\
			&   &   &   &   &   &   &   &   &   &   &   &   &   &   \\
			&   &   &   &   &   &   &   &   &   &   &   &   &   &   \\
			&   &   &   &   &   &   &   &   &   &   &   &   &   &   \\
			&   &   &   &   &   &   &   &   &   &   &   &   &   &   \\
			&   &   &   &   &   &   &   &   &   &   &   &   &   &   \\
		};
		\matrix [matrix of math nodes, right=of A](C){ 
			T1PV^{(1)}\\
			T2PV^{(1)}\\
			T3PV^{(1)}\\
			T4PV^{(1)}\\
			T5PV^{(1)}\\
			T1PV^{(2)}\\
			T2PV^{(2)}\\
			T3PV^{(2)}\\
			T4PV^{(2)}\\
			T5PV^{(2)}\\
			T1PV^{(3)}\\
			T2PV^{(3)}\\
			T3PV^{(3)}\\
			T4PV^{(3)}\\
			T5PV^{(3)}\\
		};
		\node[text centered] at (A-17-8) (L) {\huge$\mathbf{I}_{15}$};
	\end{tikzpicture}
	\caption{Structural representation of the grouped time series describing the Complete Temporal-and-PV(3) (CTPV(3)) decomposition of daily $RV$ using five intraday intervals and PV(3). $n_b=15$, $n_a=9$, $n=24$. The grouped time series is obtained by merging the hierarchies PV(3)-T and T-PV(3), described in Figures \ref{fig:PV(3)-T} and \ref{fig:T-PV(3)}, respectively.}
	\label{fig:eqCTPV(3)}
\end{figure}

\clearpage

\section*{A4: Forecasting accuracy in different test periods}

%\begin{table}
%	\centering
%	\footnotesize
%	\setlength{\tabcolsep}{5pt}
%	\input{tab/correlation.tex}
%	\caption{corr della RV con le diverse decomposioni per tutte le osservazioni}
%\end{table}

\begin{table}[H]
	\centering
	\footnotesize
	\setlength{\tabcolsep}{5pt}
	
\begin{tabular}[t]{>{}c|c>{}c|c>{}c|c>{}c|c>{}c|cc}
\toprule
\multicolumn{1}{c}{ } & \multicolumn{2}{c}{2007-2022} & \multicolumn{2}{c}{2007-2010} & \multicolumn{2}{c}{2011-2014} & \multicolumn{2}{c}{2015-2019} & \multicolumn{2}{c}{2020-2022} \\
 & MSE & QLIKE & MSE & QLIKE & MSE & QLIKE & MSE & QLIKE & MSE & QLIKE\\
\midrule
\addlinespace[0.3em]
\multicolumn{11}{c}{\textit{Panel A: DJIA index}}\\
$PV(3)$ & \textcolor{black}{\textbf{0.775}} & \textcolor{black}{0.816} & \textcolor{black}{\textbf{0.921}} & \textcolor{black}{0.997} & \textcolor{red}{1.059} & \textcolor{red}{2.691} & \textcolor{red}{2.272} & \textcolor{red}{1.073} & \textcolor{black}{\textbf{0.873}} & \textcolor{red}{6.045}\\
$PV(3)_{bu}$ & \textcolor{black}{0.922} & \textcolor{black}{0.924} & \textcolor{red}{1.036} & \textcolor{black}{0.947} & \textcolor{black}{\textbf{0.911}} & \textcolor{red}{1.023} & \textcolor{red}{1.001} & \textcolor{red}{1.076} & \textcolor{red}{1.004} & \textcolor{black}{0.874}\\
$PV(3)_{shr}$ & \textcolor{black}{0.812} & \textcolor{black}{0.833} & \textcolor{black}{0.944} & \textcolor{black}{0.934} & \textcolor{black}{0.947} & \textcolor{black}{\textbf{0.941}} & \textcolor{black}{\textbf{0.945}} & \textcolor{red}{1.047} & \textcolor{black}{0.902} & \textcolor{black}{0.902}\\
\midrule
$SV$ & \textcolor{red}{1.024} & \textcolor{red}{1.028} & \textcolor{black}{0.991} & \textcolor{red}{1.069} & \textcolor{red}{1.048} & \textcolor{black}{0.978} & \textcolor{black}{0.973} & \textcolor{black}{\textbf{0.986}} & \textcolor{black}{0.968} & \textcolor{black}{0.962}\\
$SV_{bu}$ & \textcolor{black}{0.962} & \textcolor{black}{0.960} & \textcolor{red}{1.055} & \textcolor{black}{0.966} & \textcolor{black}{0.923} & \textcolor{red}{1.024} & \textcolor{red}{1.010} & \textcolor{red}{1.069} & \textcolor{red}{1.021} & \textcolor{black}{0.897}\\
$SV_{shr}$ & \textcolor{black}{0.976} & \textcolor{black}{0.976} & \textcolor{red}{1.022} & \textcolor{black}{0.997} & \textcolor{black}{0.960} & \textcolor{black}{0.997} & \textcolor{black}{0.983} & \textcolor{red}{1.027} & \textcolor{black}{0.989} & \textcolor{black}{0.896}\\
$TPV(3)_{bu}$ & \textcolor{black}{0.799} & \textcolor{black}{0.822} & \textcolor{red}{1.148} & \textcolor{black}{0.911} & \textcolor{black}{0.916} & \textcolor{red}{1.092} & \textcolor{red}{1.090} & \textcolor{red}{1.265} & \textcolor{red}{1.075} & \textcolor{black}{0.893}\\
\midrule
$TPV(3)_{shr}$ & \textcolor{black}{0.787} & \textcolor{black}{\textbf{0.808}} & \textcolor{black}{0.992} & \textcolor{black}{\textbf{0.900}} & \textcolor{black}{0.915} & \textcolor{black}{0.967} & \textcolor{black}{0.971} & \textcolor{red}{1.095} & \textcolor{black}{0.944} & \textcolor{black}{0.874}\\
$TSV_{bu}$ & \textcolor{black}{0.801} & \textcolor{black}{0.824} & \textcolor{red}{1.151} & \textcolor{black}{0.914} & \textcolor{black}{0.923} & \textcolor{red}{1.085} & \textcolor{red}{1.085} & \textcolor{red}{1.245} & \textcolor{red}{1.074} & \textcolor{black}{0.906}\\
$TSV_{shr}$ & \textcolor{black}{0.856} & \textcolor{black}{0.869} & \textcolor{red}{1.066} & \textcolor{black}{0.930} & \textcolor{black}{0.923} & \textcolor{red}{1.020} & \textcolor{red}{1.011} & \textcolor{red}{1.109} & \textcolor{red}{1.011} & \textcolor{black}{\textbf{0.873}}\\
\addlinespace[0.3em]
\multicolumn{11}{c}{\textit{Panel B: Individual stocks}}\\
$PV(3)$ & \textcolor{black}{0.938} & \textcolor{black}{0.899} & \textcolor{black}{0.973} & \textcolor{black}{\textbf{0.836}} & \textcolor{red}{1.097} & \textcolor{red}{1.421} & \textcolor{red}{1.364} & \textcolor{red}{1.046} & \textcolor{red}{1.344} & \textcolor{red}{1.035}\\
\midrule
$PV(3)_{bu}$ & \textcolor{black}{0.953} & \textcolor{black}{0.943} & \textcolor{black}{0.993} & \textcolor{black}{0.950} & \textcolor{black}{0.994} & \textcolor{red}{1.003} & \textcolor{black}{0.900} & \textcolor{red}{1.010} & \textcolor{black}{0.990} & \textcolor{black}{0.709}\\
$PV(3)_{shr}$ & \textcolor{black}{\textbf{0.916}} & \textcolor{black}{\textbf{0.896}} & \textcolor{black}{\textbf{0.927}} & \textcolor{black}{0.874} & \textcolor{red}{1.030} & \textcolor{black}{0.939} & \textcolor{black}{0.833} & \textcolor{black}{\textbf{0.970}} & \textcolor{black}{\textbf{0.898}} & \textcolor{black}{0.704}\\
$SV$ & \textcolor{black}{0.993} & \textcolor{red}{1.016} & \textcolor{red}{1.022} & \textcolor{red}{1.005} & \textcolor{red}{1.073} & \textcolor{red}{1.124} & \textcolor{red}{1.092} & \textcolor{red}{1.076} & \textcolor{red}{1.080} & \textcolor{black}{0.963}\\
\midrule
$SV_{bu}$ & \textcolor{black}{0.967} & \textcolor{black}{0.964} & \textcolor{red}{1.000} & \textcolor{black}{0.973} & \textcolor{black}{0.998} & \textcolor{red}{1.000} & \textcolor{black}{0.921} & \textcolor{red}{1.007} & \textcolor{red}{1.006} & \textcolor{black}{0.779}\\
$SV_{shr}$ & \textcolor{black}{0.970} & \textcolor{black}{0.980} & \textcolor{black}{0.985} & \textcolor{black}{0.979} & \textcolor{red}{1.029} & \textcolor{black}{0.992} & \textcolor{black}{0.899} & \textcolor{black}{0.988} & \textcolor{black}{0.987} & \textcolor{black}{0.742}\\
$TPV(3)_{bu}$ & \textcolor{red}{1.027} & \textcolor{black}{0.982} & \textcolor{red}{1.063} & \textcolor{black}{0.859} & \textcolor{red}{1.011} & \textcolor{red}{1.029} & \textcolor{black}{0.894} & \textcolor{red}{1.134} & \textcolor{red}{1.030} & \textcolor{black}{0.641}\\
$TPV(3)_{shr}$ & \textcolor{black}{0.963} & \textcolor{black}{0.924} & \textcolor{black}{0.963} & \textcolor{black}{0.848} & \textcolor{black}{\textbf{0.991}} & \textcolor{black}{\textbf{0.923}} & \textcolor{black}{\textbf{0.805}} & \textcolor{red}{1.013} & \textcolor{black}{0.918} & \textcolor{black}{\textbf{0.606}}\\
\midrule
$TSV_{bu}$ & \textcolor{red}{1.028} & \textcolor{black}{0.991} & \textcolor{red}{1.060} & \textcolor{black}{0.867} & \textcolor{red}{1.023} & \textcolor{red}{1.019} & \textcolor{black}{0.893} & \textcolor{red}{1.119} & \textcolor{red}{1.029} & \textcolor{black}{0.656}\\
$TSV_{shr}$ & \textcolor{red}{1.009} & \textcolor{black}{0.987} & \textcolor{red}{1.014} & \textcolor{black}{0.907} & \textcolor{black}{0.997} & \textcolor{black}{0.959} & \textcolor{black}{0.850} & \textcolor{red}{1.041} & \textcolor{black}{0.991} & \textcolor{black}{0.632}\\
\bottomrule
\end{tabular}

	\caption{Accuracy of the \textbf{one-day ahead} forecasts. MSE and QLIKE ratios for the DJIA index (panel A), and geometric means of the MSE and QLIKE ratios for individual stocks (panel B) over the benchmark $HAR$ model. Values larger than one in red. The best index value in each column is highlighted in bold.
	}
\end{table}

\begin{table}
	\centering
	\footnotesize
	\setlength{\tabcolsep}{5pt}
	
\begin{tabular}[t]{>{}c|c>{}c|c>{}c|c>{}c|c>{}c|cc}
\toprule
\multicolumn{1}{c}{ } & \multicolumn{2}{c}{2007-2022} & \multicolumn{2}{c}{2007-2010} & \multicolumn{2}{c}{2011-2014} & \multicolumn{2}{c}{2015-2019} & \multicolumn{2}{c}{2020-2022} \\
 & MSE & QLIKE & MSE & QLIKE & MSE & QLIKE & MSE & QLIKE & MSE & QLIKE\\
\midrule
\addlinespace[0.3em]
\multicolumn{11}{c}{\textit{Panel A: DJIA index}}\\
$PV(3)$ & \textcolor{red}{1.047} & \textcolor{red}{1.077} & \textcolor{black}{0.973} & \textcolor{red}{1.013} & \textcolor{red}{1.163} & \textcolor{red}{1.645} & \textcolor{black}{0.977} & \textcolor{red}{1.021} & \textcolor{black}{\textbf{0.908}} & \textcolor{black}{0.942}\\
$PV(3)_{bu}$ & \textcolor{black}{0.981} & \textcolor{black}{0.957} & \textcolor{red}{1.013} & \textcolor{black}{0.979} & \textcolor{black}{0.892} & \textcolor{red}{1.015} & \textcolor{black}{0.235} & \textcolor{red}{1.058} & \textcolor{red}{1.011} & \textcolor{black}{\textbf{0.049}}\\
$PV(3)_{shr}$ & \textcolor{red}{1.010} & \textcolor{red}{1.015} & \textcolor{black}{\textbf{0.963}} & \textcolor{black}{0.975} & \textcolor{red}{1.033} & \textcolor{black}{\textbf{0.946}} & \textcolor{black}{\textbf{0.225}} & \textcolor{red}{1.030} & \textcolor{black}{0.938} & \textcolor{black}{0.051}\\
\midrule
$SV$ & \textcolor{red}{1.012} & \textcolor{red}{1.017} & \textcolor{red}{1.011} & \textcolor{black}{0.999} & \textcolor{red}{1.030} & \textcolor{black}{0.985} & \textcolor{black}{0.997} & \textcolor{black}{\textbf{0.988}} & \textcolor{black}{0.986} & \textcolor{black}{1.000}\\
$SV_{bu}$ & \textcolor{black}{0.991} & \textcolor{black}{0.972} & \textcolor{red}{1.016} & \textcolor{black}{0.985} & \textcolor{black}{0.921} & \textcolor{red}{1.014} & \textcolor{black}{0.237} & \textcolor{red}{1.049} & \textcolor{red}{1.021} & \textcolor{black}{0.051}\\
$SV_{shr}$ & \textcolor{black}{0.998} & \textcolor{black}{0.991} & \textcolor{red}{1.006} & \textcolor{black}{0.983} & \textcolor{black}{0.975} & \textcolor{black}{0.995} & \textcolor{black}{0.233} & \textcolor{red}{1.020} & \textcolor{black}{0.995} & \textcolor{black}{0.053}\\
$TPV(3)_{bu}$ & \textcolor{black}{\textbf{0.766}} & \textcolor{black}{\textbf{0.795}} & \textcolor{red}{1.078} & \textcolor{black}{0.978} & \textcolor{black}{\textbf{0.832}} & \textcolor{red}{1.067} & \textcolor{black}{0.250} & \textcolor{red}{1.231} & \textcolor{red}{1.049} & \textcolor{black}{0.051}\\
\midrule
$TPV(3)_{shr}$ & \textcolor{black}{0.909} & \textcolor{black}{0.925} & \textcolor{black}{0.984} & \textcolor{black}{\textbf{0.962}} & \textcolor{black}{0.956} & \textcolor{black}{0.956} & \textcolor{black}{0.227} & \textcolor{red}{1.061} & \textcolor{black}{0.948} & \textcolor{black}{0.050}\\
$TSV_{bu}$ & \textcolor{black}{0.774} & \textcolor{black}{0.801} & \textcolor{red}{1.072} & \textcolor{black}{0.977} & \textcolor{black}{0.833} & \textcolor{red}{1.063} & \textcolor{black}{0.249} & \textcolor{red}{1.210} & \textcolor{red}{1.045} & \textcolor{black}{0.051}\\
$TSV_{shr}$ & \textcolor{black}{0.853} & \textcolor{black}{0.878} & \textcolor{red}{1.031} & \textcolor{black}{0.967} & \textcolor{black}{0.920} & \textcolor{red}{1.009} & \textcolor{black}{0.236} & \textcolor{red}{1.094} & \textcolor{black}{0.999} & \textcolor{black}{0.050}\\
\addlinespace[0.3em]
\multicolumn{11}{c}{\textit{Panel B: Individual stocks}}\\
$PV(3)$ & \textcolor{red}{1.035} & \textcolor{black}{0.976} & \textcolor{black}{0.958} & \textcolor{black}{\textbf{0.943}} & \textcolor{black}{0.978} & \textcolor{red}{1.139} & \textcolor{red}{1.025} & \textcolor{black}{\textbf{0.955}} & \textcolor{red}{1.029} & \textcolor{black}{0.865}\\
\midrule
$PV(3)_{bu}$ & \textcolor{black}{0.981} & \textcolor{black}{0.982} & \textcolor{black}{0.999} & \textcolor{black}{0.984} & \textcolor{black}{0.990} & \textcolor{red}{1.010} & \textcolor{black}{0.895} & \textcolor{red}{1.013} & \textcolor{black}{0.984} & \textcolor{black}{0.750}\\
$PV(3)_{shr}$ & \textcolor{black}{0.998} & \textcolor{black}{\textbf{0.964}} & \textcolor{black}{\textbf{0.955}} & \textcolor{black}{0.949} & \textcolor{black}{0.971} & \textcolor{black}{\textbf{0.956}} & \textcolor{black}{0.824} & \textcolor{black}{0.965} & \textcolor{black}{0.929} & \textcolor{black}{0.662}\\
$SV$ & \textcolor{red}{1.014} & \textcolor{red}{1.008} & \textcolor{black}{0.985} & \textcolor{red}{1.001} & \textcolor{red}{1.005} & \textcolor{red}{1.004} & \textcolor{black}{0.998} & \textcolor{black}{0.982} & \textcolor{black}{0.993} & \textcolor{red}{1.025}\\
\midrule
$SV_{bu}$ & \textcolor{black}{0.991} & \textcolor{black}{0.992} & \textcolor{red}{1.001} & \textcolor{black}{0.991} & \textcolor{red}{1.000} & \textcolor{red}{1.010} & \textcolor{black}{0.977} & \textcolor{red}{1.008} & \textcolor{black}{0.999} & \textcolor{black}{0.946}\\
$SV_{shr}$ & \textcolor{black}{0.998} & \textcolor{black}{0.996} & \textcolor{black}{0.987} & \textcolor{black}{0.992} & \textcolor{red}{1.000} & \textcolor{red}{1.001} & \textcolor{black}{0.947} & \textcolor{black}{0.991} & \textcolor{black}{0.990} & \textcolor{black}{0.897}\\
$TPV(3)_{bu}$ & \textcolor{black}{0.996} & \textcolor{red}{1.039} & \textcolor{red}{1.055} & \textcolor{black}{0.994} & \textcolor{red}{1.000} & \textcolor{red}{1.083} & \textcolor{black}{0.902} & \textcolor{red}{1.123} & \textcolor{black}{0.993} & \textcolor{black}{0.655}\\
$TPV(3)_{shr}$ & \textcolor{black}{\textbf{0.973}} & \textcolor{black}{0.967} & \textcolor{black}{0.972} & \textcolor{black}{0.955} & \textcolor{black}{\textbf{0.965}} & \textcolor{black}{0.963} & \textcolor{black}{\textbf{0.793}} & \textcolor{black}{0.990} & \textcolor{black}{\textbf{0.924}} & \textcolor{black}{\textbf{0.565}}\\
\midrule
$TSV_{bu}$ & \textcolor{red}{1.004} & \textcolor{red}{1.054} & \textcolor{red}{1.048} & \textcolor{black}{0.994} & \textcolor{red}{1.021} & \textcolor{red}{1.076} & \textcolor{black}{0.908} & \textcolor{red}{1.108} & \textcolor{black}{0.994} & \textcolor{black}{0.698}\\
$TSV_{shr}$ & \textcolor{black}{0.987} & \textcolor{red}{1.007} & \textcolor{red}{1.010} & \textcolor{black}{0.987} & \textcolor{black}{0.988} & \textcolor{red}{1.018} & \textcolor{black}{0.861} & \textcolor{red}{1.043} & \textcolor{black}{0.979} & \textcolor{black}{0.646}\\
\bottomrule
\end{tabular}

	\caption{Accuracy of the \textbf{five-day ahead} forecasts. MSE and QLIKE ratios for the DJIA index (panel A), and geometric means of the MSE and QLIKE ratios for individual stocks (panel B) over the benchmark $HAR$ model. Values larger than one in red. The best index value in each column is highlighted in bold.
	}
\end{table}

\begin{table}
	\centering
	\footnotesize
	\setlength{\tabcolsep}{5pt}
	
\begin{tabular}[t]{>{}c|c>{}c|c>{}c|c>{}c|c>{}c|cc}
\toprule
\multicolumn{1}{c}{ } & \multicolumn{2}{c}{2007-2022} & \multicolumn{2}{c}{2007-2010} & \multicolumn{2}{c}{2011-2014} & \multicolumn{2}{c}{2015-2019} & \multicolumn{2}{c}{2020-2022} \\
 & MSE & QLIKE & MSE & QLIKE & MSE & QLIKE & MSE & QLIKE & MSE & QLIKE\\
\midrule
\addlinespace[0.3em]
\multicolumn{11}{c}{\textit{Panel A: DJIA index}}\\
$PV(3)$ & \textcolor{red}{1.030} & \textcolor{red}{1.025} & \textcolor{red}{1.001} & \textcolor{black}{0.988} & \textcolor{red}{1.012} & \textcolor{black}{\textbf{0.995}} & \textcolor{black}{0.962} & \textcolor{red}{1.011} & \textcolor{black}{\textbf{0.969}} & \textcolor{black}{0.954}\\
$PV(3)_{bu}$ & \textcolor{red}{1.010} & \textcolor{red}{1.009} & \textcolor{red}{1.009} & \textcolor{black}{0.988} & \textcolor{red}{1.006} & \textcolor{red}{1.011} & \textcolor{black}{0.596} & \textcolor{red}{1.021} & \textcolor{black}{1.000} & \textcolor{black}{0.460}\\
$PV(3)_{shr}$ & \textcolor{red}{1.018} & \textcolor{red}{1.015} & \textcolor{red}{1.003} & \textcolor{black}{\textbf{0.983}} & \textcolor{red}{1.005} & \textcolor{red}{1.000} & \textcolor{black}{0.596} & \textcolor{red}{1.015} & \textcolor{black}{0.976} & \textcolor{black}{0.465}\\
\midrule
$SV$ & \textcolor{red}{1.002} & \textcolor{red}{1.003} & \textcolor{black}{\textbf{0.998}} & \textcolor{red}{1.003} & \textcolor{red}{1.006} & \textcolor{red}{1.001} & \textcolor{black}{0.928} & \textcolor{black}{\textbf{0.999}} & \textcolor{black}{0.996} & \textcolor{black}{0.904}\\
$SV_{bu}$ & \textcolor{red}{1.007} & \textcolor{red}{1.005} & \textcolor{red}{1.007} & \textcolor{black}{0.991} & \textcolor{red}{\textbf{1.003}} & \textcolor{red}{1.011} & \textcolor{black}{0.748} & \textcolor{red}{1.014} & \textcolor{red}{1.004} & \textcolor{black}{0.662}\\
$SV_{shr}$ & \textcolor{red}{1.002} & \textcolor{red}{1.002} & \textcolor{red}{1.003} & \textcolor{black}{1.000} & \textcolor{red}{1.003} & \textcolor{red}{1.006} & \textcolor{black}{0.728} & \textcolor{red}{1.007} & \textcolor{black}{0.996} & \textcolor{black}{0.637}\\
$TPV(3)_{bu}$ & \textcolor{red}{1.125} & \textcolor{red}{1.103} & \textcolor{red}{1.064} & \textcolor{black}{0.991} & \textcolor{red}{1.037} & \textcolor{red}{1.054} & \textcolor{black}{0.386} & \textcolor{red}{1.106} & \textcolor{red}{1.007} & \textcolor{black}{0.169}\\
\midrule
$TPV(3)_{shr}$ & \textcolor{black}{\textbf{0.914}} & \textcolor{black}{\textbf{0.939}} & \textcolor{red}{1.057} & \textcolor{black}{0.989} & \textcolor{red}{1.012} & \textcolor{black}{0.997} & \textcolor{black}{\textbf{0.374}} & \textcolor{red}{1.032} & \textcolor{black}{0.977} & \textcolor{black}{0.169}\\
$TSV_{bu}$ & \textcolor{red}{1.121} & \textcolor{red}{1.099} & \textcolor{red}{1.055} & \textcolor{black}{0.991} & \textcolor{red}{1.035} & \textcolor{red}{1.047} & \textcolor{black}{0.385} & \textcolor{red}{1.092} & \textcolor{red}{1.006} & \textcolor{black}{0.170}\\
$TSV_{shr}$ & \textcolor{black}{0.949} & \textcolor{black}{0.965} & \textcolor{red}{1.044} & \textcolor{red}{1.005} & \textcolor{red}{1.010} & \textcolor{red}{1.014} & \textcolor{black}{0.379} & \textcolor{red}{1.053} & \textcolor{black}{0.997} & \textcolor{black}{\textbf{0.169}}\\
\addlinespace[0.3em]
\multicolumn{11}{c}{\textit{Panel B: Individual stocks}}\\
$PV(3)$ & \textcolor{red}{1.049} & \textcolor{red}{1.010} & \textcolor{black}{\textbf{0.990}} & \textcolor{black}{\textbf{0.941}} & \textcolor{black}{0.964} & \textcolor{black}{\textbf{0.980}} & \textcolor{red}{1.055} & \textcolor{black}{\textbf{0.976}} & \textcolor{red}{1.016} & \textcolor{red}{1.106}\\
\midrule
$PV(3)_{bu}$ & \textcolor{black}{0.997} & \textcolor{black}{0.995} & \textcolor{red}{1.004} & \textcolor{black}{0.980} & \textcolor{black}{0.997} & \textcolor{red}{1.009} & \textcolor{black}{0.945} & \textcolor{red}{1.008} & \textcolor{black}{0.982} & \textcolor{black}{0.899}\\
$PV(3)_{shr}$ & \textcolor{red}{1.017} & \textcolor{black}{0.994} & \textcolor{black}{0.993} & \textcolor{black}{0.949} & \textcolor{black}{0.970} & \textcolor{black}{0.983} & \textcolor{black}{0.916} & \textcolor{black}{0.988} & \textcolor{black}{0.961} & \textcolor{black}{0.865}\\
$SV$ & \textcolor{red}{1.001} & \textcolor{red}{1.001} & \textcolor{black}{0.990} & \textcolor{red}{1.005} & \textcolor{red}{1.001} & \textcolor{black}{1.000} & \textcolor{black}{0.993} & \textcolor{black}{0.988} & \textcolor{black}{0.997} & \textcolor{black}{0.994}\\
\midrule
$SV_{bu}$ & \textcolor{black}{0.998} & \textcolor{black}{0.998} & \textcolor{red}{1.001} & \textcolor{black}{0.992} & \textcolor{red}{1.001} & \textcolor{red}{1.008} & \textcolor{black}{0.971} & \textcolor{red}{1.003} & \textcolor{black}{0.995} & \textcolor{black}{0.942}\\
$SV_{shr}$ & \textcolor{black}{0.997} & \textcolor{black}{0.998} & \textcolor{black}{0.993} & \textcolor{black}{0.995} & \textcolor{red}{1.000} & \textcolor{red}{1.003} & \textcolor{black}{0.960} & \textcolor{black}{0.995} & \textcolor{black}{0.993} & \textcolor{black}{0.927}\\
$TPV(3)_{bu}$ & \textcolor{red}{1.038} & \textcolor{red}{1.012} & \textcolor{red}{1.053} & \textcolor{black}{0.955} & \textcolor{black}{1.000} & \textcolor{red}{1.058} & \textcolor{black}{0.868} & \textcolor{red}{1.065} & \textcolor{black}{0.978} & \textcolor{black}{0.712}\\
$TPV(3)_{shr}$ & \textcolor{black}{\textbf{0.982}} & \textcolor{black}{\textbf{0.965}} & \textcolor{red}{1.013} & \textcolor{black}{0.941} & \textcolor{black}{\textbf{0.960}} & \textcolor{black}{0.989} & \textcolor{black}{\textbf{0.824}} & \textcolor{black}{0.996} & \textcolor{black}{\textbf{0.956}} & \textcolor{black}{\textbf{0.672}}\\
\midrule
$TSV_{bu}$ & \textcolor{red}{1.043} & \textcolor{red}{1.019} & \textcolor{red}{1.043} & \textcolor{black}{0.960} & \textcolor{red}{1.006} & \textcolor{red}{1.051} & \textcolor{black}{0.875} & \textcolor{red}{1.055} & \textcolor{black}{0.980} & \textcolor{black}{0.730}\\
$TSV_{shr}$ & \textcolor{black}{0.994} & \textcolor{black}{0.984} & \textcolor{red}{1.018} & \textcolor{black}{0.971} & \textcolor{black}{0.990} & \textcolor{red}{1.016} & \textcolor{black}{0.853} & \textcolor{red}{1.023} & \textcolor{black}{0.979} & \textcolor{black}{0.706}\\
\bottomrule
\end{tabular}

	\caption{Accuracy of the \textbf{22-day ahead} forecasts. MSE and QLIKE ratios for the DJIA index (panel A), and geometric means of the MSE and QLIKE ratios for individual stocks (panel B) over the benchmark $HAR$ model. Values larger than one in red. The best index value in each column is highlighted in bold.
	}
\end{table}

\begin{landscape}
	\begin{table}[htb]
		\centering
		\begin{footnotesize}

\begin{tabular}[t]{>{}l|cc>{}c|ccc>{}c|cccc}
\toprule
& $RV$ & $SV$ & $PV(3)$ & $SV_{bu}$ & $TSV_{bu}$ & $PV(3)_{bu}$ & $TPV(3)_{bu}$ & $SV_{shr}$ & $TSV_{shr}$ & $PV(3)_{shr}$ & $TPV(3)_{shr}$\\
\midrule
\addlinespace[0.3em]
\multicolumn{12}{l}{\textit{Panel A: DJIA index}}\\
$MSE$ & 6.352 & 6.527 & 5.184 & 6.099 & 5.236 & 5.867 & 5.219 & 6.202 & 5.523 & 5.293 & \textbf{5.135}\\
$p$-value $dm_{RV}$ & $-$ & 0.785 & 0.168 & 0.142 & 0.119 & 0.075 & 0.121 & 0.119 & 0.118 & 0.066 & 0.074\\
$p$-value $dm_{SV}$ & $-$ & $-$ & 0.131 & 0.148 & 0.112 & 0.079 & 0.114 & 0.115 & 0.110 & \textbf{0.044} & 0.064\\
$p$-value $dm_{PV}$ & $-$ & $-$ & $-$ & 0.800 & 0.526 & 0.759 & 0.517 & 0.821 & 0.659 & 0.579 & 0.467\\
$p$-value MCS & \textbf{0.412} & \textbf{0.299} & \textbf{0.687} & \textbf{0.535} & \textbf{0.559} & \textbf{0.605} & \textbf{0.657} & \textbf{0.496} & \textbf{0.572} & \textbf{0.804} & \textbf{1.000}\\
\addlinespace[0.3em]
$QLIKE$ & 0.216 & 0.210 & 0.491 & 0.218 & 0.235 & 0.217 & 0.236 & 0.212 & 0.219 & \textbf{0.204} & 0.210\\
$p$-value $dm_{RV}$ & $-$ & \textbf{0.004} & 0.908 & 0.692 & 1.000 & 0.528 & 1.000 & 0.176 & 0.698 & \textbf{0.008} & 0.104\\
$p$-value $dm_{SV}$ & $-$ & $-$ & 0.911 & 0.988 & 1.000 & 0.956 & 1.000 & 0.796 & 0.997 & \textbf{0.031} & 0.452\\
$p$-value $dm_{PV}$ & $-$ & $-$ & $-$ & 0.098 & 0.112 & 0.097 & 0.113 & 0.093 & 0.098 & 0.087 & 0.092\\
$p$-value MCS & \textbf{0.279} & 0.119 & \textbf{0.231} & 0.029 & 0.000 & 0.117 & 0.000 & 0.061 & 0.015 & \textbf{1.000} & \textbf{0.361}\\
\midrule
\addlinespace[0.3em]
\multicolumn{12}{l}{\textit{Panel B: Individual stocks}}\\
$\overline{MSE}$ & 39.595 & 40.993 & 36.374 & 37.805 & 58.265 & 36.947 & 55.191 & 38.970 & 45.080 & \textbf{35.553} & 37.942\\
$p$-value $dm_{RV}$ & $-$ & 0 & 0 & 1 & 0 & 2 & 0 & 1 & 2 & 2 & 8\\
$p$-value $dm_{SV}$ & $-$ & $-$ & 0 & 0 & 1 & 1 & 1 & 5 & 3 & 2 & 8\\
$p$-value $dm_{PV}$ & $-$ & $-$ & $-$ & 1 & 0 & 1 & 0 & 2 & 1 & 1 & 2\\
$p$-value MCS & 25 & 23 & 26 & 25 & 23 & 24 & 25 & 23 & 25 & 26 & 26\\
\addlinespace[0.3em]
$\overline{QLIKE}$ & 0.185 & 0.214 & 0.269 & 0.170 & 0.161 & 0.165 & 0.161 & 0.165 & 0.154 & 0.151 & \textbf{0.145}\\
$p$-value $dm_{RV}$ & $-$ & 2 & 1 & 2 & 4 & 4 & 4 & 7 & 6 & 19 & 16\\
$p$-value $dm_{SV}$ & $-$ & $-$ & 1 & 6 & 5 & 7 & 5 & 10 & 7 & 20 & 17\\
$p$-value $dm_{PV}$ & $-$ & $-$ & $-$ & 7 & 7 & 8 & 7 & 8 & 8 & 10 & 8\\
$p$-value MCS & 14 & 14 & 18 & 8 & 4 & 6 & 6 & 10 & 4 & 26 & 13\\
\bottomrule
\end{tabular}
\\[0.5cm]
			\textit{Note}: The table reports the \textbf{one-step ahead} forecasting performance of the different models. The top panel shows the results for the DJIA index. Diebold-Mariano: $p$-values $< 0.05$ are highlighted in bold. MCS: $p$-values $ > 0.2$ are highlighted in bold.
			The bottom panel reports the average loss and 5\% rejection frequency of the Diebold-Mariano tests for each individual stocks.
			The one-sided tests between each forecasting model against $HAR$, $SV$, and $PV(3)$ are denoted by $dm_{HAR}$, $dm_{SV}$, and $dm_{PV(3)}$, respectively.
			MCS denotes the $p$-value of that model being in the Model Confidence Set, or the number of times that model is in the 80\% Model Confidence Set. $PV(3)$ denotes the $HAR-PV(3)$ model with 3 decompositions defined by two thresholds at 10\% and 75\%.
		\end{footnotesize}
		\caption{One-day-ahead forecasting performance: 2007-2022 (3,901 days)}
		\label{Table_2007_2022}
	\end{table}	
\end{landscape}

\begin{landscape}
	\begin{table}[htb]
		\centering
		\begin{footnotesize}

\begin{tabular}[t]{>{}l|cc>{}c|ccc>{}c|cccc}
\toprule
& $RV$ & $SV$ & $PV(3)$ & $SV_{bu}$ & $TSV_{bu}$ & $PV(3)_{bu}$ & $TPV(3)_{bu}$ & $SV_{shr}$ & $TSV_{shr}$ & $PV(3)_{shr}$ & $TPV(3)_{shr}$\\
\midrule
\addlinespace[0.3em]
\multicolumn{12}{l}{\textit{Panel A: DJIA index}}\\
$MSE$ & 5.109 & 5.195 & 5.503 & 4.966 & 4.093 & 4.888 & \textbf{4.064} & 5.064 & 4.487 & 5.185 & 4.724\\
$p$-value $dm_{RV}$ & $-$ & 0.774 & 0.833 & \textbf{0.021} & \textbf{0.002} & \textbf{0.007} & \textbf{0.002} & 0.235 & \textbf{0.000} & 0.613 & \textbf{0.041}\\
$p$-value $dm_{SV}$ & $-$ & $-$ & 0.783 & 0.054 & \textbf{0.002} & \textbf{0.023} & \textbf{0.002} & \textbf{0.049} & \textbf{0.000} & 0.485 & \textbf{0.023}\\
$p$-value $dm_{PV}$ & $-$ & $-$ & $-$ & 0.105 & \textbf{0.010} & 0.077 & \textbf{0.009} & 0.139 & \textbf{0.018} & \textbf{0.027} & \textbf{0.003}\\
$p$-value MCS & 0.194 & 0.194 & \textbf{0.203} & \textbf{0.329} & \textbf{0.308} & \textbf{0.313} & \textbf{1.000} & \textbf{0.227} & \textbf{0.226} & \textbf{0.234} & \textbf{0.358}\\
\addlinespace[0.3em]
$QLIKE$ & 0.930 & 0.927 & 0.908 & 0.220 & 0.231 & 0.219 & 0.233 & 0.217 & 0.219 & \textbf{0.209} & 0.211\\
$p$-value $dm_{RV}$ & $-$ & \textbf{0.000} & 0.441 & \textbf{0.008} & \textbf{0.009} & \textbf{0.008} & \textbf{0.009} & \textbf{0.008} & \textbf{0.008} & \textbf{0.007} & \textbf{0.007}\\
$p$-value $dm_{SV}$ & $-$ & $-$ & 0.449 & \textbf{0.008} & \textbf{0.009} & \textbf{0.008} & \textbf{0.009} & \textbf{0.008} & \textbf{0.008} & \textbf{0.007} & \textbf{0.007}\\
$p$-value $dm_{PV}$ & $-$ & $-$ & $-$ & \textbf{0.007} & \textbf{0.008} & \textbf{0.007} & \textbf{0.008} & \textbf{0.007} & \textbf{0.007} & \textbf{0.006} & \textbf{0.006}\\
$p$-value MCS & \textbf{0.496} & \textbf{0.497} & 0.165 & 0.008 & 0.000 & 0.054 & 0.000 & 0.039 & 0.006 & \textbf{1.000} & \textbf{0.218}\\
\midrule
\addlinespace[0.3em]
\multicolumn{12}{l}{\textit{Panel B: Individual stocks}}\\
$\overline{MSE}$ & 25.109 & 25.371 & 22.936 & 24.804 & 28.938 & 24.433 & 28.072 & 24.923 & 25.730 & \textbf{22.753} & 23.185\\
$p$-value $dm_{RV}$ & $-$ & 0 & 1 & 2 & 3 & 9 & 4 & 4 & 5 & 2 & 7\\
$p$-value $dm_{SV}$ & $-$ & $-$ & 1 & 2 & 4 & 4 & 5 & 4 & 7 & 3 & 5\\
$p$-value $dm_{PV}$ & $-$ & $-$ & $-$ & 0 & 0 & 0 & 0 & 0 & 0 & 0 & 1\\
$p$-value MCS & 25 & 25 & 26 & 25 & 22 & 26 & 26 & 25 & 26 & 26 & 26\\
\addlinespace[0.3em]
$\overline{QLIKE}$ & 0.186 & 0.187 & 0.201 & 0.181 & 0.162 & 0.161 & 0.161 & 0.174 & 0.153 & 0.147 & \textbf{0.141}\\
$p$-value $dm_{RV}$ & $-$ & 6 & 7 & 4 & 5 & 8 & 6 & 9 & 8 & 24 & 22\\
$p$-value $dm_{SV}$ & $-$ & $-$ & 7 & 2 & 3 & 4 & 4 & 3 & 5 & 21 & 19\\
$p$-value $dm_{PV}$ & $-$ & $-$ & $-$ & 2 & 4 & 4 & 4 & 2 & 6 & 8 & 10\\
$p$-value MCS & 11 & 13 & 22 & 11 & 6 & 9 & 7 & 11 & 7 & 23 & 22\\
\bottomrule
\end{tabular}
\\[0.5cm]
			\textit{Note}: The table reports the \textbf{five-step ahead} forecasting performance of the different models. The top panel shows the results for the DJIA index. Diebold-Mariano: $p$-values $< 0.05$ are highlighted in bold. MCS: $p$-values $ > 0.2$ are highlighted in bold.
			The bottom panel reports the average loss and 5\% rejection frequency of the Diebold-Mariano tests for each individual stocks.
			The one-sided tests between each forecasting model against $HAR$, $SV$, and $PV(3)$ are denoted by $dm_{HAR}$, $dm_{SV}$, and $dm_{PV(3)}$, respectively.
			MCS denotes the $p$-value of that model being in the Model Confidence Set, or the number of times that model is in the 80\% Model Confidence Set. $PV(3)$ denotes the $HAR-PV(3)$ model with 3 decompositions defined by two thresholds at 10\% and 75\%.
		\end{footnotesize}
		\caption{Five-day-ahead forecasting performance: 2007-2022 (3,897 days)}
		\label{Table_2007_2022_h5}
	\end{table}	
\end{landscape}

\begin{landscape}
	\begin{table}[htb]
		\centering
		\begin{footnotesize}

\begin{tabular}[t]{>{}l|cc>{}c|ccc>{}c|cccc}
\toprule
& $RV$ & $SV$ & $PV(3)$ & $SV_{bu}$ & $TSV_{bu}$ & $PV(3)_{bu}$ & $TPV(3)_{bu}$ & $SV_{shr}$ & $TSV_{shr}$ & $PV(3)_{shr}$ & $TPV(3)_{shr}$\\
\midrule
\addlinespace[0.3em]
\multicolumn{12}{l}{\textit{Panel A: DJIA index}}\\
$MSE$ & 4.424 & 4.436 & 4.533 & 4.448 & 4.861 & 4.463 & 4.879 & 4.434 & 4.267 & 4.488 & \textbf{4.153}\\
$p$-value $dm_{RV}$ & $-$ & 0.613 & 0.851 & 0.631 & 0.947 & 0.683 & 0.956 & 0.569 & 0.122 & 0.780 & \textbf{0.034}\\
$p$-value $dm_{SV}$ & $-$ & $-$ & 0.843 & 0.575 & 0.948 & 0.646 & 0.956 & 0.478 & 0.079 & 0.765 & \textbf{0.016}\\
$p$-value $dm_{PV}$ & $-$ & $-$ & $-$ & 0.209 & 0.890 & 0.257 & 0.904 & 0.158 & \textbf{0.030} & 0.225 & \textbf{0.005}\\
$p$-value MCS & \textbf{0.332} & \textbf{0.272} & \textbf{0.478} & \textbf{0.336} & \textbf{0.689} & \textbf{0.342} & \textbf{0.632} & \textbf{0.314} & 0.127 & \textbf{0.475} & \textbf{1.000}\\
\addlinespace[0.3em]
$QLIKE$ & 0.973 & 0.902 & 0.936 & 0.727 & 0.375 & 0.580 & 0.376 & 0.708 & 0.368 & 0.580 & \textbf{0.364}\\
$p$-value $dm_{RV}$ & $-$ & 0.160 & 0.150 & \textbf{0.027} & \textbf{0.000} & \textbf{0.002} & \textbf{0.000} & \textbf{0.016} & \textbf{0.000} & \textbf{0.002} & \textbf{0.000}\\
$p$-value $dm_{SV}$ & $-$ & $-$ & 0.830 & \textbf{0.049} & \textbf{0.000} & \textbf{0.004} & \textbf{0.000} & \textbf{0.028} & \textbf{0.000} & \textbf{0.003} & \textbf{0.000}\\
$p$-value $dm_{PV}$ & $-$ & $-$ & $-$ & \textbf{0.031} & \textbf{0.000} & \textbf{0.002} & \textbf{0.000} & \textbf{0.017} & \textbf{0.000} & \textbf{0.002} & \textbf{0.000}\\
$p$-value MCS & \textbf{0.450} & \textbf{0.431} & \textbf{0.396} & \textbf{0.378} & 0.079 & \textbf{0.379} & 0.100 & \textbf{0.380} & 0.001 & \textbf{0.378} & \textbf{1.000}\\
\midrule
\addlinespace[0.3em]
\multicolumn{12}{l}{\textit{Panel B: Individual stocks}}\\
$\overline{MSE}$ & 14.243 & 14.277 & 14.316 & 14.205 & 14.011 & 14.231 & 13.905 & 14.207 & 13.567 & 14.122 & \textbf{13.257}\\
$p$-value $dm_{RV}$ & $-$ & 1 & 2 & 2 & 2 & 4 & 3 & 3 & 5 & 3 & 11\\
$p$-value $dm_{SV}$ & $-$ & $-$ & 2 & 1 & 2 & 2 & 3 & 2 & 5 & 2 & 11\\
$p$-value $dm_{PV}$ & $-$ & $-$ & $-$ & 3 & 1 & 2 & 1 & 3 & 1 & 6 & 12\\
$p$-value MCS & 25 & 26 & 26 & 24 & 23 & 26 & 23 & 25 & 24 & 23 & 25\\
\addlinespace[0.3em]
$\overline{QLIKE}$ & 0.269 & 0.268 & 0.292 & 0.262 & 0.227 & 0.255 & 0.225 & 0.259 & 0.222 & 0.247 & \textbf{0.214}\\
$p$-value $dm_{RV}$ & $-$ & 5 & 6 & 4 & 12 & 10 & 12 & 6 & 13 & 22 & 22\\
$p$-value $dm_{SV}$ & $-$ & $-$ & 5 & 3 & 10 & 6 & 10 & 4 & 13 & 17 & 20\\
$p$-value $dm_{PV}$ & $-$ & $-$ & $-$ & 8 & 14 & 10 & 14 & 9 & 15 & 14 & 15\\
$p$-value MCS & 16 & 19 & 23 & 15 & 7 & 19 & 9 & 17 & 7 & 24 & 25\\
\bottomrule
\end{tabular}
\\[0.5cm]
			\textit{Note}: The table reports the \textbf{22-step ahead} forecasting performance of the different models. The top panel shows the results for the DJIA index. Diebold-Mariano: $p$-values $< 0.05$ are highlighted in bold. MCS: $p$-values $ > 0.2$ are highlighted in bold.
			The bottom panel reports the average loss and 5\% rejection frequency of the Diebold-Mariano tests for each individual stocks.
			The one-sided tests between each forecasting model against $HAR$, $SV$, and $PV(3)$ are denoted by $dm_{HAR}$, $dm_{SV}$, and $dm_{PV(3)}$, respectively.
			MCS denotes the $p$-value of that model being in the Model Confidence Set, or the number of times that model is in the 80\% Model Confidence Set. $PV(3)$ denotes the $HAR-PV(3)$ model with 3 decompositions defined by two thresholds at 10\% and 75\%.
		\end{footnotesize}
		\caption{Twenty-two-day-ahead forecasting performance: 2007-2022 (3,880 days)}
		\label{Table_2007_2022_h22}
	\end{table}	
\end{landscape}

\begin{landscape}
	\begin{table}[htb]
		\centering
		\begin{footnotesize}

\begin{tabular}[t]{>{}l|cc>{}c|ccc>{}c|cccc}
\toprule
& $RV$ & $SV$ & $PV(3)$ & $SV_{bu}$ & $TSV_{bu}$ & $PV(3)_{bu}$ & $TPV(3)_{bu}$ & $SV_{shr}$ & $TSV_{shr}$ & $PV(3)_{shr}$ & $TPV(3)_{shr}$\\
\midrule
\addlinespace[0.3em]
\multicolumn{12}{l}{\textit{Panel A: DJIA index}}\\
$MSE$ & 20.587 & 21.074 & \textbf{15.957} & 19.808 & 16.485 & 18.972 & 16.440 & 20.097 & 17.614 & 16.718 & 16.198\\
$p$-value $dm_{RV}$ & $-$ & 0.718 & 0.162 & 0.197 & 0.131 & 0.107 & 0.134 & 0.159 & 0.136 & 0.077 & 0.089\\
$p$-value $dm_{SV}$ & $-$ & $-$ & 0.135 & 0.211 & 0.132 & 0.121 & 0.134 & 0.174 & 0.137 & 0.060 & 0.084\\
$p$-value $dm_{PV}$ & $-$ & $-$ & $-$ & 0.820 & 0.567 & 0.789 & 0.560 & 0.833 & 0.697 & 0.639 & 0.541\\
$p$-value MCS & \textbf{0.546} & \textbf{0.472} & \textbf{1.000} & \textbf{0.468} & \textbf{0.385} & \textbf{0.581} & \textbf{0.382} & \textbf{0.534} & \textbf{0.468} & \textbf{0.686} & \textbf{0.490}\\
\addlinespace[0.3em]
$QLIKE$ & 0.187 & 0.183 & 0.505 & 0.192 & 0.203 & 0.192 & 0.205 & 0.187 & 0.191 & \textbf{0.176} & 0.181\\
$p$-value $dm_{RV}$ & $-$ & \textbf{0.002} & 0.873 & 0.999 & 1.000 & 0.996 & 1.000 & 0.267 & 0.995 & \textbf{0.001} & \textbf{0.016}\\
$p$-value $dm_{SV}$ & $-$ & $-$ & 0.876 & 1.000 & 1.000 & 1.000 & 1.000 & 0.999 & 1.000 & \textbf{0.024} & 0.231\\
$p$-value $dm_{PV}$ & $-$ & $-$ & $-$ & 0.131 & 0.140 & 0.131 & 0.141 & 0.127 & 0.130 & 0.119 & 0.123\\
$p$-value MCS & 0.020 & \textbf{0.269} & 0.193 & 0.000 & 0.000 & 0.000 & 0.000 & 0.022 & 0.000 & \textbf{1.000} & 0.000\\
\midrule
\addlinespace[0.3em]
\multicolumn{12}{l}{\textit{Panel B: Individual stocks}}\\
$\overline{MSE}$ & 96.815 & 97.020 & 95.505 & 93.645 & 186.003 & 92.522 & 175.154 & 94.130 & 130.389 & \textbf{91.192} & 107.195\\
$p$-value $dm_{RV}$ & $-$ & 0 & 0 & 0 & 0 & 2 & 0 & 0 & 0 & 3 & 5\\
$p$-value $dm_{SV}$ & $-$ & $-$ & 0 & 1 & 1 & 1 & 1 & 2 & 1 & 3 & 4\\
$p$-value $dm_{PV}$ & $-$ & $-$ & $-$ & 1 & 0 & 1 & 0 & 1 & 0 & 1 & 1\\
$p$-value MCS & 26 & 24 & 26 & 23 & 25 & 25 & 24 & 26 & 25 & 26 & 26\\
\addlinespace[0.3em]
$\overline{QLIKE}$ & 0.196 & 0.224 & 0.287 & 0.188 & 0.171 & 0.191 & 0.172 & 0.187 & 0.163 & 0.166 & \textbf{0.156}\\
$p$-value $dm_{RV}$ & $-$ & 4 & 0 & 0 & 1 & 1 & 1 & 3 & 1 & 13 & 9\\
$p$-value $dm_{SV}$ & $-$ & $-$ & 1 & 1 & 0 & 0 & 0 & 3 & 0 & 9 & 5\\
$p$-value $dm_{PV}$ & $-$ & $-$ & $-$ & 4 & 2 & 3 & 2 & 5 & 3 & 6 & 3\\
$p$-value MCS & 15 & 18 & 21 & 10 & 1 & 12 & 3 & 15 & 5 & 26 & 13\\
\bottomrule
\end{tabular}
\\[0.5cm]
			\textit{Note}: The table reports \textbf{the one-step ahead} forecasting performance of the different models. The top panel shows the results for the DJIA index.
			Diebold-Mariano: $p$-values $< 0.05$ are highlighted in bold.
			The bottom panel reports the average loss and 5\% rejection frequency of the Diebold-Mariano tests for each individual stocks.
			The one-sided tests between each forecasting model against $HAR$, $SV$, and $PV(3)$ models are denoted by $dm_{HAR}$, $dm_{SV}$, and $dm_{PV(3)}$, respectively.
			MCS denotes the $p$-value of that model being in the Model Confidence Set, or the number of times that model is in the 80\% Model Confidence Set. PV(3) denotes the HAR-PV(3) model with 3 decompositions defined by two thresholds at 10\% and 75\%.
		\end{footnotesize}
		\caption{Forecasting performance: 2007-2010 (1,008 days).}
		\label{Table_2007_2010}
	\end{table}
\end{landscape}

\begin{landscape}
	\begin{table}[htb]
		\centering
		\begin{footnotesize}

\begin{tabular}[t]{>{}l|cc>{}c|ccc>{}c|cccc}
\toprule
& $RV$ & $SV$ & $PV(3)$ & $SV_{bu}$ & $TSV_{bu}$ & $PV(3)_{bu}$ & $TPV(3)_{bu}$ & $SV_{shr}$ & $TSV_{shr}$ & $PV(3)_{shr}$ & $TPV(3)_{shr}$\\
\midrule
\addlinespace[0.3em]
\multicolumn{12}{l}{\textit{Panel A: DJIA index}}\\
$MSE$ & 0.433 & 0.429 & \textbf{0.399} & 0.457 & 0.498 & 0.449 & 0.497 & 0.442 & 0.462 & 0.409 & 0.429\\
$p$-value $dm_{RV}$ & $-$ & 0.339 & 0.242 & 0.991 & 0.997 & 0.942 & 0.997 & 0.995 & 0.985 & 0.100 & 0.389\\
$p$-value $dm_{SV}$ & $-$ & $-$ & 0.266 & 0.929 & 0.989 & 0.849 & 0.990 & 0.866 & 0.951 & 0.184 & 0.506\\
$p$-value $dm_{PV}$ & $-$ & $-$ & $-$ & 0.860 & 0.940 & 0.834 & 0.940 & 0.803 & 0.867 & 0.621 & 0.769\\
$p$-value MCS & \textbf{0.465} & \textbf{0.633} & \textbf{1.000} & 0.123 & 0.113 & 0.101 & 0.101 & \textbf{0.281} & 0.165 & \textbf{0.668} & \textbf{0.259}\\
\addlinespace[0.3em]
$QLIKE$ & 0.195 & \textbf{0.192} & 0.209 & 0.209 & 0.243 & 0.210 & 0.247 & 0.200 & 0.216 & 0.204 & 0.214\\
$p$-value $dm_{RV}$ & $-$ & \textbf{0.003} & 1.000 & 1.000 & 1.000 & 1.000 & 1.000 & 1.000 & 1.000 & 1.000 & 1.000\\
$p$-value $dm_{SV}$ & $-$ & $-$ & 1.000 & 1.000 & 1.000 & 1.000 & 1.000 & 1.000 & 1.000 & 1.000 & 1.000\\
$p$-value $dm_{PV}$ & $-$ & $-$ & $-$ & 0.423 & 1.000 & 0.555 & 1.000 & \textbf{0.014} & 0.948 & \textbf{0.025} & 0.914\\
$p$-value MCS & 0.009 & \textbf{1.000} & 0.036 & 0.009 & 0.000 & 0.001 & 0.000 & 0.000 & 0.000 & 0.027 & 0.000\\
\midrule
\addlinespace[0.3em]
\multicolumn{12}{l}{\textit{Panel B: Individual stocks}}\\
$\overline{MSE}$ & 1.508 & 1.568 & 1.464 & 1.504 & 1.564 & 1.492 & 1.569 & 1.494 & 1.514 & \textbf{1.404} & 1.445\\
$p$-value $dm_{RV}$ & $-$ & 1 & 4 & 3 & 1 & 5 & 1 & 7 & 2 & 9 & 6\\
$p$-value $dm_{SV}$ & $-$ & $-$ & 3 & 1 & 1 & 1 & 1 & 4 & 0 & 10 & 4\\
$p$-value $dm_{PV}$ & $-$ & $-$ & $-$ & 0 & 0 & 0 & 0 & 0 & 0 & 2 & 1\\
$p$-value MCS & 22 & 20 & 25 & 21 & 8 & 21 & 11 & 21 & 11 & 24 & 13\\
\addlinespace[0.3em]
$\overline{QLIKE}$ & 0.123 & 0.148 & 0.133 & 0.124 & 0.134 & 0.124 & 0.136 & 0.121 & 0.126 & \textbf{0.118} & 0.123\\
$p$-value $dm_{RV}$ & $-$ & 8 & 4 & 1 & 1 & 1 & 1 & 10 & 2 & 14 & 3\\
$p$-value $dm_{SV}$ & $-$ & $-$ & 5 & 3 & 1 & 3 & 1 & 5 & 1 & 9 & 2\\
$p$-value $dm_{PV}$ & $-$ & $-$ & $-$ & 6 & 2 & 6 & 2 & 9 & 5 & 14 & 6\\
$p$-value MCS & 15 & 16 & 17 & 9 & 0 & 8 & 1 & 15 & 1 & 21 & 6\\
\bottomrule
\end{tabular}
\\[0.5cm]
			\textit{Note}: The table reports the \textbf{one-step ahead} forecasting performance of the different models. The top panel shows the results for the DJIA index.
			Diebold-Mariano $p$-values $< 0.05$ are highlighted in bold.
			The bottom panel reports the average loss and 5\% rejection frequency of the Diebold-Mariano tests for each individual stocks.
			The one-sided tests between each forecasting model against RV-HAR, SV-HAR, and PV(3)-HAR are denoted by $dm_{HAR}$, $dm_{SV}$, and $dm_{PV(3)}$, respectively.
			MCS denotes the $p$-value of that model being in the Model Confidence Set, or the number of times that model is in the 80\% Model Confidence Set. PV(3) denotes the HAR-PV(3) model with 3 decompositions defined by two thresholds at 10\% and 75\%.
		\end{footnotesize}
		\caption{Forecasting performance: 2011-2014 (1,006 days).}
		\label{Table_2011_2014}
	\end{table}
\end{landscape}

\begin{landscape}
	\begin{table}[htb]
		\centering
		\begin{footnotesize}

\begin{tabular}[t]{>{}l|cc>{}c|ccc>{}c|cccc}
\toprule
& $RV$ & $SV$ & $PV(3)$ & $SV_{bu}$ & $TSV_{bu}$ & $PV(3)_{bu}$ & $TPV(3)_{bu}$ & $SV_{shr}$ & $TSV_{shr}$ & $PV(3)_{shr}$ & $TPV(3)_{shr}$\\
\midrule
\addlinespace[0.3em]
\multicolumn{12}{l}{\textit{Panel A: DJIA index}}\\
$MSE$ & 0.901 & 0.963 & 0.899 & 0.871 & 0.824 & 0.854 & 0.822 & 0.898 & 0.839 & 0.841 & \textbf{0.811}\\
$p$-value $dm_{RV}$ & $-$ & 0.830 & 0.485 & 0.145 & 0.093 & 0.113 & 0.091 & 0.384 & 0.062 & 0.103 & 0.050\\
$p$-value $dm_{SV}$ & $-$ & $-$ & 0.294 & 0.155 & 0.122 & 0.141 & 0.121 & 0.139 & 0.106 & 0.124 & 0.090\\
$p$-value $dm_{PV}$ & $-$ & $-$ & $-$ & 0.324 & 0.124 & 0.216 & 0.114 & 0.498 & 0.164 & 0.056 & \textbf{0.038}\\
$p$-value MCS & \textbf{0.606} & \textbf{0.545} & \textbf{0.560} & \textbf{0.617} & \textbf{0.290} & \textbf{0.349} & \textbf{0.220} & \textbf{0.592} & \textbf{0.830} & \textbf{0.620} & \textbf{1.000}\\
\addlinespace[0.3em]
$QLIKE$ & 0.242 & 0.234 & \textbf{0.211} & 0.247 & 0.260 & 0.243 & 0.260 & 0.239 & 0.245 & 0.218 & 0.228\\
$p$-value $dm_{RV}$ & $-$ & \textbf{0.012} & \textbf{0.008} & 0.866 & 0.997 & 0.597 & 0.999 & \textbf{0.023} & 0.956 & \textbf{0.001} & \textbf{0.010}\\
$p$-value $dm_{SV}$ & $-$ & $-$ & \textbf{0.024} & 0.951 & 0.996 & 0.893 & 0.998 & 0.977 & 0.998 & \textbf{0.006} & 0.112\\
$p$-value $dm_{PV}$ & $-$ & $-$ & $-$ & 0.987 & 0.998 & 0.984 & 0.999 & 0.986 & 0.996 & 0.875 & 0.985\\
$p$-value MCS & 0.024 & 0.037 & \textbf{1.000} & 0.053 & 0.008 & 0.089 & 0.003 & 0.062 & 0.003 & \textbf{0.230} & 0.034\\
\midrule
\addlinespace[0.3em]
\multicolumn{12}{l}{\textit{Panel B: Individual stocks}}\\
$\overline{MSE}$ & 28.790 & 28.277 & \textbf{13.400} & 25.753 & 16.027 & 24.144 & 15.524 & 26.345 & 19.594 & 17.863 & 16.205\\
$p$-value $dm_{RV}$ & $-$ & 2 & 1 & 0 & 1 & 3 & 2 & 4 & 0 & 9 & 12\\
$p$-value $dm_{SV}$ & $-$ & $-$ & 3 & 1 & 0 & 1 & 1 & 2 & 1 & 7 & 9\\
$p$-value $dm_{PV}$ & $-$ & $-$ & $-$ & 0 & 0 & 0 & 0 & 0 & 0 & 0 & 0\\
$p$-value MCS & 21 & 22 & 26 & 17 & 19 & 16 & 18 & 22 & 20 & 26 & 25\\
\addlinespace[0.3em]
$\overline{QLIKE}$ & 0.183 & 0.239 & 0.345 & 0.184 & 0.185 & 0.179 & 0.185 & 0.179 & 0.178 & \textbf{0.159} & 0.163\\
$p$-value $dm_{RV}$ & $-$ & 2 & 5 & 1 & 1 & 7 & 2 & 5 & 7 & 25 & 23\\
$p$-value $dm_{SV}$ & $-$ & $-$ & 7 & 2 & 2 & 4 & 2 & 5 & 5 & 24 & 19\\
$p$-value $dm_{PV}$ & $-$ & $-$ & $-$ & 3 & 3 & 3 & 3 & 4 & 3 & 7 & 6\\
$p$-value MCS & 1 & 5 & 20 & 3 & 4 & 3 & 3 & 4 & 2 & 24 & 9\\
\bottomrule
\end{tabular}
\\[0.5cm]
			\textit{Note}: The table reports the \textbf{one-step ahead} forecasting performance of the different models. The top panel shows the results for the DJIA index.
			Diebold-Mariano $p$-values $< 0.05$ are highlighted in bold.
			The bottom panel reports the average loss and 5\% rejection frequency of the Diebold-Mariano tests for each individual stocks.
			The one-sided tests between each forecasting model against RV-HAR, SV-HAR, and PV(3)-HAR are denoted by $dm_{HAR}$, $dm_{SV}$, and $dm_{PV(3)}$, respectively.
			MCS denotes the $p$-value of that model being in the Model Confidence Set, or the number of times that model is in the 80\% Model Confidence Set. PV(3) denotes the HAR-PV(3) model with 3 decompositions defined by two thresholds at 10\% and 75\%.
		\end{footnotesize}
		\caption{Forecasting performance: 2015-2019 (1,258 days).}
		\label{Table_2015_2019}
	\end{table}
\end{landscape}

\begin{landscape}
	\begin{table}[htb]
		\centering
		\begin{footnotesize}

\begin{tabular}[t]{>{}l|cc>{}c|ccc>{}c|cccc}
\toprule
& $RV$ & $SV$ & $PV(3)$ & $SV_{bu}$ & $TSV_{bu}$ & $PV(3)_{bu}$ & $TPV(3)_{bu}$ & $SV_{shr}$ & $TSV_{shr}$ & $PV(3)_{shr}$ & $TPV(3)_{shr}$\\
\midrule
\addlinespace[0.3em]
\multicolumn{12}{l}{\textit{Panel A: DJIA index}}\\
$MSE$ & 3.910 & 4.096 & 4.141 & 3.610 & 3.608 & \textbf{3.561} & 3.583 & 3.752 & 3.607 & 3.702 & 3.578\\
$p$-value $dm_{RV}$ & $-$ & 0.838 & 0.854 & \textbf{0.012} & 0.065 & \textbf{0.011} & 0.084 & 0.076 & \textbf{0.004} & 0.057 & \textbf{0.005}\\
$p$-value $dm_{SV}$ & $-$ & $-$ & 0.605 & \textbf{0.016} & \textbf{0.038} & \textbf{0.013} & \textbf{0.044} & \textbf{0.015} & \textbf{0.015} & \textbf{0.019} & \textbf{0.012}\\
$p$-value $dm_{PV}$ & $-$ & $-$ & $-$ & \textbf{0.017} & \textbf{0.041} & \textbf{0.012} & \textbf{0.047} & \textbf{0.028} & \textbf{0.015} & \textbf{0.009} & \textbf{0.009}\\
$p$-value MCS & \textbf{0.485} & \textbf{0.464} & 0.192 & \textbf{0.498} & \textbf{0.723} & \textbf{1.000} & \textbf{0.919} & \textbf{0.404} & \textbf{0.671} & \textbf{0.424} & \textbf{0.615}\\
\addlinespace[0.3em]
$QLIKE$ & 0.245 & 0.236 & 1.482 & 0.220 & 0.222 & 0.214 & 0.219 & 0.220 & \textbf{0.214} & 0.221 & 0.214\\
$p$-value $dm_{RV}$ & $-$ & 0.206 & 0.847 & 0.160 & 0.210 & 0.128 & 0.187 & 0.153 & 0.126 & 0.176 & 0.130\\
$p$-value $dm_{SV}$ & $-$ & $-$ & 0.847 & 0.133 & 0.220 & 0.090 & 0.184 & 0.120 & 0.090 & 0.158 & 0.095\\
$p$-value $dm_{PV}$ & $-$ & $-$ & $-$ & 0.153 & 0.154 & 0.152 & 0.153 & 0.153 & 0.152 & 0.153 & 0.152\\
$p$-value MCS & \textbf{0.758} & \textbf{0.684} & \textbf{0.360} & \textbf{0.680} & \textbf{0.561} & \textbf{0.982} & \textbf{0.587} & \textbf{0.802} & \textbf{1.000} & \textbf{0.522} & \textbf{0.870}\\
\midrule
\addlinespace[0.3em]
\multicolumn{12}{l}{\textit{Panel B: Individual stocks}}\\
$\overline{MSE}$ & 30.422 & 39.697 & 43.397 & 30.482 & 28.721 & 30.197 & \textbf{28.037} & 35.764 & 29.017 & 36.386 & 28.804\\
$p$-value $dm_{RV}$ & $-$ & 0 & 0 & 1 & 1 & 3 & 1 & 0 & 2 & 0 & 5\\
$p$-value $dm_{SV}$ & $-$ & $-$ & 1 & 3 & 1 & 4 & 1 & 3 & 2 & 2 & 3\\
$p$-value $dm_{PV}$ & $-$ & $-$ & $-$ & 1 & 1 & 1 & 1 & 2 & 2 & 2 & 2\\
$p$-value MCS & 24 & 26 & 25 & 25 & 25 & 25 & 24 & 26 & 26 & 24 & 26\\
\addlinespace[0.3em]
$\overline{QLIKE}$ & 0.269 & 0.256 & 0.305 & 0.186 & 0.138 & 0.163 & 0.135 & 0.172 & 0.133 & 0.162 & \textbf{0.127}\\
$p$-value $dm_{RV}$ & $-$ & 0 & 4 & 2 & 2 & 3 & 2 & 1 & 3 & 10 & 10\\
$p$-value $dm_{SV}$ & $-$ & $-$ & 9 & 6 & 4 & 7 & 4 & 6 & 7 & 13 & 13\\
$p$-value $dm_{PV}$ & $-$ & $-$ & $-$ & 3 & 4 & 5 & 4 & 5 & 5 & 9 & 8\\
$p$-value MCS & 22 & 20 & 23 & 22 & 18 & 18 & 21 & 20 & 19 & 21 & 24\\
\bottomrule
\end{tabular}
\\[0.5cm]
			\textit{Note}: The table reports the \textbf{one-step ahead} forecasting performance of the different models. The top panel shows the results for the DJIA index.
			Diebold-Mariano $p$-values $< 0.05$ are highlighted in bold.
			The bottom panel reports the average loss and 5\% rejection frequency of the Diebold-Mariano tests for each individual stocks.
			The one-sided tests between each forecasting model against RV-HAR, SV-HAR, and PV(3)-HAR are denoted by $dm_{HAR}$, $dm_{SV}$, and $dm_{PV(3)}$, respectively.
			MCS denotes the $p$-value of that model being in the Model Confidence Set, or the number of times that model is in the 80\% Model Confidence Set. PV(3) denotes the HAR-PV(3) model with 3 decompositions defined by two thresholds at 10\% and 75\%.
		\end{footnotesize}
		\caption{Forecasting performance: 2020-2022 (629 days).}
		\label{Table_2020_2022}
	\end{table}
\end{landscape}

\begin{figure}[H]
	\centering
	\includegraphics[width=\linewidth]{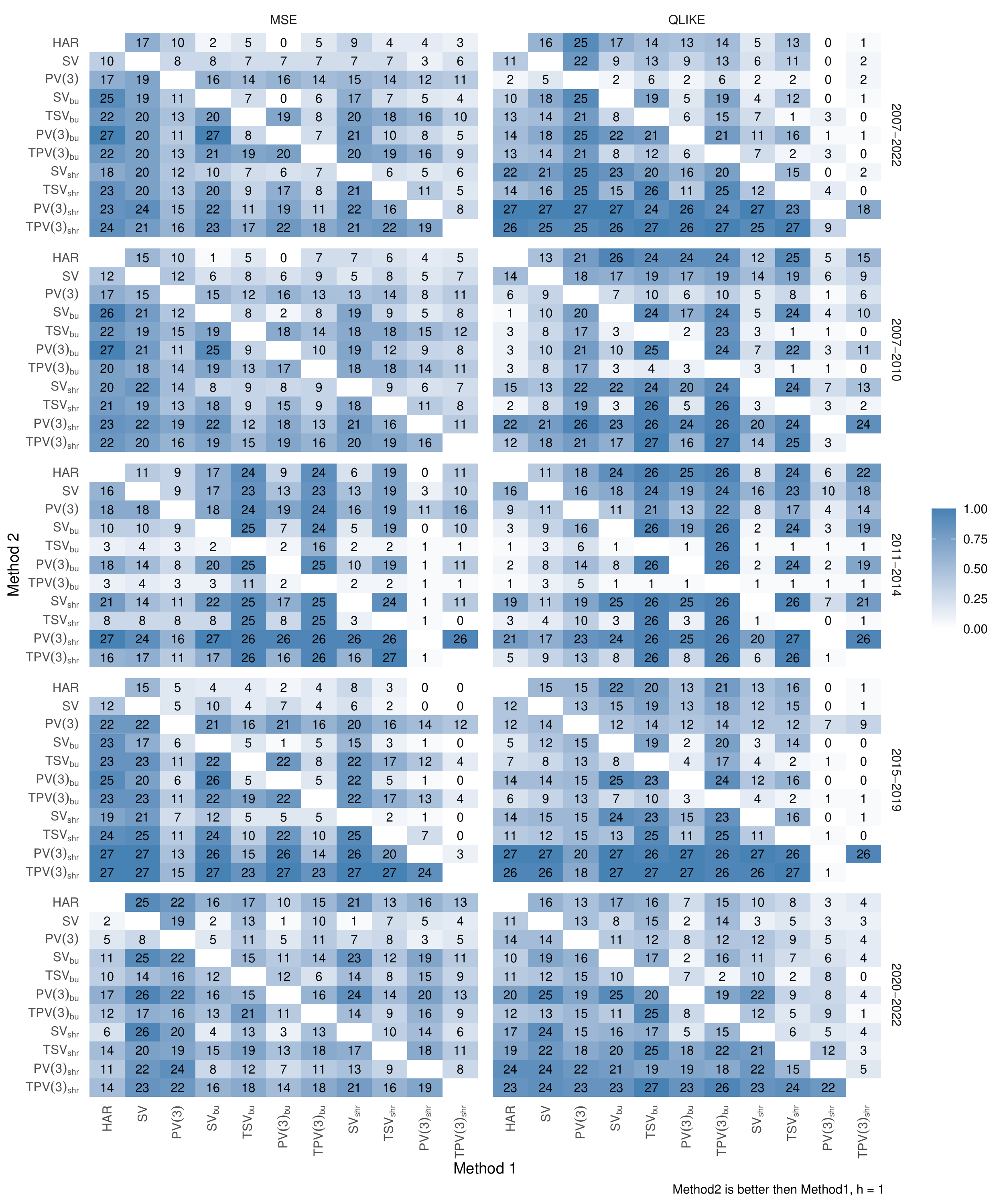}
	\caption{Qualitative evaluation of the \textbf{one-step ahead} forecasting accuracy. Each cell reports the number of times the forecasting model in the row outperforms the model in the column. Different test periods, from the top: 2007-2010, 2011-2014, 2015-2019, 2020-2022.}
\end{figure}

\begin{figure}[H]
	\centering
	\includegraphics[width=\linewidth]{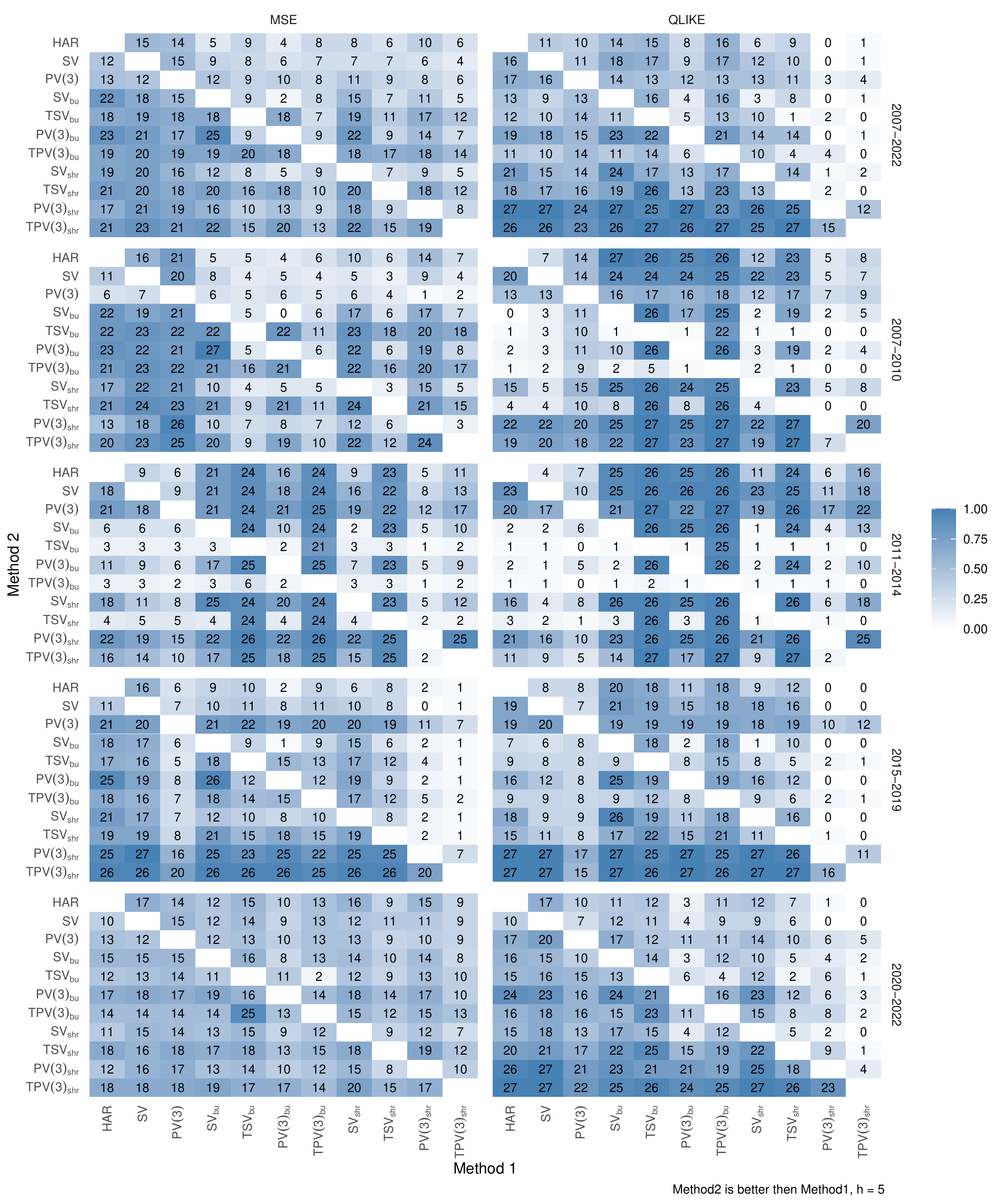}
	\caption{Qualitative evaluation of the \textbf{five-step ahead} forecasting accuracy. Each cell reports the number of times the forecasting model in the row outperforms the model in the column. Different test periods, from the top: 2007-2010, 2011-2014, 2015-2019, 2020-2022.}
\end{figure}

\begin{figure}[H]
	\centering
	\includegraphics[width=\linewidth]{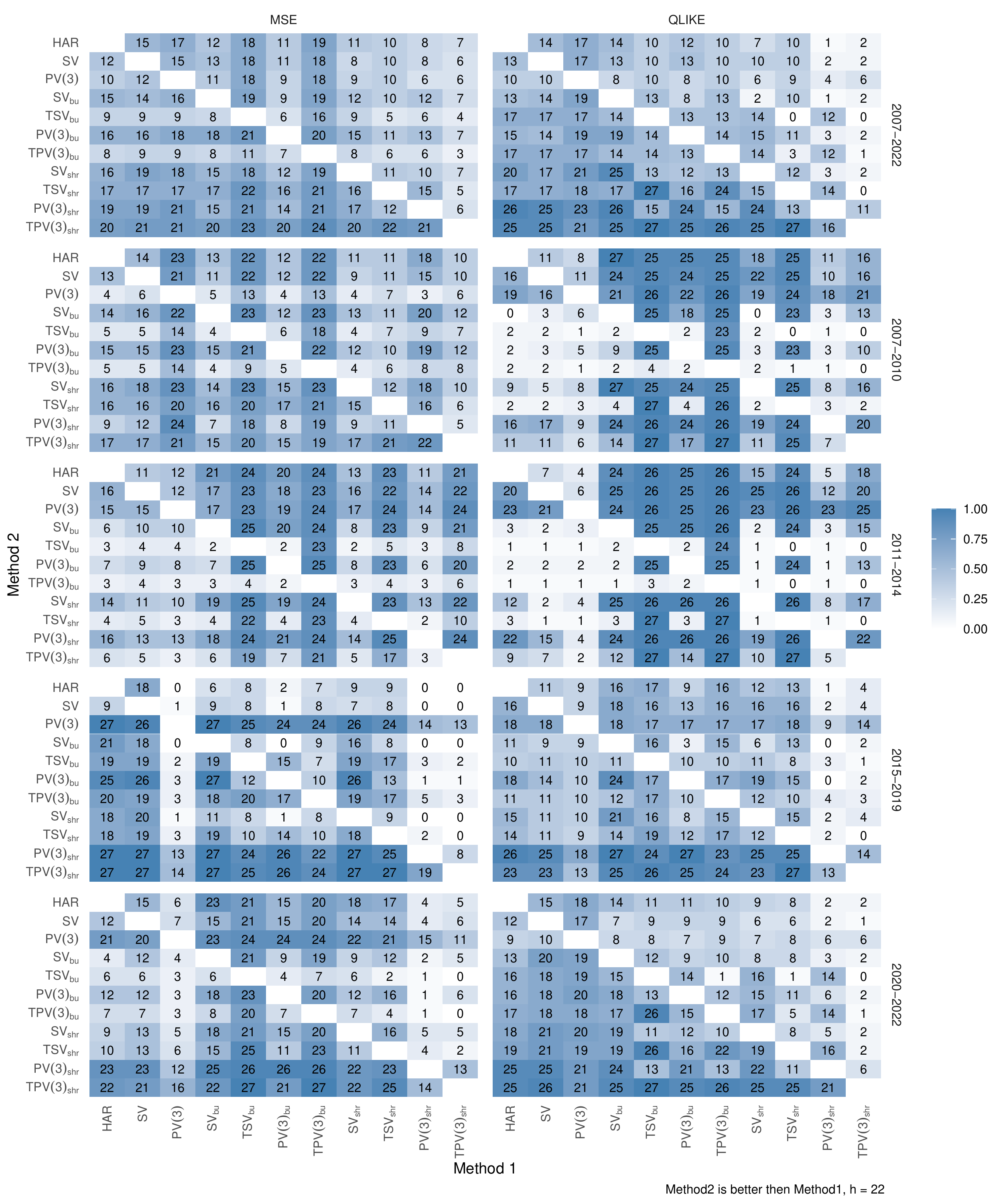}
	\caption{Qualitative evaluation of the \textbf{22-step ahead} forecasting accuracy. Each cell reports the number of times the forecasting model in the row outperforms the model in the column. Different test periods, from the top: 2007-2010, 2011-2014, 2015-2019, 2020-2022.}
\end{figure}